\documentclass[12pt,a4paper,openright]{report}
\usepackage{amsmath}
\usepackage{subcaption}
\usepackage{amsmath}
\usepackage{mathtools}
\usepackage{amsfonts}
\usepackage{amssymb}
\usepackage{fontenc}
\usepackage{makecell}
\usepackage[T1]{fontenc}
\usepackage{pifont}
\usepackage[utf8]{inputenc}
\usepackage[frenchb]{babel}
\usepackage{lipsum}
\usepackage{bbold}
\usepackage{pdfpages}
\usepackage{setspace}
\usepackage{float}
\usepackage{url}
\usepackage{wasysym}
\usepackage{tcolorbox}
\usepackage{pst-node}
\usepackage{fancybox}
\usepackage{lmodern}
\usepackage{graphicx,wrapfig}
\usepackage{titlesec}
\usepackage{sidecap}
\usepackage[tikz]{bclogo}
\usepackage{varwidth}
\usepackage[Conny]{fncychap} 
\usepackage{epigraph}
\usepackage[vlined,algoruled]{algorithm2e}
\usepackage{lipsum}
\usepackage{graphicx}
\usepackage{caption}
\usepackage{listings}
\usepackage{lipsum}
\usepackage{hyperref}
\hypersetup{
	colorlinks=true,
	linkcolor=blue,
	filecolor=magenta,      
	urlcolor=cyan,
	pdftitle={Overleaf Example},
	pdfpagemode=FullScreen,
	citecolor=blue
}
\usepackage[toc]{appendix}
\usepackage[titles]{tocloft}
\usepackage{amsmath}
\usepackage{tcolorbox}
\usepackage{braket}
\usetikzlibrary{calc}
\usepackage{minitoc}
\usepackage{braket}
\usepackage[twoside,left=1.7cm,right=1.7cm,top=1.5cm,bottom=1.5cm]{geometry}
\mtcsetrules{*}{off} 
\mtcsettitle{minitoc}{\vrule height0ptwidth0pt depth15pt\relax Sommaire}
\usepackage{silence} 
\usepackage{setspace} 
\usepackage{fancyhdr}
\usepackage{enumitem}
\usepackage{inputenc}
\usepackage[T1]{fontenc}
\makeatletter
\newcommand*{\rom}[1]{\expandafter\@slowromancap\romannumeral #1@}
\makeatother
\usepackage{fancyhdr} 

\newcounter{subsubsubsection}[subsubsection]
\renewcommand\thesubsubsubsection{\thesubsubsection\arabic{subsubsubsection}}
\titleformat{\subsubsubsection}
{\normalfont\normalsize\bfseries}{\thesubsubsubsection}{1em}{}
\titlespacing*{\subsubsubsection}
{0pt}{3.25ex plus 1ex minus .2ex}{1.5ex plus .2ex}
\makeatletter
\renewcommand\paragraph{\@startsection{paragraph}{5}{\z@}%
	{3.25ex \@plus1ex \@minus.2ex}%
	{-1em}%
	{\normalfont\normalsize\bfseries}}
\renewcommand\subparagraph{\@startsection{subparagraph}{6}{\parindent}%
	{3.25ex \@plus1ex \@minus .2ex}%
	{-1em}%
	{\normalfont\normalsize\bfseries}}
\def\toclevel@subsubsubsection{4}
\def\toclevel@paragraph{5}
\def\toclevel@paragraph{6}
\def\l@subsubsubsection{\@dottedtocline{4}{7em}{4em}}
\def\l@paragraph{\@dottedtocline{5}{10em}{5em}}
\def\l@subparagraph{\@dottedtocline{6}{14em}{6em}}
\makeatother

\newcounter{example}[section]

\usepackage{amsmath}
\makeatletter
\renewcommand{\section}{\@startsection {section}{1}{\z@}%
	{-4.5ex \@plus -1ex \@minus -.2ex}%
	{2.3ex \@plus.2ex}%
	{\normalfont\normalsize\sffamily\bfseries}}
\makeatother
\makeatletter
\renewcommand{\subsection}{\@startsection {subsection}{1}{\z@}%
	{-4.5ex \@plus -1ex \@minus -.2ex}%
	{2.3ex \@plus.2ex}%
	{\normalfont\normalsize\sffamily\bfseries}}
\makeatother

\newcolumntype{R}[1]{>{\raggedleft\arraybackslash }b{#1}}
\newcolumntype{L}[1]{>{\raggedright\arraybackslash }b{#1}}
\newcolumntype{C}[1]{>{\centering\arraybackslash }b{#1}}
\usepackage{enumitem}
\newlist{abbrv}{itemize}{1}
\setlist[abbrv,1]{label=,labelwidth=1in,align=parleft,itemsep=0.1\baselineskip,leftmargin=!}
\usepackage{longtable}

\newcommand{\chaptertoc}[1]{\chapter*{#1}
	\addcontentsline{toc}{chapter}{#1}
	\markboth{\slshape\MakeUppercase{#1}}{\slshape\MakeUppercase{#1}}}
\providecommand{\openone}{\leavevmode\hbox{\small1\kern0pt\normalsize1}}
\makeatletter
\patchcmd{\@makechapterhead}{\vspace*{50\p@}}{\vspace*{25\p@}}{}{}
\patchcmd{\@makeschapterhead}{\vspace*{50\p@}}{\vspace*{25\p@}}{}{}
\makeatother

\usepackage{lettrine}
\usepackage{blindtext} 
\usepackage{epigraph}
\usepackage{xintexpr}

\title{Epigraph example}
\author{Overleaf}

\usepackage{transparent}
\usepackage{eso-pic}
\usepackage{tikz}
\usetikzlibrary{calc}

\AddToShipoutPicture*{
	\unitlength=0.5cm
	\put(2,1){
		\parbox[b][\paperheight]{\paperwidth}{%
			\vfill
			\centering
			{\transparent{10}\includegraphics[width=1.5\textwidth]{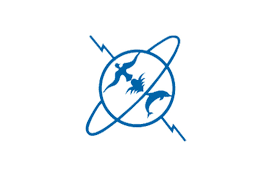}}
			
		}
	}
}
\usepackage[object=vectorian]{pgfornament} 
\usepackage{lipsum,tikz}

\dominitoc
\begin{document}

	\begin{titlepage}
		\begin{tikzpicture}[overlay,remember picture]
			\draw [line width=2pt,rounded corners=7pt]
			($ (current page.north west) + (.2cm,-.2cm) $)
			rectangle
			($ (current page.south east) + (-.2cm,.2cm) $);
			\draw [line width=1pt,rounded corners=7pt]
			($ (current page.north west) + (.3cm,-.3cm) $)
			rectangle
			($ (current page.south east) + (-.3cm,.3cm) $);
		\end{tikzpicture}
		
		\begin{figure}
			\begin{center}
				\includegraphics[width=0.6\linewidth]{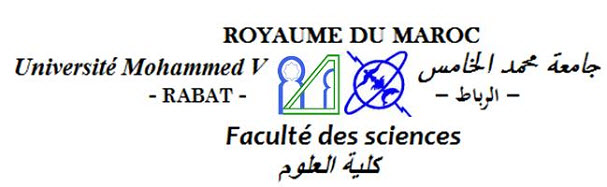}
			\end{center}		
		\end{figure}
		\vspace{-2.2cm} 	
		\begin{center}
			\begin{minipage}{18cm}
				\begin{center}
					{\textcolor{blue}{\fontfamily{pnc}{\selectfont
								{ \textit{CENTRE D’ETUDES DOCTORALES - SCIENCES ET TECHNOLOGIES}}
								\vskip .2cm
								\noindent\hrule height 2pt\vskip 0.2ex\nobreak 
								{\textcolor{green}{
										\noindent\hrule height 2pt \vskip 0.2ex}}
					}}}
				\end{center}
				\begin{flushright}
					\textbf{N° d'ordre:} 3838
				\end{flushright}
			\end{minipage}
			\end{center}
		\begin{center}
			{\Huge \textbf{THÈSE}}\\

			En vue de l’obtention du : {\large \textbf{\textit{DOCTORAT}}}
		\end{center}
	
		\begin{minipage}{18cm}
			
			{\textcolor{black}{\fontfamily{pnc}{\selectfont
	{\textbf{Structure de Recherche} : {{Physique des Hautes Énergies- Modélisation et Simulation}}\\
	\textbf{Discipline} :{ {Physique}}\\
	\textbf{Spécialité} : {{Physique mathématique et Information  quantique  }}}}
					\vspace{0.2cm}
					{\textcolor{blue}{
							\noindent\hrule height 2pt\vskip 0.2ex\nobreak} 
						{\textcolor{green}{
								\noindent\hrule height 2pt \vskip 0.2ex}}
			}}}
		\end{minipage}
		
		\begin{minipage}{17cm}
			\vspace{0.15cm}
			\begin{center}
				\textbf{Présentée et soutenue le : 23/09/2023 par :}
				\normalsize  \\ \vspace{0.03cm}
				\large \textbf{{\Large \underline{Hamid SAKHOUF}}}
			\end{center}
		\end{minipage}
		\vspace{0.28cm}
		\begin{center}
			\begin{minipage}{17cm}
				\begin{center}
					{\small \textbf{Quantum information processing with superconducting circuits:\\
								\small  realizing and  characterizing  quantum gates and algorithms in open quantum systems}}
				\end{center}
			\end{minipage}
		\end{center}
	
		\begin{center}
			\large \textbf{JURY}
		\end{center}
	\vspace{-1cm}
		\begin{center}
			\begin{tabular}{l l l c}
    \hspace{-1cm}	{\textbf{Mohammed  Loulidi}}& &{PES},{Université Mohammed \rom{5}, Faculté } &  {Président/Rapporteur}\\
	\vspace*{0.12cm}
		\hspace{-1cm}	{\textbf{}}& {} &{des sciences-Rabat.} &  {}\\
	\vspace*{0.12cm}
	\hspace{-1cm}	{\textbf{Lahoucine Bahmad}}& &{PES},{Université Mohammed \rom{5}, Faculté} &\hspace{0.3cm} {Rapporteur/Examinateur}\\	
	\vspace*{0.12cm}
	\hspace{-1cm}	{\textbf{}}& {} &{des sciences-Rabat.} &  {}\\
	\vspace*{0.12cm}
	\hspace{-1cm}	{\textbf{Abderrahim El Allati }}& &{PH},{Université Abdelmalek Essaadi, Faculté} &  {Rapporteur/Examinateur}\\
	\vspace*{0.12cm}
	\hspace{-1cm}	{\textbf{}}& {} &{ des sciences et  techniques- Al-Hoceima} &  {}\\
	\vspace*{0.12cm}
	\hspace{-1cm}	{\textbf{ Mohammed DAOUD}}& &{PES},{Université Ibn Tofail, Faculté } &\hspace{0.4cm} {Directeur de thèse}\\
	\vspace*{0.12cm}
	\hspace{-1cm}	{\textbf{}}& {} &{des sciences-Kénitra.} &  {}\\
	\vspace*{0.12cm}
	\hspace{-1cm}	
		{\textbf{Rachid Ahl Laamara }}& &{PES},{Université Mohammed \rom{5}, Faculté } &  {Directeur de thèse}\\
			\hspace{-1cm}	{\textbf{}}& {} &{ des Sciences-Rabat.} &  {}\\
			\end{tabular}
		\end{center}

		\begin{center}
			\begin{minipage}{15cm}
				\begin{center}
					{\textcolor{black}{\fontfamily{pnc}{\selectfont
								{ Année Universitaire : {2022/2023}}
								\vspace{0.2cm}
								\noindent\hrule height 1.79pt\vskip 0.2ex\nobreak
					}}}
				\end{center}
			\end{minipage}
		
			{\large \XBox} Faculté des Sciences, avenue Ibn Battouta, BP. 1014 RP, Rabat –Maroc\\
			\phone \hspace*{0.2cm}00212(0) 37 77 18 76, \hspace*{0.1cm} {\large \bell}Fax:\hspace*{0.1cm} 00212(0) 37 77 42 61 ; http://www.fsr.um5.ac.ma
		\end{center}
	\end{titlepage}
\newgeometry{left=2cm,right=2cm,top=2.5cm,bottom=2.5cm}
	\pagenumbering{roman} 
	\chaptertoc{Dedication}
	\vspace{-2cm}

\textit{First and  formost, I want  to express my deep appreciation to my family. I am immensely grateful to my parents, siblings, and extended family for their unwavering support, encouragement, and motivation throughout my academic journey. Their belief in me and their constant presence in my life were the driving forces behind the completion of this thesis.}

\textit{In addition,  I would also like to thank all members, past and present, of the LPHE-MS, with whom I have had the pleasure to interact, for the many valuable discussions and shared moments.}
\newpage
\chaptertoc{Acknowledgment }
\vspace{-2cm}

\textit{During the preparation of this thesis, many things changed in my life, both personally and scientifically. Before diving into the content of this thesis, I would like to express my gratitude and thanks to those who have been present in this period of my life and have contributed directly or indirectly to this final work.}

\textit{First, I would like to thank \textbf{Mr. El Hassan SAIDI} Professor at the Faculty of Science of Rabat for giving me the opportunity to enroll in a Master's degree in mathematical physics, which in turn allowed me to enroll in the Ph.D. program at the Laboratory of High Energy Physics, Modelisation, and Simulation (LPHE-MS). I appreciate their advice and encouragement throughout this work. My sincere thanks for their availability as well as the efforts they have always made for the success of several activities of the LPHE-MS.}

\textit{In addition to the director of LPHE-MS laboratory, I would like to  warmly thank my supervisor \textbf{Mr. Rachid AHL LAAMARA}, professor at the Faculty of Sciences of Rabat  for welcoming me to the LPHE-MS. His continuous encouragement gave me the strength and courage to complete this thesis as well as for his teaching of the group theory course and Lie algebra course.}

\textit{Additionally,  I would like to warmly thank my supervisor \textbf{Mr. Mohammed DAOUD} Professor at faculty of science of Kenitra for his magnificent guidance throughout my graduate studies. Without him, this thesis would never have been possible. I appreciate his continuous support and guidance with formidable motivation and support in my studies for my Master’s and Ph.D.}

\textit{My regards also go to \textbf{Mr. Mohammed  LOULIDI}, Professor at the Faculty of Science of Rabat, who accepted the presidency of the jury also agreeing to be a reporter of my modest work. I also acknowledge him for his help, discussion, encouragement, and motivation.}

\textit{Let me also thank the rapporteurs of the thesis : I will start with \textbf{Mr. Lahoucine BAHMAD}, Professor at the Faculty of Science of Rabat.  I would like to thank him warmly for  accepting  to be a reporter of my modest work. Yours sincerely.}

\textit{My acknowledgment also goes to \textbf{Mr.  Abderrahim  El ALLATI}, PH, Faculty of Sciences and Techniques Al-Hoceima, for accepting to be the reporter of my thesis.  I thank him for his interest and responsibility as an examiner of this work and  for his bearing the hardships of traveling from Al-Hoceima to Rabat to be with the jury members.}

\newpage

\chaptertoc{Résumé}
	\vspace{-3cm}	\lettrine[lines=2]C {ette} thèse se concentre sur le traitement de l'information quantique à l'aide d'un dispositif supraconducteur, en particulier sur la réalisation de portes quantiques et d'algorithmes dans des systèmes quantiques ouverts. Un tel dispositif est construit par des qubits supraconducteurs de type transmon couplés à un résonateur supraconducteur. Pour la réalisation des portes quantiques et des algorithmes, une approche en une seule  étape est utilisée. Nous proposons des schémas plus rapides et plus efficaces pour réaliser des portes de $X$-rotation  et des portes d'enchevêtrement pour deux et trois qubits. Au cours de ces opérations, le nombre de photons du résonateur est annulé en raison du fort champ de micro-ondes ajouté. Elles ne nécessitent pas que le résonateur soit initialement préparé dans l'état de vide et les schémas est insensible à la désintégration du résonateur.  En outre, la robustesse de ces opérations est démontrée en incluant l'effet de la décohérence des systèmes transmon et la désintégration du résonateur dans une équation maîtresse, ce qui permet d'obtenir une grande fidélité dans la simulation quantique.  En outre, en utilisant les portes de $X$-rotation  mises en œuvre ainsi que les portes de phase, nous présentons une autre façon de mettre en œuvre l'algorithme de Grover pour deux et trois qubits qui ne nécessite pas une série de portes simples. En outre, nous démontrons en simulant numériquement l'utilisation de la tomographie des processus quantiques pour caractériser pleinement la performance d'une porte d'enchevêtrement à un seul coup pour deux et trois qubits et nous obtenons des fidélités de processus supérieures à 93$\%$.  Ces portes sont utilisées pour créer des états intriqués de Bell et de Greenberger-Horne-Zeilinger (GHZ).\\
	\underline{\bf Mots clés:} Traitement de l'information et calcul quantique, Portes quantiques, Algorithme de recherche de Grover, Portes d'intrication, Etats de Bell et GHZ, circuits supraconducteurs, Haute fidélité.\\
	
\newpage
\chaptertoc{Abstract}
\vspace{-3cm}
	\lettrine[lines=2]T{his}    thesis focuses on quantum information processing using the superconducting device, especially, on realizing quantum gates and algorithms in open quantum systems. Such a  device is constructed by transmon-type superconducting qubits coupled to a superconducting resonator. For the realization of quantum gates and algorithms, a one-step approach is used. We suggest faster and more efficient schemes for realizing     $X$-rotation and entangling gates for two and three qubits. During these operations, the resonator photon number is canceled owing to the strong microwave field added. They do not require the resonator to be initially prepared in the vacuum state and the scheme is insensitive to resonator decay.  Furthermore, the robustness of these operations is demonstrated by including the effect of the  decoherence of transmon systems and the resonator decay in a master equation, and as a result   high fidelity will be achieved in quantum simulation.  In addition,  using the implemented x-rotation gates as well as the phase gates, we present an alternative way for implementing Grover’s algorithm for two and three qubits, which does not require a series of single gates. As well, we demonstrate by a numerical simulation the use of quantum process tomography to fully characterize the performance of a single-shot entangling gate for two and three qubits and obtaining  process fidelities greater than 93$\%$.  These gates are used to create   Bell and Greenberger-Horne-Zeilinger (GHZ) entangled states.    
	
	\underline{\bf Keywords:} Quantum  information  processing  and computation, Quantum  gates, Grover's search  algorithm, Entangling  gates, Bell and GHZ states, High  fidelity, superconducting  circuits.\\ 

\newpage
\chaptertoc{Résumé  détaillé}
\vspace{-3cm}

Cette thèse se concentre sur la réalisation et la
caractérisation de portes quantiques et d'algorithmes à
l'aide d'une approche en une seule étape basée sur un
dispositif quantique supraconducteur. Un tel dispositif
quantique est construit en couplant des qubits
supraconducteurs de type transmon à un résonateur
supraconducteur, piloté par un champ classique. À l'aide
de ce dispositif, nous proposons un schéma rapide pour
préparer un registre quantique comprenant de nombreux
états de base en superposition quantique en réalisant la
porte de rotation X pour deux qubits ou plus, ne
nécessitant qu'une seule opération. De plus, en utilisant la
même approche, nous réalisons la porte d'intrication en
une seule étape pour deux qubits ou plus. Ces portes sont
utilisées pour créer un état totalement intriqué tel qu'un
état de Bell à deux qubits et un état GHZ à trois qubits.
En utilisant une simulation strictement numérique basée
sur des hypothèses réalistes concernant les paramètres du
système, nous étudions la dynamique quantique de ces
portes et démontrons que ces schémas peuvent être mis
en œuvre avec une grande fidélité.

De plus, en utilisant les portes de rotation X mises
en œuvre, nous présentons une alternative pour mettre en
œuvre l'algorithme de Grover à deux et trois qubits, qui
ne nécessite pas une série de portes individuelles. Notre
implémentation a montré que le schéma proposé permet
de trouver efficacement l'état correct avec une fidélité
élevée d'environ 90$\%$. Ces schémas présentent les
avantages suivants : (i) ils ne nécessitent pas la
combinaison des portes de base, par exemple les portes
individuelles et à deux qubits, (ii) le résonateur n'a pas
besoin d'être initialement préparé dans l'état de vide, (iii)
nos schémas sont insensibles à la dégradation de la cavité, ce qui facilite les expériences pratiques. De plus,
nous démontrons, en simulant numériquement
l'utilisation de la Tomographie du Processus Quantique
(TPQ) pour caractériser pleinement les performances
d'une porte d'interaction en une seule étape pour deux
qubits ou plus, que nous obtenons des fidélités de
processus supérieures à 93$\%$.

Dans le premier chapitre, nous discuterons
brièvement des bases du traitement et du calcul de
l'information quantique. Cela comprend les aspects
mathématiques et géométriques des bits quantiques, ainsi
que la description de la façon dont ils peuvent être
manipulés et de la différence entre un ordinateur
quantique et son équivalent classique. Nous identifierons
également un ensemble universel de portes logiques
quantiques, comprenant des portes individuelles et multi-
qubits, ainsi que leur contribution à la construction
d'algorithmes quantiques, tels que les algorithmes de
Grover et Deutsch-Jozsa, les protocoles d'intrication
quantique et la préparation d'états de superposition.
Dans le deuxième chapitre, nous aborderons la mise
en œuvre physique des bits quantiques en utilisant des
circuits électriques pour réaliser des qubits
supraconducteurs. Nous passerons en revue certains
types de qubits tels que la boîte à paires de Cooper (qubit
de charge) et le qubit transmon qui est utilisé dans cette
thèse, ainsi que l'oscillateur LC harmonique quantique en
tant que composant essentiel dans électrodynamique
quantique en circuit(circuit QED). Nous décrirons
également l'interaction entre les qubits supraconducteurs
et la ligne de transmission supraconductrice ou
l'oscillateur LC (le champ électromagnétique) à
l'intérieur de la cavité de la QED des circuits, qui est une version similaire de l’électrodynamique quantique en
cavité (cavité QED) classique qui sera également
abordée. De plus, nous discuterons du régime dispersif
nécessaire pour le traitement de l'information quantique.
Enfin, nous terminerons ce chapitre en associant la
circuit QED au traitement de l'information quantique,
où nous identifierons les portes à qubit unique, ainsi que
les portes C-Phase et iSWAP.
Le chapitre III de cette thèse vise à réaliser la porte
de rotation X pour deux et trois qubits, qui peut être
utilisée pour préparer des états de superposition et
d'autres portes appelées portes d'intrication, utilisées pour
générer des états totalement intriqués tels que les états de
Bell et GHZ. Leur mise en œuvre est basée sur un
dispositif quantique comprenant des qubits transmon
couplés capacitivement à un résonateur, piloté par un
champ micro-onde, en utilisant une approche en une
seule étape. La robustesse de ces portes est démontrée en
prenant en compte l'effet de la décohérence des systèmes
transmon et de la dégradation du résonateur dans un
formalisme d'équation maîtresse. Pour des hypothèses
réalistes concernant les paramètres du système transmon-
résonateur, une grande fidélité sera obtenue lors de la
simulation quantique. Au cours de ces opérations, le
nombre de photonsdu résonateur est annulé en raison du
fort champ de micro-ondes ajouté. Elles ne nécessitent
pas que le résonateur soit initialement préparé dans l’état
de vide et les schémas est insensible à la désintégration
du
résonateur. Enfin, nous discuterons de la faisabilité
expérimentale des schémas proposés et les comparerons
aux derniers résultats expérimentaux dans le domaine des
dispositifs supraconducteurs.

\underline{\bf Mots clés:} Traitement de l'information et calcul quantique, Portes quantiques, Algorithme de recherche de Grover, Portes d'intrication, Etats de Bell et GHZ, circuits supraconducteurs, Haute fidélité.\\
 \newgeometry{left=2cm,right=2cm,top=2cm,bottom=2cm}

	\chaptertoc{List of publications}
	\section*{{\large This thesis is based in part on the following published articles:}}
	\vspace{0.7cm}
	\begin*{}{}
		\makeatletter
	\providecommand \@ifxundefined [1]{%
		\@ifx{#1\undefined}
	}%
	\providecommand \@ifnum [1]{%
		\ifnum #1\expandafter \@firstoftwo
		\else \expandafter \@secondoftwo
		\fi
	}%
	\providecommand \@ifx [1]{%
		\ifx #1\expandafter \@firstoftwo
		\else \expandafter \@secondoftwo
		\fi
	}%
	\providecommand \natexlab [1]{#1}%
	\providecommand \enquote  [1]{``#1''}%
	\providecommand \bibnamefont  [1]{#1}%
	\providecommand \bibfnamefont [1]{#1}%
	\providecommand \citenamefont [1]{#1}%
	\providecommand \href@noop [0]{\@secondoftwo}%
	\providecommand \href [0]{\begingroup \@sanitize@url \@href}%
	\providecommand \@href[1]{\@@startlink{#1}\@@href}%
	\providecommand \@@href[1]{\endgroup#1\@@endlink}%
	\providecommand \@sanitize@url [0]{\catcode `\\12\catcode `\$12\catcode
		`\&12\catcode `\#12\catcode `\^12\catcode `\_12\catcode `\%12\relax}%
	\providecommand \@@startlink[1]{}%
	\providecommand \@@endlink[0]{}%
	\providecommand \url  [0]{\begingroup\@sanitize@url \@url }%
	\providecommand \@url [1]{\endgroup\@href {#1}{\urlprefix }}%
	\providecommand \urlprefix  [0]{URL }%
	\providecommand \Eprint [0]{\href }%
	\providecommand \doibase [0]{http://dx.doi.org/}%
	\providecommand \selectlanguage [0]{\@gobble}%
	\providecommand \bibinfo  [0]{\@secondoftwo}%
	\providecommand \bibfield  [0]{\@secondoftwo}%
	\providecommand \translation [1]{[#1]}%
	\providecommand \BibitemOpen [0]{}%
	\providecommand \bibitemStop [0]{}%
	\providecommand \bibitemNoStop [0]{.\EOS\space}%
	\providecommand \EOS [0]{\spacefactor3000\relax}%
	\providecommand \BibitemShut  [1]{\csname bibitem#1\endcsname}%
	\let\auto@bib@innerbib\@empty
	\BibitemOpen
	\bibfield  {author} {\bibinfo {author} $\bullet$ { \bibnamefont{Hamid Sakhouf, }\
	\bibfnamefont {Mohammed Daoud,} \bibfnamefont {Rachid Ahl Laamara ,}}\ }\bibfield  {title} {\enquote {\bibinfo {title} {Quantum process tomography of the single-shot entangling gate with superconducting qubits},}\ }\href {\doibase 10.1088/1361-6455/acc916} {\bibfield  {journal} {\bibinfo  {journal} {Journal of Physics B: Atomic, Molecular and Optical Physics}\ }\textbf {\bibinfo {volume} {56}},\ \bibinfo {pages} {105501} (\bibinfo {year} {2023})}\BibitemShut
	{NoStop}%
	
	\vspace{0.5cm}
	\bibfield  {author} {\bibinfo {author} $\bullet$ { \bibnamefont{Hamid Sakhouf, }\
			\bibfnamefont {Mohammed Daoud,} \bibfnamefont {Rachid Ahl Laamara ,}}\ }\bibfield  {title} {\enquote {\bibinfo {title} {Simple scheme for implementing the Grover search algorithm with superconducting qubits},}\ }\href {\doibase 10.1088/1361-6455/ac24ad} {\bibfield  {journal} {\bibinfo  {journal} {Journal of Physics B: Atomic, Molecular and Optical Physics}\ }\textbf {\bibinfo {volume} {54}},\ \bibinfo {pages} {175501} (\bibinfo {year} {2021})}\BibitemShut
	{NoStop}%
	
	\vspace{0.5cm}
	\bibfield  {author} {\bibinfo {author} $\bullet$ { \bibnamefont{Hamid Sakhouf, }\
			\bibfnamefont {Mohammed Daoud,} \bibfnamefont {Rachid Ahl Laamara ,}}\ }\bibfield  {title} {\enquote {\bibinfo {title} {Implementation of Grover’s Search Algorithm in the QED Circuit for Two Superconducting Qubits},}\ }\href {\doibase 10.1007/s10773-020-04602-1} {\bibfield  {journal} {\bibinfo  {journal} {International Journal of Theoretical Physics}\ }\textbf {\bibinfo {volume} {59}},\ \bibinfo {pages} {3436–3448} (\bibinfo {year} {2020})}\BibitemShut
	{NoStop}%
\end*{}

		\vspace{0.7cm}
		\begin*{}{}
		\makeatletter
		\providecommand \@ifxundefined [1]{%
			\@ifx{#1\undefined}
		}%
		\providecommand \@ifnum [1]{%
			\ifnum #1\expandafter \@firstoftwo
			\else \expandafter \@secondoftwo
			\fi
		}%
		\providecommand \@ifx [1]{%
			\ifx #1\expandafter \@firstoftwo
			\else \expandafter \@secondoftwo
			\fi
		}%
		\providecommand \natexlab [1]{#1}%
		\providecommand \enquote  [1]{``#1''}%
		\providecommand \bibnamefont  [1]{#1}%
		\providecommand \bibfnamefont [1]{#1}%
		\providecommand \citenamefont [1]{#1}%
		\providecommand \href@noop [0]{\@secondoftwo}%
		\providecommand \href [0]{\begingroup \@sanitize@url \@href}%
		\providecommand \@href[1]{\@@startlink{#1}\@@href}%
		\providecommand \@@href[1]{\endgroup#1\@@endlink}%
		\providecommand \@sanitize@url [0]{\catcode `\\12\catcode `\$12\catcode
			`\&12\catcode `\#12\catcode `\^12\catcode `\_12\catcode `\%12\relax}%
		\providecommand \@@startlink[1]{}%
		\providecommand \@@endlink[0]{}%
		\providecommand \url  [0]{\begingroup\@sanitize@url \@url }%
		\providecommand \@url [1]{\endgroup\@href {#1}{\urlprefix }}%
		\providecommand \urlprefix  [0]{URL }%
		\providecommand \Eprint [0]{\href }%
		\providecommand \doibase [0]{http://dx.doi.org/}%
		\providecommand \selectlanguage [0]{\@gobble}%
		\providecommand \bibinfo  [0]{\@secondoftwo}%
		\providecommand \bibfield  [0]{\@secondoftwo}%
		\providecommand \translation [1]{[#1]}%
		\providecommand \BibitemOpen [0]{}%
		\providecommand \bibitemStop [0]{}%
		\providecommand \bibitemNoStop [0]{.\EOS\space}%
		\providecommand \EOS [0]{\spacefactor3000\relax}%
		\providecommand \BibitemShut  [1]{\csname bibitem#1\endcsname}%
		\let\auto@bib@innerbib\@empty
		\end*{}

	 \newgeometry{left=1.7cm,right=1.7cm,top=2.5cm,bottom=2.5cm}
	 \listoffigures
	 \addstarredchapter{Liste of figures}
	 \adjustmtc

	 \newgeometry{left=1.7cm,right=1.7cm,top=2.5cm,bottom=2.5cm}
	\chaptertoc{{List of Abbreviations}}
	\renewcommand*{\arraystretch}{1.37}
	\begin{longtable}{@{}l @{\hspace{5mm}} l }
		QIP   \hspace{3cm}              &Quantum information processing\\
		NMR  \hspace{3cm}              &nuclear magnetic resonance\\
		QED   \hspace{3cm}             &quantum electrodynamics\\
		CP or C-phase   \hspace{3cm}             &conditional phase \\
		GHZ  \hspace{3cm}             &Greenberger-Horne-Zeilinger\\
		CNOT   \hspace{3cm}             &Controlled Not \\
		QPT    \hspace{3cm}             &Quantum process tomography\\
		SQUID    \hspace{3cm}             &superconducting quantum interference device\\
		RWA     \hspace{3cm}            &rotating wave approximation\\
		
	\end{longtable}

	\chaptertoc{{List of Symbols}}
			\vspace{-2cm}
	\renewcommand*{\arraystretch}{1.37}
	\begin{longtable}{@{}l @{\hspace{5mm}} l }
	iSWAP &\hspace{3cm} i-swap gate.\\	
	X, Y, Z& \hspace{3cm}	single-qubit Pauli operator, also defned as ${\sigma _{x,y,z}}$.\\
	$\sqrt {iSWAP} $ & \hspace{3cm} square-root of i-swap gate.\\	
	${R_i}\left( \theta  \right)$ & \hspace{3cm} rotation around the axis $i$ by angle $\theta $.\\
	$H$ & \hspace{3cm} Hadamard gate.\\
	$Toffoli$ & \hspace{3cm} controlled-controlled-X gate.\\
	$Fredkin$ & \hspace{3cm} controlled SWAP operation.\\
	${I_c}$ & \hspace{3cm} the critical current of the junction\\
	${\Phi _0}$& \hspace{3cm}  the magnetic flux quantum\\
	$L$& \hspace{3cm} the Lagrangian of a Josephson junction\\
	${E_C}$ &\hspace{3cm} the charging energy of a Cooper pair\\
	${E_J}$&\hspace{3cm} the Josephson tunneling energy\\
	$p$&\hspace{3cm} the canonical momentum operator \\
	${U_T}$&\hspace{3cm}  the potential energy of the SQUID\\
	${C_J}$&\hspace{3cm} capacitance of the   Josephson junction \\
		$\alpha $& \hspace{3cm} anharmonicity of transmon qubit\\

	${\omega _q}$ & \hspace{3cm} the qubit transition frequency\\
	$g$& \hspace{3cm} vacuum Rabi coupling frequency\\
	${\omega _r}$& \hspace{3cm} frequency of the resonator\\
	${T_1}$& \hspace{3cm} qubit relaxation time\\
	${T_2}$& \hspace{3cm} qubit dephasing time\\
	$ {\Omega _R}$& \hspace{3cm} Rabi  frequency\\
	${\omega _d}$ &\hspace{3cm} microwave feld frequency\\
    $\bigotimes, \quad \bigoplus$& \hspace{3cm}  tensor product, direct sum\\
	
	\end{longtable}
		\newgeometry{left=1.7cm,right=1.7cm,top=2.5cm,bottom=2.5cm}

		\renewcommand\contentsname{Contents}
		\tableofcontents
		\addstarredchapter{Contents} 
		\adjustmtc
		
	\thispagestyle{empty}
	\newpage\pagenumbering{arabic} 
	\renewcommand{\thechapter}{\Roman{chapter}}
	\setcounter{chapter}{-1}
		 \newgeometry{left=1.7cm,right=1.8cm,top=3.05cm,bottom=3.05cm}

	\chapter{General Introduction}
\epigraph{\textit{\textbf{{ "Computers are physical objects, and computations are physical processes.
				What computers can or cannot compute is determined by the laws of physics
				alone, and not by pure mathematics."}}}}{\textit{David Deutsch}}
	Quantum information is an emerging field that includes several specialties such as physics, engineering, chemistry, computer science, and mathematics. The main aim of this integration is to know how the realization of a quantum computer and the use of such a computer to perform specific calculations are much faster than with a computer operating in a standard way (classical computer)\cite{1,2}. It captures all of the operations regarding quantum information processing exploiting the properties of quantum mechanics such as the superposition of states, entanglement, and interference.
	
	The important general   idea of quantum information processing (QIP)   is to  realize  a quantum  device-  so-called a quantum computer  which  has started just about the last century,  exactly in the early 1980s when Feynman showed that a classical machine would be unable to efficiently simulate a quantum system\cite{3},  where the idea is that a classical bit can be replaced by a two-level quantum system.  Similar to a classical machine which can store its information in strings of bits that can either be  $0$ or $1$,  a quantum computer uses a unit of quantum information called a   quantum bit or qubit,
	which is a two-level quantum system.  Still, unlike in classical physics, the qubit can exist in a superposition of its computational basis states $\left| 0 \right\rangle $ and $\left| 1 \right\rangle $ which can be formed by the ground and excited states of an electron in an atom for  example, which  represents the phenomenon of quantum superposition. Some examples mentioned in the physical realization are spin, photons, ion traps, quantum dots, nuclear magnetic resonance (NMR), superconducting qubits, etc.
	
	Quantum information processing and computation is an application of quantum information that includes quantum gates, quantum algorithms, quantum error-correcting codes, quantum cryptography, etc,  and represents a revolutionary field of information processing. The first simple quantum algorithm using quantum mechanics to solve essentially the   determining problem of whether a coin is  fair or biased more efficiently than any classical  algorithm  was proposed by David Deutsch in 1985\cite{4}. In addition,  Peter Shor 1994  invented an efficient quantum algorithm that factorizes a large integer number\cite{5}, which is exponentially more efficient than its classical counterpart in terms of time.    Another famous algorithm worth mentioning in this area was proposed by Lov Grover in 1996\cite{6}, addresses searching in an unsorted large database which is a hard mathematical problem in the classical approach. The computation time of this search  algorithm equals
	the square root of the fastest classical algorithm time.
	
In the following,  in the context of $QIP$ the quantum algorithms have been demonstrated with many physical systems. For instance, the Deutsch-Jozsa algorithm has  been   realized in NMR\cite{7,8,9,10,11,12}, trapped ion\cite{13}, photonic\cite{14} systems, and superconducting devices\cite{15,16,17}, as well as for  further  challenging Shor's algorithm has only been demonstrated for the factorization of the number $15$ 
with NMR techniques\cite{18} and photons \cite{19,20,21}. In the last decades,  following the context of the building blocks of a quantum device, the experimental realization of quantum error correction protocols has been made in NMR\cite{22}, and then in linear optics \cite{23}, trapped ions\cite{24,25}, and also superconducting circuits\cite{26,27,28}.
	
	Another interesting quantum algorithm in terms of physical realizations and considered a good candidate for demonstrating the power of quantum information processing and computation in view of its being easy to implement is Grover's  algorithm. Grover's algorithm  is known as a search  algorithm,  which  is a quantum  method that is  used to  search  through  unstructured 
	databases using iteration steps that are fewer than classical algorithms to determine the unique input to a black box function (quantum Oracle)  by increasing its probability that it will be measured. This algorithm  has been theoretically and experimentally  realized with  its two- and  multi-qubit case using  the  NMR system\cite{29,30},  trapped ions\cite{31,32,33,34},  cavity quantum electrodynamics (cavity QED)\cite{35,36,37,38}. Recently,   Grover’s search algorithm was experimentally realized using a scalable trapped atomic ion system for three qubits and it was executed for one iteration\cite{39}. 
	
	In  these physical systems such as NMR,  trapped ions, and photonic, the   properties of the qubits are naturally occurring. In contrast,  the superconducting qubits are made up of artificial electronic structures that consist of several circuit components, which are capacitors, inductors, and the Josephson junction characterized by its inductance and capacitance. Different of these elements allow the designing and engineering of their properties to a large extent, such as the energy level structure and the coupling of the qubits to their environment.  Whereas some of these properties can be tuned using magnetic fields. Several types of superconducting qubits were proposed in different configurations using Josephson junctions such  as  flux qubit\cite{40}, phase qubit\cite{41}, charge qubit\cite{42}, as well as the transmon qubit as an interesting case in this thesis\cite{43}, which is a modified version of the charge qubit.

	It is  worth noticing that the experimental realization for two-qubit  Grover's algorithm has been demonstrated using a quantum device\cite{16}, including transmon-type superconducting qubits coupled to a quantum electrodynamics circuit (circuit  QED).   Circuit  QED setup  as proposed by Blais et al. \cite{44} and then realized by   Wallraff et  al. in the experiment\cite{45}, studying    the interaction between single microwave photons and superconducting qubits  behaving as artificial atoms, and a similar version  of cavity quantum  electrodynamics (cavity QED), which  is the fundamental  coupling between a photon  and  a two-level system. Additionally,  many  experimental realizations  have been successfully  made with  superconducting  qubits in  QED circuit in  the context  of QIP: demonstration of a  single qubit gate\cite{46}, generation and measurement of two- and multi-qubit entanglement\cite{48,49,50,51,52},  realizing multi-qubit quantum error correction\cite{26,27,28}. 
	
Entanglement is the most interesting aspect of quantum theory\cite{66}, and is   an essential resource for QIP. Therefore,  the creation and demonstration of  maximally entangled states are central for quantum communication and QIP, in particular, a few examples that require entangling quantum systems in Bell and Greenberger-Horne-Zeilinger (GHZ) states, quantum computation\cite{1,67},  quantum error correction\cite{68,69}, quantum sense\cite{70}, and quantum metrology\cite{71}. In the past decade, significant progress  has been witnessed in entanglement    engineering with superconducting devices: the preparation and measurement of two-qubit entangled state(Bell  states) based on  the combination of  single rotation and  controlled-phase gates, was realized by  L. DiCarlo et  al.  in the experiment\cite{16}. In addition,   the preparation of W- and GHZ-maximally-entangled states for three and more qubits is based on the decomposition into sequences of a single rotation and two-qubit controlled phase or iSWAP gates which were realized in the experiments\cite{48,49,50}.
	
	To fully characterize the performance    of these quantum  algorithms and gates,   it is necessary to use
	quantum process tomography (QPT)\cite{1, 72,73}.  QPT is an essential tool for reliable QIP, it allows us to predict the evolution of a quantum state as it propagates through an imperfect quantum gate. In real experiments, where the interaction between the quantum system and the environment cannot be neglected, it is not possible to describe the evolution of a quantum state by a unitary operator.  Thus, QPT is a useful method for experimentally describing a complete implementation of quantum gates. This method is used to characterize the systems of one- and two-qubit quantum gates in\cite{74,75,76,77,78,79,80}. QPT  has been used with superconducting systems for the two-qubit entangling gate to characterize a square root i-SWAP gate\cite{62}, and for three-qubit operations to characterize Toffoli and C-Phase gates\cite{63,26}.
	
	Although the building blocks of quantum algorithms are
	constructed  using  basic quantum  logic gates as single- and two-qubit  gates\cite{81,82,83,84,85,86}, however, building a multiqubit gate using only quantum basic gates will be difficult owing to the number of basic gates drastically increasing with the growing number of qubits.   Therefore,  the use of the direct implementation of a multiqubit gate offers more efficiency than the combination of basic gates. In addition,   the realization of a multi-qubit gate using only a one-step operation replaces the complex combination of one- and two-qubit gates. Over the past years, using the direct implementation approach of the multiqubit gate has drawn much interest. In this context,  single-shot implementation of a Toffoli gate of three qubits has been experimentally demonstrated in different physical systems\cite{87,88,90}. In addition, based on superconducting qubits coupled to a resonator in circuit QED, many schemes have been previously proposed for the one-step realization of multi-qubit gates\cite{91,92,93,94,95,96,97,98,99,100,101,102}.
	\subsection*{Outline of the thesis}
	
	This thesis focuses on the realization,  and characterization of quantum gates and algorithms using a one-step approach based on the superconducting quantum device.  Such a quantum device is constructed by transmon-type superconducting qubits coupled to a superconducting resonator, driven by a classical field.  Using  it,    we propose a fast scheme to prepare a quantum register  involving many of the basic states in quantum superposition by  realizing  
	the $X$-rotation gate for two and more qubits requiring only one-step operation.  In addition,  using the same approach we realize the single-shot entangling gate for two and more qubits,  these gates are used to create a maximally entangled state such as a two-qubit Bell-state and a three-qubit GHZ-state.  Using strictly numerical simulation under realistic assumptions about system parameters, we investigate the quantum dynamics of these gates and show that these schemes can be implemented with high fidelity. 
	
	Using the implemented x-rotation gates,   we present an alternative way to implement the two- and three-qubit Grover’s algorithm, which does not require a series of single gates.  Our implementation showed that the proposed scheme allows  for efficiently finding the correct state with a high $\succ 90\%$ fidelity. As well,  we demonstrate by numerically simulating the use of QPT to fully characterize the performance of a single-shot  entangling gate for two and more qubits and obtaining   process fidelity greater than  93$\%$.
	
	The structure of this thesis is organized as follows:
	
	In the first chapter \ref{Ch. 1}, we will  briefly  discuss some of the basics of quantum information processing and  computing.  This includes the mathematical and geometric properties  of quantum bits as well as the  description of how they can be manipulated and how a quantum computer differs from its classical counterpart. We will also identify a universal set of quantum logic gates,  involving single- and multi-qubit gates, and how they contribute to building quantum algorithms, for instance, the Grover and Deutsch  Jozsa algorithms,  quantum entanglement protocols, and the preparation of  superposition states.
	
	In the second chapter \ref{Ch. 2}, we  will  discuss the physical implementation of quantum bits using electrical circuits to realize superconducting qubits, and we  will  review some types of these qubits such as the Cooper-pair box   (charge qubit), and the transmon qubit which is used in this thesis, as well as the quantum harmonic $LC$ oscillator as an essential  component in circuit  QED. There will also be  a description about the interaction  between   superconducting  qubits and a superconducting transmission line or LC oscillator (  the electromagnetic field)  inside a  circuit  QED that  is a  similar version   to a  cavity  QED which is   also   discussed. As well,  the dispersive regime needed for QIP is discussed. Finally, we will conclude this chapter by associating circuit QED (Quantum Electrodynamics) with QIP. In this section, we will identify the single-qubit gates as well as C-Phase and iSWAP gates, which have been experimentally realized using superconducting devices\cite{16,62}.
	
	Chapter \ref{Ch. 3} of this thesis is aimed to realizing an  $X$-rotation  gate for two and three qubits which  can be used for  preparing superposition states and others  called entangling  gates used  for generating maximally  entangled states such  as Bell and GHZ states, and thier  implementation is  based on a  quantum  device comprising  transmon  qubits capacitively coupled with a  resonator  driven by a microwave field using  a one-step  approach.  The robustness of these gates  is demonstrated by including the effect of the decoherence of the transmon systems and the resonator decay  in a master equation formalism. For  realistic assumptions about the  transmon-resonator system parameters,  high fidelity  will  be  achieved on a quantum  simulation. In the end, we will discuss the experimental feasibility of the proposed schemes and compare them with the latest experimental results within superconducting devices.
	
	In the following,  based on the implemented  X-rotation gates in chapter \ref{Ch. 3} and C-Phase gate  we will propose   practical and faster  schemes for implementing a two- and  three-qubit Grover’s search
	with an implementation time signifcantly shorter than the coherences time  of superconducting qubits in chapter \ref{Ch. 4}. There will also be a brief discussion about some   thoughts regarding the the possibility of experimental realization of the proposed schemes. In addition to the  characterization of the single-shot entatngling  gates for two  and three qubits in the chapter \ref{Ch. 3}  we here fully characterize these gates using QPT demonstrated by  nemurical simulation and obtaining   process fidelities greater than 93$\%$.
	
	Finally, this thesis will end with a conclusion and perspectives.
	
	\newgeometry{left=1.7cm,right=1.8cm,top=2.55cm,bottom=2.55cm}
	\chapter{Fundamental concepts of quantum information processing  and computation}\label{Ch. 1}

	  This chapter is intended to briefly give the fundamental concepts of quantum computation based on the book of Nielsen and
	  Chuang\cite{1}, including the mathematical and geometric properties of a qubit or a quantum bit. In addition, it describes the quantum gates for single-  and multi-qubit, as well as the discussion of universal quantum gates needed to accomplish the arbitrary quantum computation.   At the end of this chapter,  we also discuss  quantum algorithms exploiting quantum superposition and entanglement to perform specific calculations. We first discuss the quantum parallelism, which characterizes a method allowing  quantum computers to evaluate a function f(x) for a large number of x-values simultaneously. We will then  use  quantum parallelism and other laws of quantum mechanics to move on to discuss some quantum algorithms such as  Grover and the Deatsch-Jozsa algorithm. 
	\section{Quantum bit}
	In a  classical computer, the fundamental unit of information is the so-called binary digit or bit and can take one of two possible values $0$ or $1$.  These values  physically correspond for instance to whether   a capacitor is being charged or discharged. Similarly, in a quantum computer  two-level quantum systems can form the quantum bit or qubit.
	\subsection{Definition}
The fundamental unit of quantum information is the quantum bit or, more simply, qubit, which can  be  in  states  $\left| 0 \right\rangle $ and $\left| 1 \right\rangle $. These quantum states can be  formed by any quantum system  with two different energy levels, for example of an appropriate system is  the spin of an electron in a static magnetic field. Unlike classical computing in which the bit must be either 0 or 1, the qubit    can be prepared in what is called a " quantum superposition" of  both states  $\left| 0 \right\rangle $ and $\left| 1 \right\rangle $  at the same time, in quantum  computing. Mathematically, such a superposition state is  represented by
\begin{equation}\label{E1}
\left| \psi  \right\rangle  = \alpha \left( {\begin{array}{*{20}{c}}
	1\\
	0
	\end{array}} \right) + \beta \left( {\begin{array}{*{20}{c}}
	0\\
	1
	\end{array}} \right),
\end{equation}
where the numbers $\alpha $ and $\beta $ are complex numbers,  which satisfy  the normalization condition  ${\left| \alpha  \right|^2} + {\left| \beta  \right|^2} = 1$.  While the orthonormal basis states $\left| 0 \right\rangle$  and $\left| 1 \right\rangle $ are known as 
computational basis states. The measurement outcome of the qubit is  either $0$, with the probability ${\left| \alpha  \right|^2}$ or $1$ , with the probability ${\left| \beta  \right|^2}$.
\begin{figure}[H]
	\centering \includegraphics[scale=0.5]{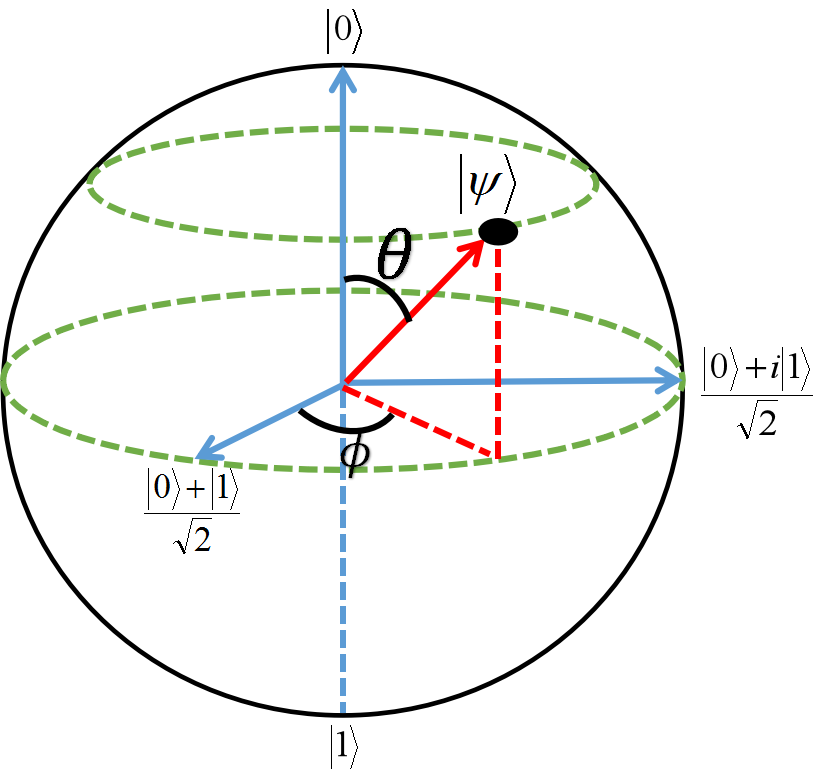}
	\caption{ Bloch sphere representation of a qubit.}
	\label{F1}
\end{figure}

 It is useful to geometrically represent the qubit state for a better understanding of the action of quantum operations on the state. To find   such a visual interpretation,  without losing any generality we may rewrite equation (\ref{E1}) as 
 \begin{equation}\label{E2}
 \left| \psi  \right\rangle  = \cos \left( {\frac{\theta }{2}} \right)\left| 0 \right\rangle  + {e^{i\phi }}\sin \left( {\frac{\theta }{2}} \right)\left| 1 \right\rangle,
 \end{equation}
  where  $\theta  \in \left[ {0,\pi } \right] $ and  $\phi  \in \left[ {0,2\pi } \right] $.  The ground and excited states $\left| 0 \right\rangle $ and $\left| 1 \right\rangle $ can  represent the South and North poles of a sphere, respectively, and thus each point on this so-called Bloch sphere represents a superposition state (see figure \ref{F1}).
	\subsection{Many qubits or quantum register}
	A classical register  is a set of $n$ bits that, in a sense,  stores all numbers between $0$ and  ${2^n} - 1$.  A quantum register, like a classical register, just is a system comprising many qubits. In the case of three quantum bits, this yields the computational basis states involving the eight possible states $\left\{ {\left| {000} \right\rangle ,\left| {001} \right\rangle ,\left| {010} \right\rangle ,\left| {011} \right\rangle ,\left| {100} \right\rangle ,\left| {101} \right\rangle ,\left| {110} \right\rangle ,\left| {111} \right\rangle } \right\}$ (where $\left| {ijk} \right\rangle  = \left| i \right\rangle  \otimes \left| j \right\rangle  \otimes \left| k \right\rangle $). Like a single qubit, the three-qubit state is given by the linear combination of the eight basis states
	\begin{eqnarray}\label{E3}
	\left| \psi  \right\rangle  &=& {\alpha _{000}}\left| {000} \right\rangle  + {\alpha _{001}}\left| {001} \right\rangle  + {\alpha _{010}}\left| {010} \right\rangle  + {\alpha _{011}}\left| {011} \right\rangle  + {\alpha _{100}}\left| {100} \right\rangle\nonumber\\  &+& {\alpha _{101}}\left| {101} \right\rangle  + {\alpha _{110}}\left| {110} \right\rangle  + {\alpha _{111}}\left| {111} \right\rangle, 
	\end{eqnarray}
	where ${{\alpha _{ijk}}}$ are the complex numbers which satisfy the normalization condition ${\sum\nolimits_{ijk} {\left| {{\alpha _{ijk}}} \right|} ^2} = 1$ ($ijk = 000,001,010,011,100,101,110,111$). An important property of qubits is that they can be entangled with other qubits. Three quantum systems can be entangled if the state of the total system cannot be written as a Kronecker product of its parts. In the context of quantum computing, three qubits can be entangled in two different ways by the states
	\begin{equation}\label{E4}
		\left| {GHZ} \right\rangle  = \frac{1}{{\sqrt 2 }}\left( {\left| {000} \right\rangle  + \left| {111} \right\rangle } \right),
	\end{equation}
	\begin{equation}\label{E5}
		\left| W \right\rangle  = \frac{1}{{\sqrt 3 }}\left( {\left| {001} \right\rangle  + \left| {010} \right\rangle  + \left| {100} \right\rangle } \right),
	\end{equation}
	
	\section{Quantum  gates and circuits}
	
	Besides storing information, we also mention that  the  processing of quantum information is useful  to build quantum computing, in order to move information from one qubit to another. Such information  processing  is carried out by the action of quantum gates on  qubits or   
	quantum registers. In analogy to classical computing which through in it the circuit model is very useful for computing processes and is also used to design and construct computing hardware using different types of classical logic gates acting on some binary input, for example,  the NAND, NOR, AND, and  OR  gates, whereas quantum computers can be interpreted in terms of a quantum circuit model comprising quantum gates, which  are  applied to quantum register. In quantum computers, we have two different types of quantum gates: single qubit and multi-qubit gates.  Single qubit gates operate only on one qubit, while   multi-qubit gates
	operate on two or more  qubits. We will give examples of these types of operations in detail in the rest of this section.
	\subsection{Single-qubit gates}
	 An n-qubit gate can be represented as ${2^n} \times {2^n}$ unitary matrices.  Due to  the single-qubit gate operating only on  one qubit, thus  single-qubit gates can be represented by matrices of size ${2} \times {2}$.
	 \section*{Rotation gates }
	
	Generally,  the single quantum operation can be visualized by using the Bloch sphere (shown in fgure \ref{F1}) as a rotation of the state around the x,y, and z axes.  Some of the single qubit gates which are  used to generate the rotation about the Bloch sphere axes   can be described using Pauli  matrices as  follows: 
\begin{eqnarray}\label{E6}
	X = {\sigma _x} = \left( {\begin{array}{*{20}{c}}
		0&1\\
		1&0
		\end{array}} \right),
\end{eqnarray}
\begin{eqnarray}\label{E7}
	Y = {\sigma _y} = \left( {\begin{array}{*{20}{c}}
		0&{ - i}\\
		i&0
		\end{array}} \right),
\end{eqnarray}
\begin{eqnarray}\label{E8}
	Z = {\sigma _z} = \left( {\begin{array}{*{20}{c}}
		1&0\\
		0&{ - 1}
		\end{array}} \right).
\end{eqnarray}

When we apply  these single-qubit gates to the computational basis states $\left| 0 \right\rangle $ and $\left| 1 \right\rangle $, we easily get
	 \begin{eqnarray}\label{E9}
	 X\left| 0 \right\rangle  = \left| 1 \right\rangle, X\left| 1 \right\rangle  = \left| 0 \right\rangle, 
	 \end{eqnarray}
	 \begin{eqnarray}\label{E10}
	 Y\left| 0 \right\rangle  = i\left| 1 \right\rangle, Y\left| 1 \right\rangle  =  - i\left| 0 \right\rangle, 
	 \end{eqnarray}
	 \begin{eqnarray}\label{E11}
	 Z\left| 0 \right\rangle  = \left| 0 \right\rangle, Z\left| 1 \right\rangle  =  - \left| 1 \right\rangle. 
	 \end{eqnarray}
	 
	 Besides these rotation operations, we  would like to mention the  other important  single-qubit gates, which   are also rotation  operations, which are  given by 
	 \begin{eqnarray}\label{E12}
	 {R_x}\left( \theta  \right) = {e^{ - i\theta {{{\sigma _x}} \mathord{\left/
	 				{\vphantom {{{\sigma _x}} 2}} \right.
	 				\kern-\nulldelimiterspace} 2}}} = \left( {\begin{array}{*{20}{c}}
	 	{\cos \left( {\frac{\theta }{2}} \right)}&{ - i\sin \left( {\frac{\theta }{2}} \right)}\\
	 	{ - i\sin \left( {\frac{\theta }{2}} \right)}&{\cos \left( {\frac{\theta }{2}} \right)}
	 	\end{array}} \right),  
	 \end{eqnarray} 
	 \begin{eqnarray}\label{E13}
	 {R_y}\left( \theta  \right) = {e^{ - i\theta {{{\sigma _y}} \mathord{\left/
	 				{\vphantom {{{\sigma _y}} 2}} \right.
	 				\kern-\nulldelimiterspace} 2}}} = \left( {\begin{array}{*{20}{c}}
	 	{\cos \left( {\frac{\theta }{2}} \right)}&{ - \sin \left( {\frac{\theta }{2}} \right)}\\
	 	{\sin \left( {\frac{\theta }{2}} \right)}&{\cos \left( {\frac{\theta }{2}} \right)}
	 	\end{array}} \right),  
	 \end{eqnarray} 
	 
	 \begin{eqnarray}\label{E14}
	 {R_z}\left( \theta  \right) = {e^{ - i\theta {{\sigma z} \mathord{\left/
	 				{\vphantom {{\sigma z} 2}} \right.
	 				\kern-\nulldelimiterspace} 2}}} = \left( {\begin{array}{*{20}{c}}
	 	{{e^{ - i{\theta  \mathord{\left/
	 						{\vphantom {\theta  2}} \right.
	 						\kern-\nulldelimiterspace} 2}}}}&0\\
	 	0&{{e^{i{\theta  \mathord{\left/
	 						{\vphantom {\theta  2}} \right.
	 						\kern-\nulldelimiterspace} 2}}}}
	 	\end{array}} \right),
	 \end{eqnarray}
	 where $\theta$ is the angle of rotation. Based on these gates, we can identify some other important  single-qubit operations just by performing rotations of $\theta  =  \mp \pi $ and $\theta  =  \mp {\pi  \mathord{\left/
	 		{\vphantom {\pi  2}} \right.
	 		\kern-\nulldelimiterspace} 2}$, often mentioned  as $\pi$-pulses and $\mp {\pi  \mathord{\left/
	 		{\vphantom {\pi  2}} \right.
	 		\kern-\nulldelimiterspace} 2}$-pulses, respectively.  We can identify the Pauli  matrices by  choosing $\theta  = \pi $, ${R_x}\left( \pi  \right) =  - i{\sigma _x}$, ${R_y}\left( \pi  \right) =  - i{\sigma _y}$, ${R_z}\left( \pi  \right) =  - i{\sigma _z}$.  Similarly,
	 when choosing $\theta  = {\pi  \mathord{\left/
	 		{\vphantom {\pi  2}} \right.
	 		\kern-\nulldelimiterspace} 2}$  in the equations (\ref{E12}) and (\ref{E13}), these operations  map a state initially  prepared in  the ground ($\left| 0 \right\rangle $)  or an excited ($\left| 1 \right\rangle $) state into an equal superposition of the  states $\left| 0 \right\rangle $ and $\left| 1 \right\rangle $ as follows 
	
	\begin{eqnarray}\label{E15}
{R_x}\left( {\frac{\pi }{2}} \right)\left| 0 \right\rangle  = \frac{1}{{\sqrt 2 }}\left( {\left| 0 \right\rangle  - i\left| 1 \right\rangle } \right),{R_x}\left( {\frac{\pi }{2}} \right)\left| 1 \right\rangle  = \frac{{ - i}}{{\sqrt 2 }}\left( {\left| 0 \right\rangle  + i\left| 1 \right\rangle } \right),
	\end{eqnarray} 
	\begin{eqnarray}\label{E16}
	{R_y}\left( {\frac{\pi }{2}} \right)\left| 0 \right\rangle  = \frac{1}{{\sqrt 2 }}\left( {\left| 0 \right\rangle  +\left| 1 \right\rangle } \right),{R_y}\left( {\frac{\pi }{2}} \right)\left| 1 \right\rangle  = \frac{{ 1}}{{\sqrt 2 }}\left( {\left| 0 \right\rangle  -\left| 1 \right\rangle } \right).
	\end{eqnarray}
	
	 In addition, it is worth mentioning that the similar single-qubit gate of these operations, the Hadamard gate enables qubit interference which is necessary for quantum algorithms and has no classical analog. It can be described by a  unitary  matrix 
	  \begin{eqnarray}\label{E17}
	  H = \frac{1}{{\sqrt 2 }}\left( {\begin{array}{*{20}{c}}
	  	1&1\\
	  	1&{ - 1}
	  	\end{array}} \right).
	  \end{eqnarray}
	  
	On the other  hand,  one can usually  rewrite the Hadamard gate by decomposing it into  sequences of     rotation operations around the x  and y  axes in the Bloch sphere in the following form:
	\begin{eqnarray}\label{E18}
	H = {e^{i\frac{\pi }{2}}}{R_x}\left( \pi  \right){R_y}\left( {\frac{\pi }{2}} \right).
	\end{eqnarray}
	
	 Finally, we have to mention other important single qubit gates,  S and T gates, which  are so-called the phase gates. These gates rotate the state vector around the z-axis on the Bloch sphere by changing  the dynamic phase of the state
	 
	 \begin{eqnarray}\label{E19}
	 S = \left( {\begin{array}{*{20}{c}}
	 	1&0\\
	 	0&i
	 	\end{array}} \right),
	 \end{eqnarray}
	 \begin{eqnarray}\label{E20}
	 T = \left( {\begin{array}{*{20}{c}}
	 	1&0\\
	 	0&{{e^{i\frac{\pi }{4}}}}
	 	\end{array}} \right).
	 \end{eqnarray} 
	
	 For more details on how to implement the rotation gates for single- and multi-qubit see chapters \ref{Ch. 2} and \ref{Ch. 3}.

	 \subsection{Multi-qubit gates and universilaty}
	 
    \section*{Multi-qubit gates}
\begin{figure}[H]
		\centering \includegraphics[scale=0.5]{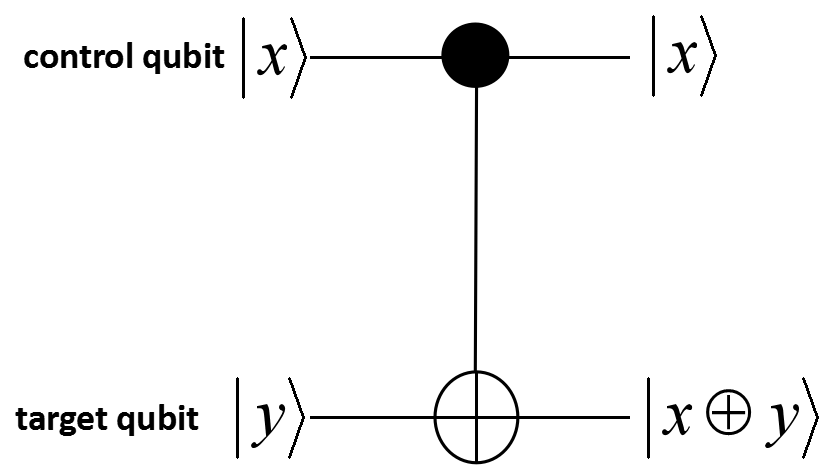}
		\caption{ Quantum circuit representation for   CNOT gate. Here  $x,y=0$ or $1$ and $ \oplus $ denotes the addition  modulo2,  while the vertical line between the control
			and target qubit represents  the circuit  diagram of CNOT gate.}
		\label{F2}
	\end{figure}

	 In addition to single-qubit gates that can be applied to individual qubits, there are also other important multi-qubit gates known as controlled-unitary gates.  In most of these gates,  some qubits play a target role while some qubits play the control role.  Such controlled-unitary  operations are  performed   on the target qubit which will change depending on the state of the control qubit.  One of the most  famous  two-qubit gates  is the controlled-not operation($CNOT$). The action  of the controlled-not operation is to  invert the state of the target qubit if the control qubit is $\left| 1 \right\rangle $ , and   the state of the target qubit  does not change    if the control qubit is $\left| 0 \right\rangle $. Therefore,  the quantum  circuit (Figure (\ref{F2}))  and  matrix representation of this gate in the computational basis states of two-qubits, $\left| {0,0} \right\rangle $, $\left| {0,1} \right\rangle $, $\left| {1,0} \right\rangle $, and  $\left| {1,1} \right\rangle $
	 \begin{eqnarray}\label{E21}
	 CNOT = \left( {\begin{array}{*{20}{c}}
	 	1&0&0&0\\
	 	0&1&0&0\\
	 	0&0&0&1\\
	 	0&0&1&0
	 	\end{array}} \right).
	 \end{eqnarray}
	
	This gate is often used to generate the maximally  entangled state, the Bell states defined by  $\left| {{\Phi ^ \pm }} \right\rangle  = {{\left( {\left| {00} \right\rangle  \pm \left| {11} \right\rangle } \right)} \mathord{\left/
			{\vphantom {{\left( {\left| {00} \right\rangle  \pm \left| {11} \right\rangle } \right)} {\sqrt 2 }}} \right.
			\kern-\nulldelimiterspace} {\sqrt 2 }}$ and $\left| {{\psi ^ \pm }} \right\rangle  = {{\left( {\left| {01} \right\rangle  \pm \left| {10} \right\rangle } \right)} \mathord{\left/
			{\vphantom {{\left( {\left| {01} \right\rangle  \pm \left| {10} \right\rangle } \right)} {\sqrt 2 }}} \right.
			\kern-\nulldelimiterspace} {\sqrt 2 }}$, and thus CNOT is an entangling gate.
	
	Another controlled-unitary operation of a two-qubit gate is called the  CPhase ($CP$) gate which performs a controlled-Z operation. This is done if  two  qubits are in $\left| 1 \right\rangle $. Thus, the matrix representation of this gate (CP) is given by
	\begin{eqnarray}\label{E22}
	CP = \left( {\begin{array}{*{20}{c}}
		1&0&0&0\\
		0&1&0&0\\
		0&0&1&0\\
		0&0&0&{ - 1}
		\end{array}} \right).
	\end{eqnarray}
	
	We note that the CNOT gate can be realized by  combining  $CP$ gate and two Hadamard gates on both qubits. 
	
	 Other important  two-qubit  gates are $iSWAP$ and $\sqrt {iSWAP} $, which are perfect  to  generate maximally  entangled states, and are defined by 
	  \begin{eqnarray}\label{E23}
	  iSWAP = \left( {\begin{array}{*{20}{c}}
	  	1&0&0&0\\
	  	0&0&i&0\\
	  	0&i&0&0\\
	  	0&0&0&1
	  	\end{array}} \right) and ~~\sqrt {iSWAP}  = \left( {\begin{array}{*{20}{c}}
	  	1&0&0&0\\
	  	0&{\frac{1}{{\sqrt 2 }}}&{ \frac{i}{{\sqrt 2 }}}&0\\
	  	0&{\frac{i}{{\sqrt 2 }}}&{\frac{1}{{\sqrt 2 }}}&0\\
	  	0&0&0&1
	  	\end{array}} \right).  
	  \end{eqnarray}
	  
	   An example of a controlled-unitary operation for a three-qubit gate is a Toffoli gate,  which is essentially a controlled-controlled-X gate,   flipping  the state of the target qubit only  if both of the first two  control   qubits are in $\left| 1 \right\rangle $
	    \begin{eqnarray}\label{E24}
	    Toffoli = \left( {\begin{array}{*{20}{c}}
	    	1&0&0&0&0&0&0&0\\
	    	0&1&0&0&0&0&0&0\\
	    	0&0&1&0&0&0&0&0\\
	    	0&0&0&1&0&0&0&0\\
	    	0&0&0&0&1&0&0&0\\
	    	0&0&0&0&0&1&0&0\\
	    	0&0&0&0&0&0&0&1\\
	    	0&0&0&0&0&0&1&0
	    	\end{array}} \right).
	    \end{eqnarray}
	    
	     In addition, we also have another three-qubit gate that is the Fredkin gate, which performs a controlled SWAP operation,  which exchanges the states of two target qubits  if only  the first  control qubit is in state $\left| 1 \right\rangle $ 
	     \begin{eqnarray}\label{E25}
	     Fredkin = \left( {\begin{array}{*{20}{c}}
	     	1&0&0&0&0&0&0&0\\
	     	0&1&0&0&0&0&0&0\\
	     	0&0&1&0&0&0&0&0\\
	     	0&0&0&1&0&0&0&0\\
	     	0&0&0&0&1&0&0&0\\
	     	0&0&0&0&0&0&1&0\\
	     	0&0&0&0&0&1&0&0\\
	     	0&0&0&0&0&0&0&1
	     	\end{array}} \right).
	     \end{eqnarray}
	     
	      \section*{ Universilaty}
	      The concept of universality is associated with the ability to comprise any computational algorithm with a  set of simple gates. For instance, in classical computing,  the MAND gate is  a  gate that can be implemented using a combination of AND and NOT gates and  thus the NAND gate is  referred to as
	      a universal gate. Similar to  universal  classical logic operations, there is also  a set of quantum gates that are universal, such  combinations of some gates can realize complex algorithms in quantum computing. However, unlike classical computing in which a single gate is necessary, the university in quantum counterpart can only be achieved by combining the arbitrary single qubit gates and cNOT gate, as well as the set of single qubit gates and Toffoli gate. In
	      general,  at least one multi-qubit gate is required in a universal set for quantum computing.
	      
	      \subsection{Quantum  circuit  for two- and three-qubit  entanglement }
\begin{figure}[H]
	       	\centering \includegraphics[scale=0.5]{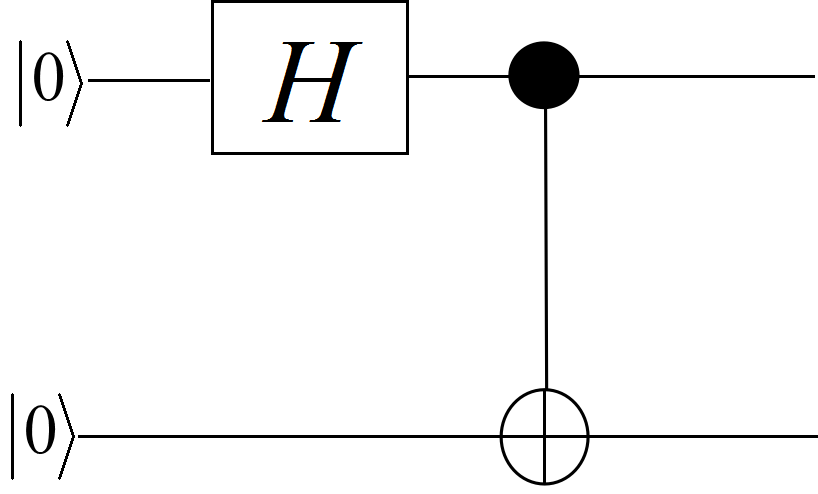}
	       	\caption{ Quantum  circuit for creating  the Bell state.}
	       	\label{F3}
	       \end{figure}

	      A quantum circuit is a set of gates applied to quantum registers or qubits, it is used to design and construct computing hardware. As an example, we here  discuss  the two  quantum circuits that can be used to generate maximally  entangled states, the Bell state ($\left| {{\Phi ^ + }} \right\rangle  = {{\left( {\left| {00} \right\rangle  + \left| {11} \right\rangle } \right)} \mathord{\left/
	      		{\vphantom {{\left( {\left| {00} \right\rangle  + \left| {11} \right\rangle } \right)} {\sqrt 2 }}} \right.
	      		\kern-\nulldelimiterspace} {\sqrt 2 }}$), and the GHZ-state given in the equation  (\ref{E4}).  In order to  generate the Bell state, the single and two-qubit gates  are required in the corresponding  quantum  circuit as shown in  Fig. (\ref{F3}). This circuit  is read   from left to right. First, the single-qubit gate on the control qubit is applied as depicted in Fig (\ref{F3}), followed by  a CNOT gate to  generate the entanglement.
	       
\begin{figure}[H]
	      	\centering \includegraphics[scale=0.5]{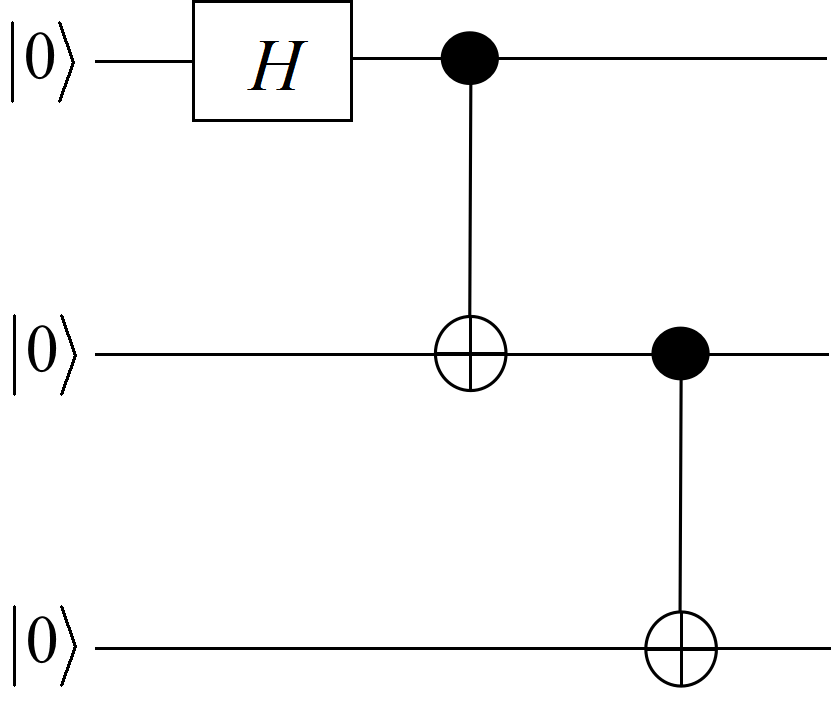}
	      	\caption{ Quantum  circuit for creating  the GHZ state.}
	      	\label{F4}
	      \end{figure}

	      In addition, the GHZ protocol can be generated by using the single qubit gate and two CNOT gates,  as shown in the circuit  diagram  in Figure (\ref{F4}). Starting in ground state $\left| {000} \right\rangle $  for all three qubits, we first perform  a rotation on the first qubit (with $\theta  = {\pi  \mathord{\left/
	      		{\vphantom {\pi  2}} \right.
	      		\kern-\nulldelimiterspace} 2}$) to put the whole system  in state ${{\left( {\left| {000} \right\rangle  + \left| {100} \right\rangle } \right)} \mathord{\left/
	      		{\vphantom {{\left( {\left| {000} \right\rangle  + \left| {100} \right\rangle } \right)} {\sqrt 2 }}} \right.
	      		\kern-\nulldelimiterspace} {\sqrt 2 }}$. To create GHZ-state, we need to map the second component of this state ${\left| {100} \right\rangle }$ to ${\left| {111} \right\rangle }$ without changing the first component. This is exactly  done  by applying two CNOT gates  to flip the second and third qubit  conditioned on the state of the first  qubit. These CNOTs can be applied in the following way; using the first qubit as the control qubit for the first gate and the second qubit as the control qubit for another gate.

    \section{Quantum algorithms}
	            With access to a large number of quantum gates, we can now build more complex quantum algorithms circuits by combining these gates. The power of these algorithms lies in the fundamental laws of quantum mechanics such as superposition and entanglement, which leads to performing calculations more efficiently than their classical counterparts. In  this section,  we will   describe, in general, some  quantum  algorithms and their circuits diagram,  they  are implementable in  most  quantum devices.

	            \subsection{Quantum parallelism  needed in an algorithm}
	            
	            One of the important difference  between  quantum  and classical  computers is that quantum  computers can  store their information in superposition states. The ability of quantum  computing  to accomplish a quantum  operation  on all states at
	            the same time, which is referred to as quantum parallelism, leads to a significant speed-up advantage over classical ones. Bellow,  we explain in details   how quantum parallelism  works.

	             In general, we consider  a quantum  register involving  $n+m$ qubits which  takes $n$ qubits as  control  qubits and $m$  qubits as a target  qubits. Furthermore, let us  assume $f\left( x \right):{\left\{ {0,1} \right\}^n} \to {\left\{ {0,1} \right\}^m}$ is a  Boolean function and  we suppose that  we have an unitary  transformation ${U_f}$ of size $\left( {n + m} \right) \times \left( {n + m} \right)$ in which  we compute the function $f(x)$  such that its action on the state  ${\left| x \right\rangle _n}{\left| y \right\rangle _m}$ gives ${\left| x \right\rangle _n}{\left| {y \oplus f\left( x \right)} \right\rangle _m}$, using a set of quantum  gates.  Instead of applying  ${U_f}$   on each   individual input  state, we first create a superposition of all input states ${\left| x \right\rangle _n}$ using the Hadamard gate as follows
	             \begin{eqnarray}\label{E26}
	             {H^{\left( 1 \right)}} \otimes {H^{\left( 2 \right)}} \otimes ...{H^{\left( n \right)}}{\left| 0 \right\rangle _n} = \frac{1}{{\sqrt {{2^n}} }}\sum\limits_x {{{\left| x \right\rangle }_n}}. 
	             \end{eqnarray}
	             
\begin{figure}[H]
	              	\centering \includegraphics[scale=0.5]{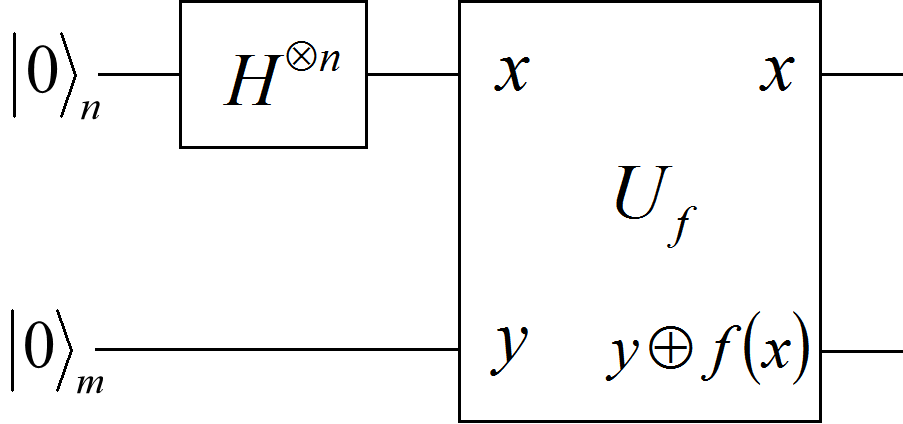}
	              	\caption{ Quantum circuit for evaluating  $f(x)$. ${U_f}$ is the quantum circuit which transforms 
	              		input states ${\left| x \right\rangle _n}{\left| y \right\rangle _m}$ to ${\left| x \right\rangle _n}{\left| {y \oplus f\left( x \right)} \right\rangle _m}$.}
	              	\label{F5}
	              \end{figure}

	              As shown in Figure (\ref{F5}),  we then apply  the unitary transformation ${U_f}$ on this superposition state and all qubits of the target  register in the state $\left| 0 \right\rangle $
	              \begin{eqnarray}\label{E27}
	              {U_f}\left( {\frac{1}{{\sqrt {{2^n}} }}\sum\limits_x {{{\left| x \right\rangle }_n}{{\left| 0 \right\rangle }_m}} } \right) = \frac{1}{{\sqrt {{2^n}} }}\sum\limits_x {{{\left| x \right\rangle }_n}{{\left| {f\left( x \right)} \right\rangle }_m}}. 
	              \end{eqnarray}
	      
	 \subsection{Grover quantum algorithm }
	 Another class of quantum algorithms is for searching with quantum computing. Consider the problem of searching for a specific element through an unordered database of N elements.  To classically  do this, one can take up  to  N evaluations  because  all elements of the database  list must be checked. Thus, the classical  algorithm  uses $O(N)$ operations. However,   this problem can be solved significantly more efficiently in quantum computing. Quantum search algorithm, known as Grover’s algorithm, which enables one to search through an unordered list with a high probability of success, requiring only $O\left( {\sqrt N } \right)$ operations.  In this section, we describe Grover's algorithm steps step-by-step.  We then represent the geometric visualization of this algorithm to better understand its action. 
	 
	 \section*{Steps and procedure }
	 
\begin{figure}[H]
	    	\centering \includegraphics[scale=0.5]{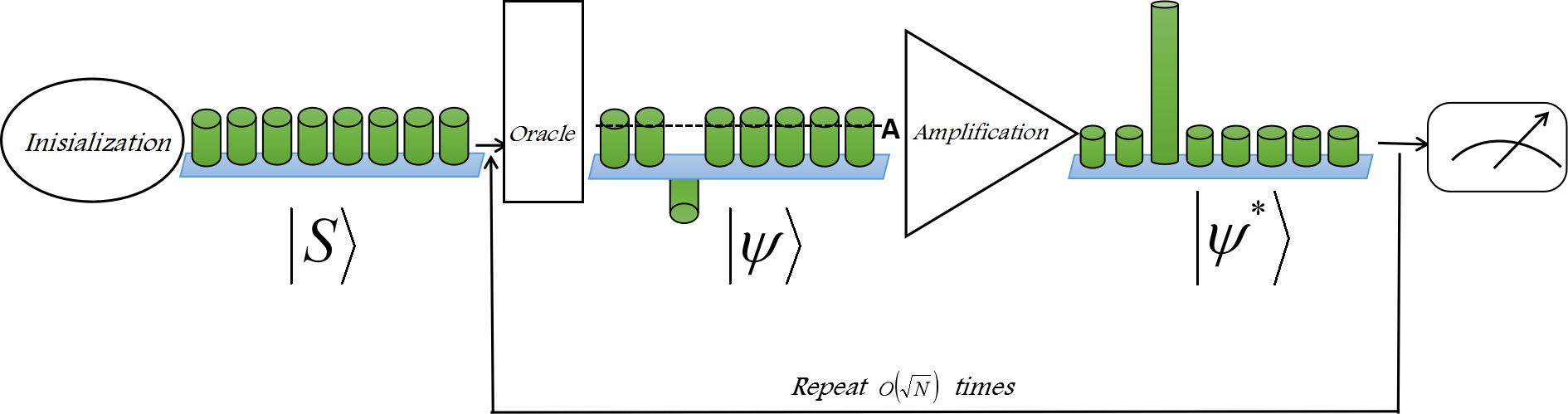}
	    	\caption{ Evolution of  amplitudes for each state during the Grover search algorithm.}
	    	\label{F6}
	    \end{figure}

	  Suppose we have an unordered database of $N = {2^n}$ elements (where $n$ denotes the number of qubits), represented with a ket,  $\left\{ {\left| 0 \right\rangle ,\left| 1 \right\rangle ,...\left| N \right\rangle } \right\}$. The general Grover algorithm implementation is divided into four steps as follows(see figure \ref{F6}): 
	  \section*{Step$1$:  Initialization}
	  We  initialize all  qubits,  by   creating  an equal superposition of all states, that  were in the state $\left| 0 \right\rangle $. This done   by applying the $n$-qubit Hadamard
	  \begin{eqnarray}\label{E28}
	  \left| S \right\rangle  = \frac{1}{{\sqrt {{2^n}} }}\sum\limits_{x = 1}^{{2^n}} {\left| x \right\rangle }. 
	  \end{eqnarray}
	  
	   \section*{Step$2$:  Applying  the oracle transformation}
	   
	   An oracle can be viewed as a black box that performs a quantum gate on a  quantum register that is not easily specified by universal gates. In Grover's algorithm,  an oracle can work by applying a unitary transformation ${U_O}$ to the initialization qubit ($\left| s \right\rangle $) such that   it flips the sign of $x $ if and only if $x $ is a state we are looking for or the correct state,  with the following  properties 
	   \begin{eqnarray}\label{E129}
	   \left| x \right\rangle \mathop  \to \limits^{{U_O}} {\left( { - 1} \right)^{f\left( x \right)}}\left| x \right\rangle, 
	   \end{eqnarray}
	   where $f(x)=1$ if $x$ is the correct quantum  state and $f(x)=0$
	   otherwise. The resulting state after applying an oracle  is
	   \begin{eqnarray}\label{E30}
	   \left| \psi  \right\rangle  =  - {\alpha _x}\left| x \right\rangle  + {\alpha _{{x^ * }}}\sum\limits_{{x^ * } = 1,{x^ * } \ne x}^{{2^n}} {\left| {{x^ * }} \right\rangle }.   
	   \end{eqnarray}
	   
	   \section*{Step $3$:  Amplification} 
	   The amplification, sometimes known as the inversion about mean operation,  is applied to the result quantum state after the application of an oracle, and given by
	   \begin{eqnarray}\label{E31}
	   {U_\phi } = 2\left| S \right\rangle \left\langle S \right| - I.  
	   \end{eqnarray}

	   This stage  performs a reflection of the mean, thus increasing the amplitude of the  correct state. Then,  the resulting state is
	   \begin{eqnarray}\label{E32}
	   \left| {{\psi ^ * }} \right\rangle  = (2A + {\alpha _x})\left| x \right\rangle  + (2A - {\alpha _{{x^ * }}})\sum\limits_{{x^ * } = 1,{x^ * } \ne x}^{{2^n}} {\left| {{x^ * }} \right\rangle } ,
	   \end{eqnarray} 
	   where $A$ is the amplitude of the maen vector $\left| \psi  \right\rangle $ ($A = {1 \mathord{\left/
	   		{\vphantom {1 {{2^n}\left\{ { - {\alpha _x} + \left( {{2^n} - 1} \right){\alpha _{{x^ * }}}} \right\}}}} \right.
	   		\kern-\nulldelimiterspace} {{2^n}\left\{ { - {\alpha _x} + \left( {{2^n} - 1} \right){\alpha _{{x^ * }}}} \right\}}}$).
	   \section*{Step $4$: measurement}
	   Finally, the algorithm output state  is measured.
	    
	    \section*{Geometric visualization}
	    
\begin{figure*}
	     	\centering \includegraphics[scale=0.5]{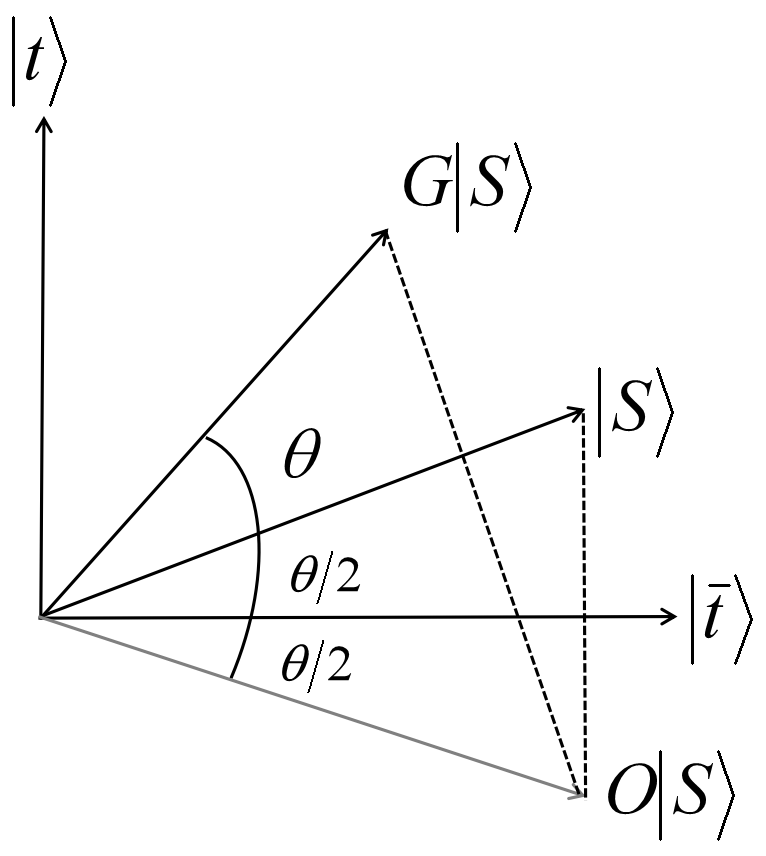}
	     	\caption{ Geometric representation of Grover's algorithm for a single Grover iteration, $G$. }
	     	\label{F7}
	     \end{figure*}

	    We  note that  $G = \left( {2\left| S \right\rangle \left\langle S \right| - I} \right)O$,  called the Grover iteration  and  it   must be repeated many  times owing to  the nature of the algorithm. We will here show that  this operator can be considered a rotation in the two-dimensional space spanned by the starting vector $\left| S \right\rangle$(See Fig. \ref{F7}).  We redefine the state vector in equation (\ref{E28}) as follows 
	    \begin{eqnarray}\label{E33}
	    \left| S \right\rangle  = \sqrt {\frac{M}{N}} \left| t \right\rangle  + \sqrt {\frac{{N - M}}{N}} \left| {\bar t} \right\rangle,   
	    \end{eqnarray}
	    with $\left| t \right\rangle $ and $\left| {\bar t} \right\rangle $ are the normalized
	    states ($\left| t \right\rangle  = \frac{1}{{\sqrt M }}\left| x \right\rangle $ and $\left| {\bar t} \right\rangle  = \frac{1}{{\sqrt {N - M} }}\sum\limits_{{x^ * }} {\left| {{x^ * }} \right\rangle }  $).
	    
	    The effect of $G$ can be realized by applying successively the oracle operation $O$ that  performs a reflection about the state $\left| t \right\rangle $ in the two-dimensional space defined by  $\left| t \right\rangle $ and $\left| {\bar t} \right\rangle $ and then the the amlification ${U_\phi } = 2\left| S \right\rangle \left\langle S \right| - I$ which  performs a reflection in the plane defined by  $\left| t \right\rangle $ and $\left| {\bar t} \right\rangle $, about the vector $\left| S \right\rangle $. Let's set $\cos ({\theta  \mathord{\left/
	    		{\vphantom {\theta  2}} \right.
	    		\kern-\nulldelimiterspace} 2}) = \sqrt {{{\left( {N - M} \right)} \mathord{\left/
	    			{\vphantom {{\left( {N - M} \right)} N}} \right.
	    			\kern-\nulldelimiterspace} N}} $, and thus the equation (\ref{E33}) can be rewritten as $\left| S \right\rangle  = \sin \left( {{\theta  \mathord{\left/
	    			{\vphantom {\theta  2}} \right.
	    			\kern-\nulldelimiterspace} 2}} \right)\left| t \right\rangle  + \cos ({\theta  \mathord{\left/
	    		{\vphantom {\theta  2}} \right.
	    		\kern-\nulldelimiterspace} 2})\left| {\bar t} \right\rangle $.
	     
	      \subsection{Deutsch-Jozsa quantum algorithm}
	      One of the  first and simple quantum algorithms that show that quantum computing works better than classical computing is the Deutsch-Jozsa  algorithm\cite{103}, it demonstrates exponential
	      acceleration compared to a classical algorithm. A generalization of this algorithm  to  functions  $f\left( x \right):{\left\{ {0,1} \right\}^n} \to {\left\{ {0,1} \right\}}$   was given by Deutsch and Jozsa and improved by Cleve, Ekert, Macchiavello, and Mosca\cite{104}. In the Deutsch-Jozsa algorithm,  the function $f$ is   either constant in which  $2^n$ values are all either equal to $0$ or to $1$ or balanced in which exactly half of the values are equal to $0$ and the other half to $1$.
	      
\begin{figure}[H]
	         	\centering \includegraphics[scale=0.5]{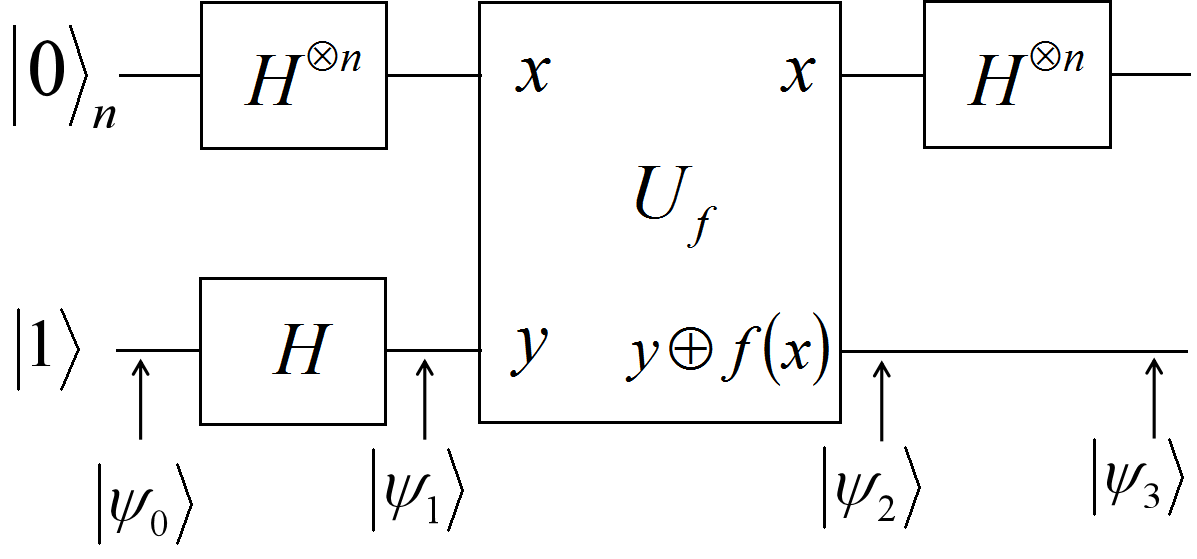}
	         	\caption{ Quantum circuit implementing the general Deutsch-Jozsa algorithm.}
	         	\label{F8}
	         \end{figure}

	      The Deutsch -Jozsa algorithm is divided in general into four steps. These steps are depicted in figure (\ref{F8}),   let  us follow the numbered  states according to this circuit:
	      \section*{Step $1$} 
	      Deutsch-Jozsa algorithm  starts with a quantum register of n qubits,  we  initialize all the qubits of the first  register in $\left| 1 \right\rangle $  and of the second register in $\left| 1 \right\rangle $ 
	      \begin{eqnarray}\label{E34}
	      \left| {{\psi _0}} \right\rangle  = {\left| 0 \right\rangle ^{ \otimes n}}\left| 1 \right\rangle.
	      \end{eqnarray}
	      \section*{Step $2$}
	      We apply the Hadamard operator to each qubit of the registers, we get 
	      \begin{eqnarray}\label{E35}
	      \left| {{\psi _1}} \right\rangle  = \frac{1}{{\sqrt {{2^n}} }}\sum\limits_x {{{\left| x \right\rangle }_n}} \left[ {\frac{{\left| 0 \right\rangle  - \left| 1 \right\rangle }}{{\sqrt 2 }}} \right].
	      \end{eqnarray}
	       \section*{Step $3$}
	       
	       Next step, the function $f$ is evaluated using ${U_f}:\left| x \right\rangle \left| y \right\rangle  \to \left| x \right\rangle \left| {y \oplus f(x)} \right\rangle $, giving
	       \begin{eqnarray}\label{E1136}
	       \left| {{\psi _2}} \right\rangle  = \frac{1}{{\sqrt {{2^n}} }}{\sum\limits_x {\left( { - 1} \right)} ^{f(x)}}{\left| x \right\rangle _n}\left[ {\frac{{\left| 0 \right\rangle  - \left| 1 \right\rangle }}{{\sqrt 2 }}} \right].
	       \end{eqnarray}
	        \section*{Step $4$}
	        The last step is to apply the Hadamard operator on a state $\left| x \right\rangle $ and the measurement.\\
	        We  separately   check  the cases $x = 0$ and $x = 1$,  we then  see the Hadamard transform   for  a single qubit $H\left| x \right\rangle  = {{\sum\limits_x {{{\left( { - 1} \right)}^{xz}}\left| z \right\rangle } } \mathord{\left/
	        		{\vphantom {{\sum\limits_x {{{\left( { - 1} \right)}^{xz}}\left| z \right\rangle } } {\sqrt 2 }}} \right.
	        		\kern-\nulldelimiterspace} {\sqrt 2 }}$, and thus for $n$-qubits
	        \begin{eqnarray}\label{E37}
	        {H^{ \otimes n}}{\left| x \right\rangle _n} = \frac{1}{{\sqrt {{2^n}} }}\sum\limits_z {{{\left( { - 1} \right)}^{xz}}{{\left| z \right\rangle }_n}}. 
	        \end{eqnarray}
	        
	        So by  using this equation and (\ref{E1136}), we easily  find 
	        \begin{eqnarray}\label{E38}
	        \left| {{\psi _3}} \right\rangle  = \frac{1}{{\sqrt {{2^n}} }}\sum\limits_z {{{\left( { - 1} \right)}^{xz + f\left( x \right)}}{{\left| z \right\rangle }_n}} \left[ {\frac{{\left| 0 \right\rangle  - \left| 1 \right\rangle }}{{\sqrt 2 }}} \right] .
	        \end{eqnarray}
	    
	   \section{Chapter summary}
	   In this chapter, we have discussed the fundamental notions needed for quantum information processing such as the mathematical and geometric of a  qubit, quantum superposition, and entanglement as well as single qubit gates. Such notions and entangling two-qubit gates are required to build any simple quantum processor. We have also seen how some two- and multi-qubit gates, for instance, the $CNOT$,  $CZ$, $iSWAP$ , $\sqrt {SWAP} $ and $Toffoli$,  when combined with the single rotation gates,   can be realized as a universal set for quantum computing. They permit the operation of some simple algorithms, such as the   Grover search and  Deutsch-Jozsa. Over the next  chapters,  we will suggest the implementation of some of these concepts with superconducting qubit devices.

	\chapter{Quantum superconducting  circuits} \label{Ch. 2}

Physically, Quantum bit or qubit implementations have taken many forms, for instance, nuclear spins,   trapped-ions,  photons,  and superconducting qubits. In this chapter, we will describe how  we use can  electrical circuits to realize  superconducting qubits such as charge qubits and transmon qubits.  An essential building block to realizing these superconducting qubits is the Josephson junction, the fundamental properties of which are discussed. We will also describe the interaction  between  light  and matter inside the  cavity QED   and its analog called the circuit QED,  in which   the cavity is replaced by a superconducting transmission line resonator capacitively coupled to a Cooper box which plays the role of the two-level atom or the superconducting qubit in the context of quantum information.   We then discuss how the coupling between transmon-type superconducting qubits and the transmission line resonator, can be used to realize the single and  two-qubit gates.   
\section{Superconducting qubits  }
One of the most popular and promising candidates for realizing scalable universal quantum computing is superconducting qubits, based on Josephson effect. The general principle of operation of these qubits includes a Josephson junction, which is characterized by its nonlinear inductance and its capacitance. Here, we introduce the Josephson junction as the key device for realizing superconducting qubits. We then derive the Hamiltonian of the Cooper pair box(Charge qubit) and the transmon-type superconducting, using the quantization of electrical circuits to perform canonical quantization of our circuits.

	\subsection{The Josephson Junction }
\begin{figure}[H]
		\centering \includegraphics[scale=0.5]{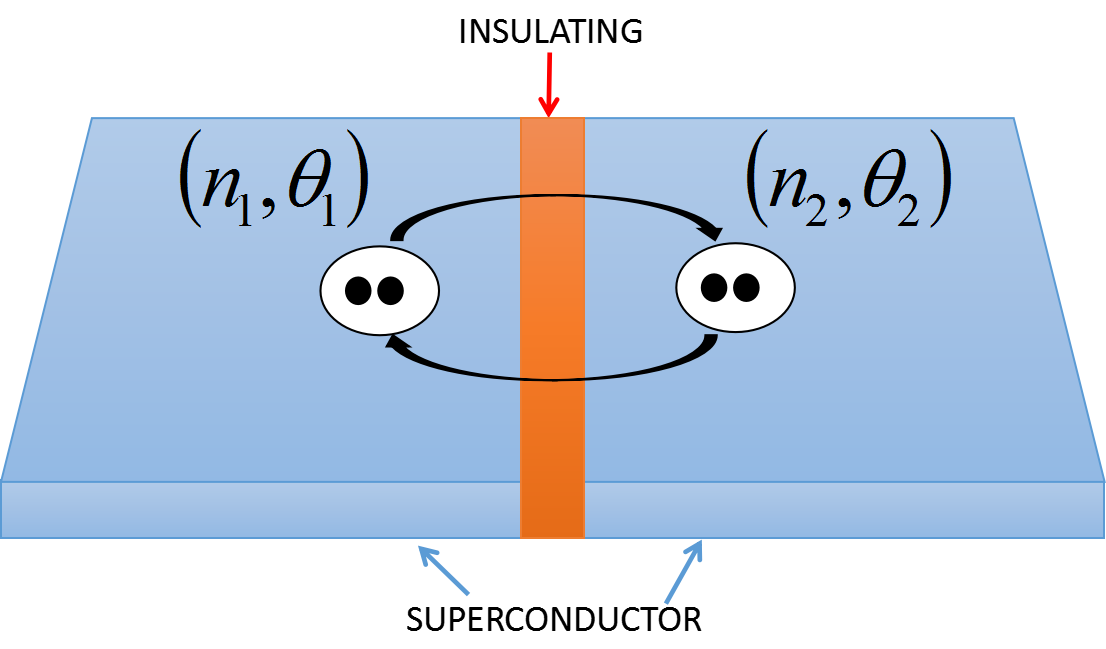}
		\caption{ Schematic of a Josephson tunnel junction. It consists of two superconductors separated by an insulating barrier. The two sides of the junction are characterized by the number of Cooper pairs ${{n_1}}$, ${{n_2}}$ and the phases ${{\theta _1}}$, ${{\theta _2}}$. The charge ${n_2} - {n_1}$ and phase  difference ${\theta _2} - {\theta _1}$ between the two metals are the essential parameters describing the behavior of the junction.}
		\label{F21}
	\end{figure}

	Josephson junctions  are based on the Josephson effect\cite{105,106}. A Josephson junction consists of two superconductors separated by an insulating barrier which limits the flow of supercurrent between the two superconductors, as shown in the figure(\ref{F21}).  Cooper pairs can coherently tunnel from one superconductor to the other, with the supercurrent  $I$ that is given by
	\begin{eqnarray}\label{E21}
	I = {I_c}\sin \left( \phi  \right), 
	\end{eqnarray}
	where ${I_c}$ is the critical current of the junction, and $\phi  = {\theta _2} - {\theta _1}$ is a  phase diﬀerence across the junction, which  is related to  the potential $V$ between the two superconductors  according  to 
	\begin{eqnarray}\label{E22}
	\frac{{d\phi }}{{dt}} = \frac{{2\pi }}{{{\Phi _0}}}V ,
	\end{eqnarray}  
	where ${\Phi _0} = {h \mathord{\left/
			{\vphantom {h {2e}}} \right.
			\kern-\nulldelimiterspace} {2e}} = {2,07.10^{ - 15}}.T.{m^{ - 15}}$ is the magnetic flux quantum.
	\section*{Equivalent electric circuit of the Josephson junction}
	
\begin{figure}[H]
			\centering \includegraphics[scale=0.7]{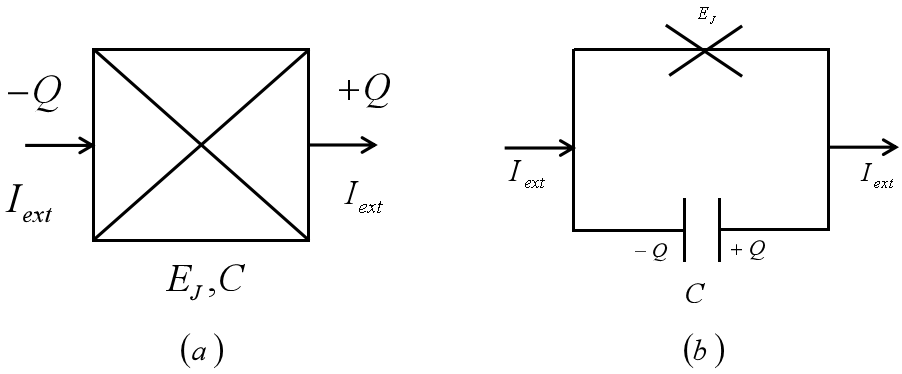}
			\caption{ Schematic of a Josephson tunnel junction. (a) The symbol of the Josephson junction (Cross in a box)
				.(b) The equivalent electrical circuit of the Josephson junction.}
			\label{F22}
		\end{figure}

	Physically,   the Josephson junction is symbolized by a cross in a box and can be modeled as an ideal Josephson element with the Josephson energy ${E_J}$ shunted by a  capacitance ${C_J}$\cite{107},  as shown in Figure (\ref{F22}). We introduce the number of Cooper pairs $N$ by $Q = 2eN$, with $Q$ is the electrode charge, and the voltage across the junction is given by $Q = CV$. We  can then determine the dynamics of a  phase difference $\phi $  by using the Kirchhoff’s rule for the circuit shown in Fig. (\ref{F22}.b) to  give the differential equation 
	\begin{eqnarray}\label{E23}
	{I_{ext}} &=& {I_c}\sin \phi  + C\frac{{dV}}{{dt}}\nonumber\\
	&=& {I_c}\sin \phi  + \frac{h}{{2e}}C\ddot \phi. 
	\end{eqnarray}
	 From this equation, we readily find the Lagrangian of a  Josephson junction 
	 
	 \begin{eqnarray}\label{E24}
	 L = \frac{1}{2}\frac{{C{\hbar ^2}}}{{4{e^2}}}{\ddot \phi ^2} + {I_c}\frac{\hbar }{{2e}}\cos \phi  + {I_{ext}}\frac{\hbar }{{2e}}\phi  = K - U,
	 \end{eqnarray}
	 where $K = {1 \mathord{\left/
	 		{\vphantom {1 2}} \right.
	 		\kern-\nulldelimiterspace} 2}C{V^2}$ is the kinetic term, and the potential energy of the system $U$ is given by 
	 \begin{eqnarray}\label{E25}
	 U =  - \frac{{{I_c}\hbar }}{{2e}}\cos \phi  - {I_{ext}}\frac{\hbar }{{2e}}\phi. 
	 \end{eqnarray}
		The dynamics of the circuit  (Hamiltonian $H$)  is associated with the Lagrangian $L$ through 
		\begin{eqnarray}\label{E26}
		H &=& p\dot \phi  - L\nonumber\\
		&=& \frac{1}{2}\frac{{{E_C}}}{{{\hbar ^2}}}{p^2} - p\cos \phi  - {E_J}\frac{{{I_{ext}}}}{{{I_c}}}\phi,  
		\end{eqnarray}
		where:\\
		- $p =  - \frac{{\partial H}}{{\partial \dot \phi }} = \frac{{C{\hbar ^2}}}{{4{e^2}}}\dot \phi  = \frac{\hbar }{{2e}}Q$ is the canonical momentum operator conjugate to $\phi $.\\
		- ${E_C} = \frac{{{{\left( {2e} \right)}^2}}}{C}$ is the charging energy of a Cooper pair, i.e. the energy required to increase the number of Cooper pairs in the box.\\
		- ${E_J} = \frac{{{I_0}\hbar }}{{2e}}$ is the Josephson tunneling energy.\\
		And the Hamiltonian equations of motion
		\begin{eqnarray}\label{E27}
		\dot \phi  = \frac{{\partial H}}{{\partial p}}, \dot p =  - \frac{{\partial H}}{{\partial \phi }}.
		\end{eqnarray}
		
		\subsection{SQUID}
\begin{figure}[H]
			\centering \includegraphics[scale=1]{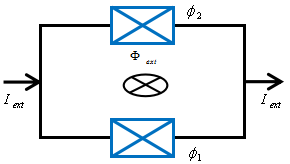}
			\caption{ Schematic of a SQUID. The two Josephson junctions form a loop, which is crossed by a flux magnetic field. }
			\label{F23}
		\end{figure}

		There is a variant of the Josephson junction called "SQUID" \cite{108}((Superconducting Quantum Interference Device), which consists of two ideally identical Josephson junctions in parallel coupled to a current source as shown in figure (\ref{F23}).\\ 
		According to the relation of Stockes, the magnetic flux through the loop in the presence of a magnetic field which  describes  the  vector potential $\mathop A\limits^ \to  \left( r \right)$  is expressed by
		\begin{eqnarray}\label{E28}
		{\Phi _{ext}} = \oint_C {\mathop A\limits^ \to  } \left( r \right)dr.
		\end{eqnarray}
		As known, the phase variation between two points in the case of a closed superconducting loop is given by
		\begin{eqnarray}\label{E29}
		{\phi _2} - {\phi _1} = \oint_C {\mathop A\limits^ \to  } \left( r \right)dr = 2k\pi, 
		\end{eqnarray}
		and thus by combining the equations (\ref{E28}) and (\ref{E29}), the magnetic flux  ${\Phi _{ext}}$ can be expressed as follows
		\begin{eqnarray}\label{E210}
		{\Phi _{ext}} = \frac{{{\Phi _0}}}{{2\pi }}\left( {{\phi _2} - {\phi _1}} \right).
		\end{eqnarray} 
		The Hamiltonian (\ref{E26}) becomes
		\begin{eqnarray}\label{E211}
		H = \frac{1}{2}\frac{{{E_C}}}{{{\hbar ^2}}}{p^2} - {U_T},
		\end{eqnarray}
		with ${U_T} = {U_1} + {U_2}$: is the potential energy of the SQUID (the sum of the Josephson energies of the two junctions), which is given by
		\begin{eqnarray}\label{E212}
		{U_T} =  - {E_J}\left( {\cos \left( {{\phi _1}} \right) + \cos \left( {{\phi _2}} \right)} \right) =  - 2{E_J}\cos \left( {\frac{{{\phi _1} + {\phi _1}}}{2}} \right)\cos \left( {\frac{{{\phi _2} -{\phi _1}}}{2}} \right).
		\end{eqnarray}
		We assume that the phase  $\phi  = {{\left( {{\phi _1} + {\phi _2}} \right)} \mathord{\left/
				{\vphantom {{\left( {{\phi _1} + {\phi _2}} \right)} 2}} \right.
				\kern-\nulldelimiterspace} 2}$. Then the potential term is written in the
		form
		\begin{eqnarray}\label{E213}
		{U_T} &=&  - {E_J}\cos \left( {\pi \frac{{{\Phi _{ext}}}}{{{\Phi _0}}}} \right)\cos \left(\phi  \right)\nonumber\\
		&=&  - {E_J}\left( {{\Phi _{ext}}} \right)\cos \left( \phi \right),
		\end{eqnarray}
		where the effective Josephson energy  ${E_J}\left( {{\Phi _{ext}}} \right) =  - {E_J}\cos \left( {\pi {{{\Phi _{ext}}} \mathord{\left/
					{\vphantom {{{\Phi _{ext}}} {{\Phi _0}}}} \right.
					\kern-\nulldelimiterspace} {{\Phi _0}}}} \right)$ can be tuned via a magnetic flux ${{\Phi _{ext}}}$ applied through the loop of the SQUID.

	\subsection{Charge qubit( The Cooper pair box)}
	In the classical case, superconductivity is attributed to electron pairs, which can exist within the island: the pairs can either be inside or outside the island. However, if the tunnel barrier is large enough, as dictated by the uncertainty principle, it becomes possible for a pair to simultaneously exist both inside and outside the island, creating a superposition of $0$ and $1$.
	
\begin{figure*}
		\centering \includegraphics[scale=0.5]{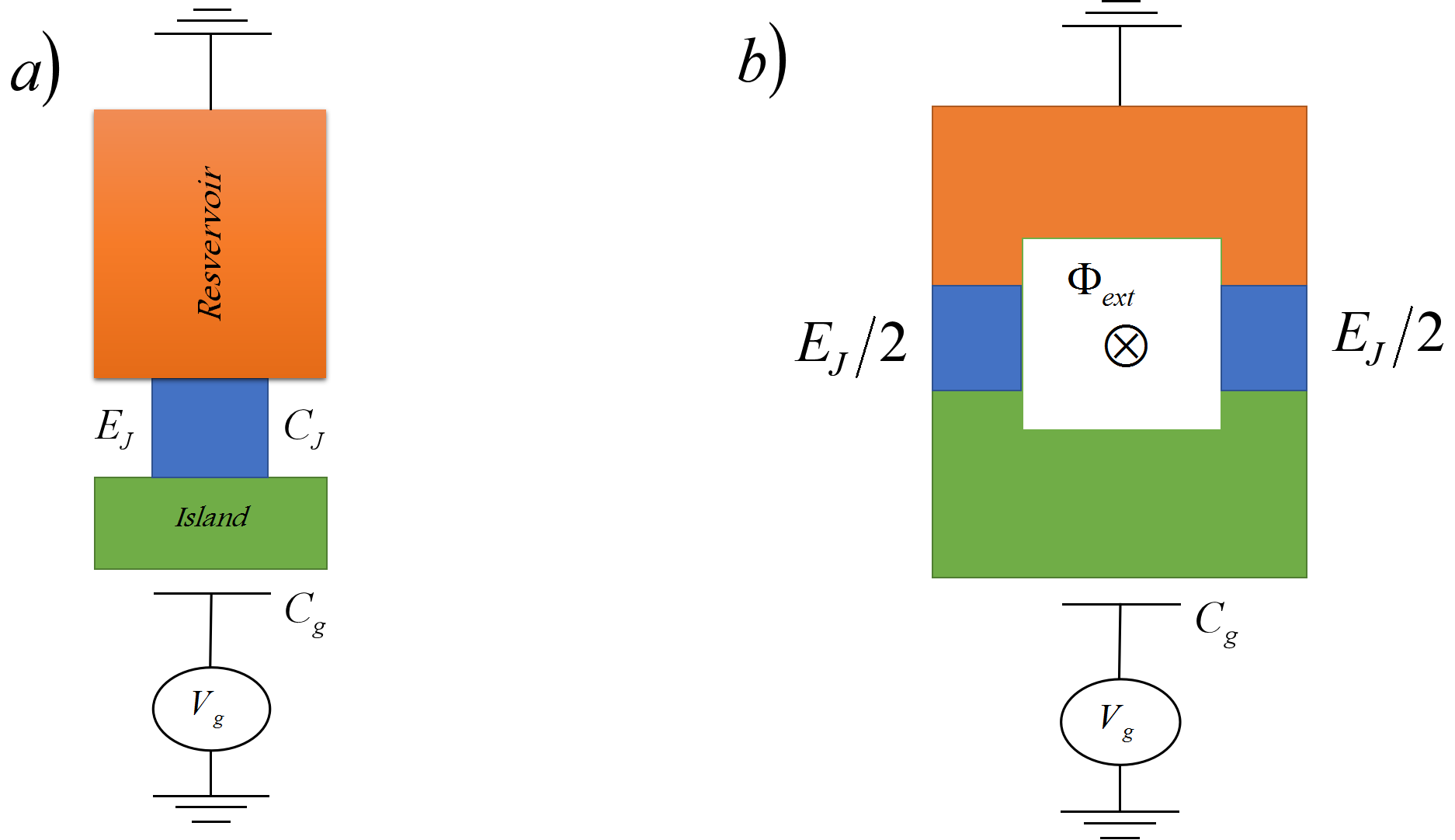}
		\caption{ a) Schematic of the Cooper pair box consisting of a Josephson junction (blue) which connects an island (green) with a reservoir (orange). b) The Josephson junction is replaced by  a SQUID in  which    the qubit  becomes more controllable by a magnetic flux.  }
		\label{F24}
	\end{figure*}

	The possible quantum bit, known as the "Cooper pair box," was first theoretically proposed in ref.\cite{109} and later experimentally achieved  in ref. \cite{110}. Such the Cooper pair box consists  of a superconducting island (green), which is connected via a Josephson  junction (blue) with capacitance $C_J$ and Josephson energy ${E_J}$ to a large superconducting electrode (orange) as shown  in figure \ref{F24}. The quantum variable is the number of superconducting pairs that are attracted to the island by a gate potential ${V_g}$  applied to the gate capacitance ${C_g}$.
	
	\section*{Electric circuit  corresponding to  the Cooper  pair  box }
\begin{figure}[H]
		\centering \includegraphics[scale=0.5]{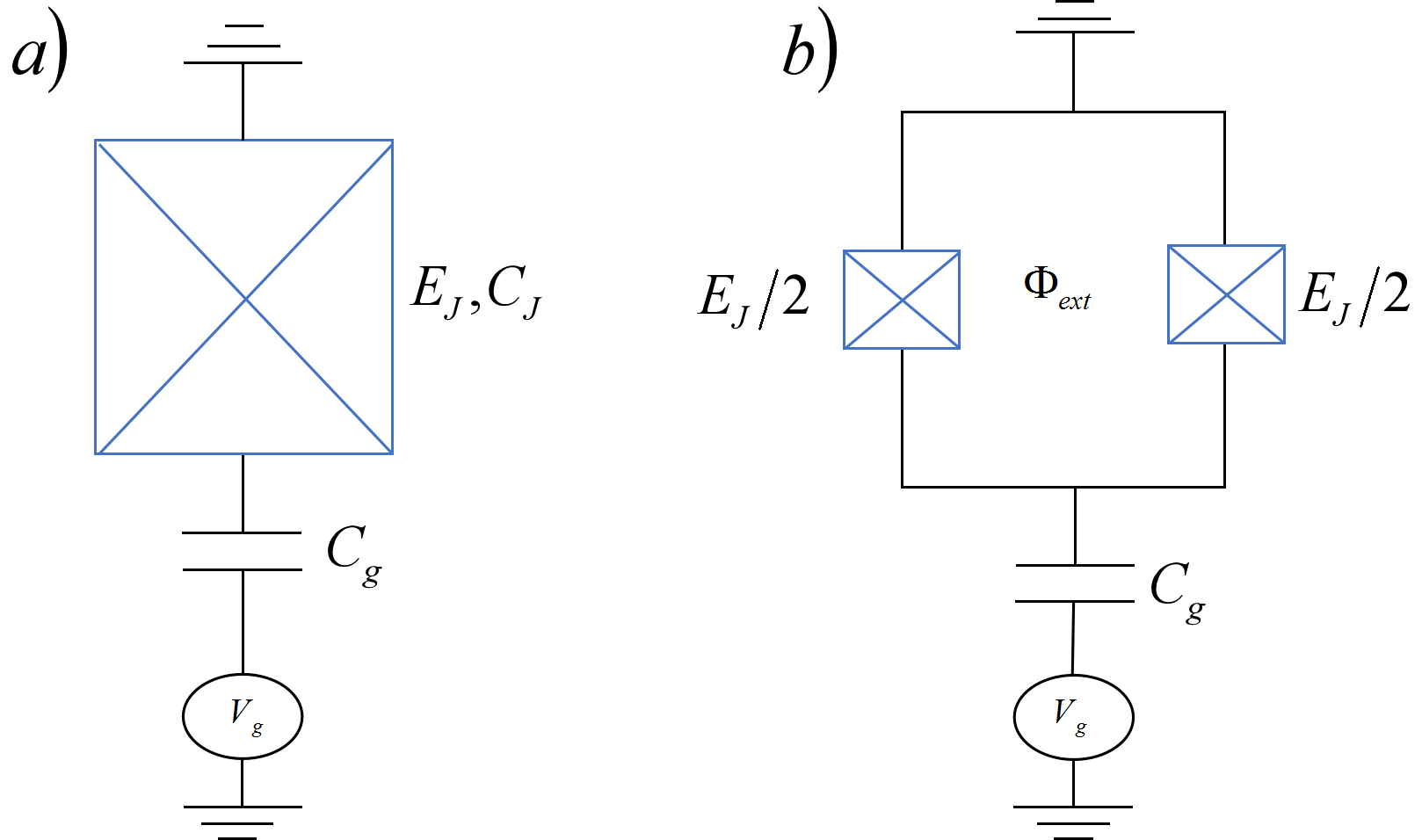}
		\caption{ a) Circuit diagram corresponding to the Cooper pair box schematic in Fig. (\ref{F24}a). b) Circuit diagram corresponding to the Cooper pair box schematic in Fig. (\ref{F24}b).  }
		\label{F25}
	\end{figure}

	The superconducting circuit consists  either of a Josephson junction having a Josephson energy ${E_J}$ and an intrinsic capacitance ${C_J}$  connected to a source of potential via a capacitance ${C_g}$  as shown
	in the figure (\ref{F25}a) or the loop SQUID in figure(\ref{F25}b).
	
	We apply on the figure (\ref{F25}) the same procedure that  we have  seen in the previous paragraphs (the Kirchhoff's rule), we can easily identify the following Lagrangian
	\begin{eqnarray}\label{E214}
	L = K -U,
	\end{eqnarray}
	where $U =  - {E_J}\cos \phi $  is the potential energy, which is already determined in the previous sections and the kinetic energy $K$  is  the sum of the capacitive energies of the circuit, which  is given by 
	 \begin{eqnarray}\label{E215}
	 K &=& \frac{1}{2}{C_J}V_J^2 + \frac{1}{2}{C_g}{\left( {{C_J} - {V_g}} \right)^2}\nonumber\\
	 &=& \frac{{{C_J} + {C_g}}}{2}{\left( {{V_J} - \frac{{{C_g}}}{{{C_J} + {C_g}}}{V_g}} \right)^2} + \frac{1}{2}{\left( {{C_g} - \frac{{C_g^2}}{{{C_J} + {C_g}}}} \right)^2}.
	 \end{eqnarray} 
	 
	 By using the  relation in equation (\ref{E22}) and omitting a constant term (independent of $\phi $ and ${\dot \phi }$) in equation (\ref{E23}) because it does not influence the  system dynamics, the Lagrangian
	 can be written as
	 \begin{eqnarray}\label{E216}
	 L = \frac{{{C_\Sigma }}}{2}{\left( {\frac{\hbar }{{2e}}\dot \phi  - \frac{{{C_g}}}{{{C_\Sigma }}}{V_g}} \right)^2} - {E_J}\cos \phi,
	 \end{eqnarray}
	where ${{C_\Sigma } = {C_J} + {C_g}}$. Here the kinetic energy  can be controlled  by the external gate
	voltage, whereas the potential energy, is however fixed by the energy  $E_J$.
	
	As shown in figure (\ref{F25}b), the single Josephson junction is replaced by a loop  SQUID which consists of two identical junctions (${E_{J,1}} = {E_{J,2}} = {{{E_J}} \mathord{\left/
			{\vphantom {{{E_J}} 2}} \right.
			\kern-\nulldelimiterspace} 2}$),   in order to add additional control for potential energy via a magnetic flux.  The potential  term in the Lagrangian (\ref{E216}) can be replaced by  the potential term (\ref{E216}), the corresponding  Lagrangian is then  given  by
	\begin{eqnarray}\label{E217}
	L = \frac{{{C_\Sigma }}}{2}{\left( {\frac{\hbar }{{2e}}\dot \phi  - \frac{{{C_g}}}{{{C_\Sigma }}}{V_g}} \right)^2} - {E_J}\left( {{\Phi _{ext}}} \right)\cos \phi.
	\end{eqnarray}
	
	The canonical momentum operator corresponding to (\ref{E215}) with respect to $\dot \phi $  is given  by $p = {{\partial K} \mathord{\left/
			{\vphantom {{\partial K} {\partial \dot \phi  = \left( {{\hbar  \mathord{\left/
									{\vphantom {\hbar  {2e}}} \right.
									\kern-\nulldelimiterspace} {2e}}} \right)Q}}} \right.
			\kern-\nulldelimiterspace} {\partial \dot \phi  = \left( {{\hbar  \mathord{\left/
						{\vphantom {\hbar  {2e}}} \right.
						\kern-\nulldelimiterspace} {2e}}} \right)Q}}$,  (with $Q = \left( {{C_J} + {C_g}} \right){V_J} - {C_g}{V_g} =  - 2eN$),  which  implies $\left( {{C_J} + {C_g}} \right)V_J^2 = {{{{{\left( {2e} \right)}^2}} \mathord{\left/
				{\vphantom {{{{\left( {2e} \right)}^2}} {\left( {{C_J} + {C_g}} \right)\left( {N - {{{C_g}{N_g}} \mathord{\left/
										{\vphantom {{{C_g}{N_g}} {2e}}} \right.
										\kern-\nulldelimiterspace} {2e}}} \right)}}} \right.
				\kern-\nulldelimiterspace} {\left( {{C_J} + {C_g}} \right)\left( {N - {{{C_g}{N_g}} \mathord{\left/
							{\vphantom {{{C_g}{N_g}} {2e}}} \right.
							\kern-\nulldelimiterspace} {2e}}} \right)}}^2}$.
	
	 Finally,  we can identify the Hamiltonian of the following system:  
	 \begin{eqnarray}\label{E218}
	 H = \frac{{{E_c}}}{2}{\left( {N - {N_g}} \right)^2} - {E_J}\left( {{\Phi _{ext}}} \right)\cos \phi, 
	 \end{eqnarray} 
	 where ${E_c} = {{{{(2e)}^2}} \mathord{\left/
	 		{\vphantom {{{{(2e)}^2}} {\left( {{C_J} + {C_g}} \right)}}} \right.
	 		\kern-\nulldelimiterspace} {\left( {{C_J} + {C_g}} \right)}}$ and ${N_g} = {{{C_g}{V_g}} \mathord{\left/
	 		{\vphantom {{{C_g}{V_g}} {2e}}} \right.
	 		\kern-\nulldelimiterspace} {2e}}$.
	 
	 We consider that  the electrostatic energy is dominant compared to the Josephson energy(${E_C} \gg {E_J}$), it is called the charge qubit regime. To quantitatively describe the Cooper pair box, we define the operator $\hat N$ which is associated with the number of Cooper pairs  in the base $\left| N \right\rangle $ as follows 
	 \begin{eqnarray}\label{E219}
	 \hat N = \sum\limits_{N \in Z} {N\left| N \right\rangle \left\langle N \right|}.
	 \end{eqnarray}
	  
	The phase difference $\phi $ across the Josephson junction presents fluctuations related  to $N$ by the following uncertainty relation: $\Delta N.\Delta \phi  = {1 \mathord{\left/
			{\vphantom {1 2}} \right.
			\kern-\nulldelimiterspace} 2}$, assuming that the fluctuations of the number of excess Cooper pairs  $N$ is  very small to that of  $\phi $, the canonical commutation relation $\left[ {N,\phi } \right] = i$ and $N = i{\partial  \mathord{\left/
			{\vphantom {\partial  {\partial \phi }}} \right.
			\kern-\nulldelimiterspace} {\partial \phi }}$, and thus  the basis $\left| \phi  \right\rangle $ can be introduced as
	\begin{eqnarray}\label{E220}
	\left| \phi  \right\rangle  = \sum\limits_{N \in Z} {{e^{ - iN\phi }}\left| N \right\rangle }, 
	\end{eqnarray}
and by using the Fourier transform\cite{111}, the number states $\left| N \right\rangle $ can be laid to the phase difference ${\left| \phi  \right\rangle }$ states
	 \begin{eqnarray}\label{E221}
	 \left| N \right\rangle  = \frac{1}{{2\pi }}\int_0^{2\pi } {d\phi {e^{iN\phi }}\left| \phi  \right\rangle }.
	 \end{eqnarray}
  
	 From equation (\ref{E221}), we notice the following identity ${e^{i\phi }}\left| N \right\rangle  \approx \left| {N + 1} \right\rangle $, and we have $\cos \phi  = {{\left( {{e^{i\phi }} + {e^{ - i\phi }}} \right)} \mathord{\left/
	 		{\vphantom {{\left( {{e^{i\phi }} + {e^{ - i\phi }}} \right)} 2}} \right.
	 		\kern-\nulldelimiterspace} 2} = {{\left( {\left| N \right\rangle \left\langle {N + 1} \right| + \left| {N + 1} \right\rangle \left\langle N \right|} \right)} \mathord{\left/
	 		{\vphantom {{\left( {\left| N \right\rangle \left\langle {N + 1} \right| + \left| {N + 1} \right\rangle \left\langle N \right|} \right)} 2}} \right.
	 		\kern-\nulldelimiterspace} 2}$. We can then write the Hamiltonian of the Cooper pair box in the matrix form
	 \begin{eqnarray}\label{E222}
	 H = \frac{{{E_C}}}{2}\sum\limits_N {{{\left( {N - {N_g}} \right)}^2}\left| N \right\rangle \left\langle {N + 1} \right| - \frac{{{E_J}}}{2}\sum\limits_N {\left( {\left| N \right\rangle \left\langle {N + 1} \right| + \left| {N + 1} \right\rangle \left\langle N \right|} \right)} }.
	 \end{eqnarray}
	 
\begin{figure}[H]
	\centering \includegraphics[scale=0.5]{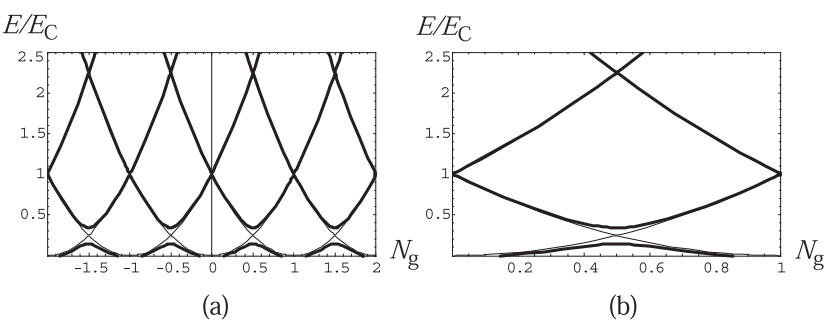}
	\caption{ Energy spectrum of  the charge qubit  for the ratio as a function of gate charge $N_g$. The  ${E \mathord{\left/
				{\vphantom {E {{E_C}}}} \right.
				\kern-\nulldelimiterspace} {{E_C}}}$ is lower,   the first two levels are isolated from the third for values of $N_g$ near   the half-integer. a)  the    bold  curve designates the energy spectrum of the Hamiltonian with $E_C$ = 0.2, the fine curve with ${E \mathord{\left/
				{\vphantom {E {{E_C}}}} \right.
				\kern-\nulldelimiterspace} {{E_C}}} = 0$. b) the energy spectrum for $0 \le {N_g} \le 1$.  }
	\label{F26}
\end{figure}

	 We can diagonalize the Hamiltonian (\ref{E222}) and obtain the  energy spectrum as a function of the gate charge $N_g$, which will be composed of a set of parabolas centered around the integer values of  $N_g$. We see that for half-integer values of $N_g$ (${N_g} = N + {1 \mathord{\left/
	 		{\vphantom {1 2}} \right.
	 		\kern-\nulldelimiterspace} 2}$ , $N \in {\rm Z}$) the state of charge is twice degenerate, more precisely the states $\left| N \right\rangle $ and $\left| {N + 1} \right\rangle $ have the same energy. The Cooper box in the ${E_C} \gg {E_J}$ regime is a good two-level system, especially if we skew the grid voltage so that we have $N_g$ $=$ 0.5. At this point, the quantum states are the superposition  of the  symmetric and anti-symmetric of a state corresponding to   $N$ Cooper pairs and  $N + 1$ pairs of Cooper on the superconducting island, we have $\left| 0 \right\rangle  = {{\left( {\left| {N + 1} \right\rangle  - \left| N \right\rangle } \right)} \mathord{\left/
	 		{\vphantom {{\left( {\left| {N + 1} \right\rangle  - \left| N \right\rangle } \right)} {\sqrt 2 }}} \right.
	 		\kern-\nulldelimiterspace} {\sqrt 2 }}$ and $\left| 1 \right\rangle  = {{\left( {\left| {N + 1} \right\rangle  + \left| N \right\rangle } \right)} \mathord{\left/
	 		{\vphantom {{\left( {\left| {N + 1} \right\rangle  + \left| N \right\rangle } \right)} {\sqrt 2 }}} \right.
	 		\kern-\nulldelimiterspace} {\sqrt 2 }}$ as indicated in  figure(\ref{F26}).
	 
	 The energy levels of the Hamiltonian (\ref{E222}) being very anharmonic, we can restrict to only two levels and thus we can consider the circuit as a quantum bit.
	 
	  For $N = 0$ the Hamiltonian (\ref{E222}) becomes 
	  \begin{eqnarray}\label{E223}
	  H = \frac{{{E_C}}}{2}N_g^2\left| 0 \right\rangle \left\langle 0 \right| - \frac{{{E_J}}}{2}\left( {\left| 0 \right\rangle \left\langle 1 \right| + \left| 1 \right\rangle \left\langle 0 \right|} \right).
	  \end{eqnarray}
	  
	  For $N=1$ 
	  \begin{eqnarray}\label{E224}
	  H = \frac{{{E_C}}}{2}{\left( {1 - {N_g}} \right)^2}\left| 1 \right\rangle \left\langle 1 \right| - \frac{{{E_J}}}{2}\left( {\left| 1 \right\rangle \left\langle 2 \right| + \left| 2 \right\rangle \left\langle 1 \right|} \right).
	  \end{eqnarray}
	 
	 From equations (\ref{E223}) and (\ref{E224}), one can determine the elements of the matrix of (\ref{E222}) in the states of charge $\left\{ {\left| 0 \right\rangle ,\left| 1 \right\rangle } \right\}$ as follows:
	 
	 \begin{eqnarray}\label{E225}
	 H =  - \frac{{{E_C}}}{2}\left( {1 - 2{N_g}} \right){\sigma _z} - \frac{{{E_J}}}{2}{\sigma _x},
	 \end{eqnarray}
	 where ${\sigma _z}$ and ${\sigma _x}$ are the Pauli matrices. The Hamiltonian of the charge qubit is therefore written in the following form
	  \begin{eqnarray}\label{E226}
	  H =  - \frac{1}{2}{B_z}{\sigma _z} - \frac{1}{2}{B_x}{\sigma _x}, with~{B_z} = \frac{{{E_C}}}{2}\left( {1 - 2{N_g}} \right); {B_x} = {E_J}.
	  \end{eqnarray}
	  
	  The corresponding  eigenvalues and   eigenstates   of this Hamiltonian are given by
	  \begin{eqnarray}\label{E227}
	  {E_0} =  - \frac{1}{2}\sqrt {B_z^2 + B_x^2};~{E_1} = \frac{1}{2}\sqrt {B_z^2 + B_x^2},
	  \end{eqnarray}
	  \begin{eqnarray}\label{E228}
	  \left|  +  \right\rangle  = \cos \theta \left|  \uparrow  \right\rangle  + \sin \theta \left|  \downarrow  \right\rangle;~\left|  -  \right\rangle  =  - \sin \theta \left|  \uparrow  \right\rangle  + \cos \theta \left|  \downarrow  \right\rangle,
	  \end{eqnarray}
	  with the expression of $\theta $ is given by the following relation $\theta  = \left( {{1 \mathord{\left/
	  			{\vphantom {1 2}} \right.
	  			\kern-\nulldelimiterspace} 2}} \right){\tan ^{ - 1}}\left[ {{{2{E_J}} \mathord{\left/
	  			{\vphantom {{2{E_J}} {{E_C}\left( {1 - 2{N_g}} \right)}}} \right.
	  			\kern-\nulldelimiterspace} {{E_C}\left( {1 - 2{N_g}} \right)}}} \right] = \left( {{1 \mathord{\left/
	  			{\vphantom {1 2}} \right.
	  			\kern-\nulldelimiterspace} 2}} \right){\tan ^{ - 1}}\left[ {{{{B_x}} \mathord{\left/
	  			{\vphantom {{{B_x}} {{B_z}}}} \right.
	  			\kern-\nulldelimiterspace} {{B_z}}}} \right]$.
	 \subsection{Transmon qubit }
\begin{figure}[H]
	 	\centering \includegraphics[scale=0.5]{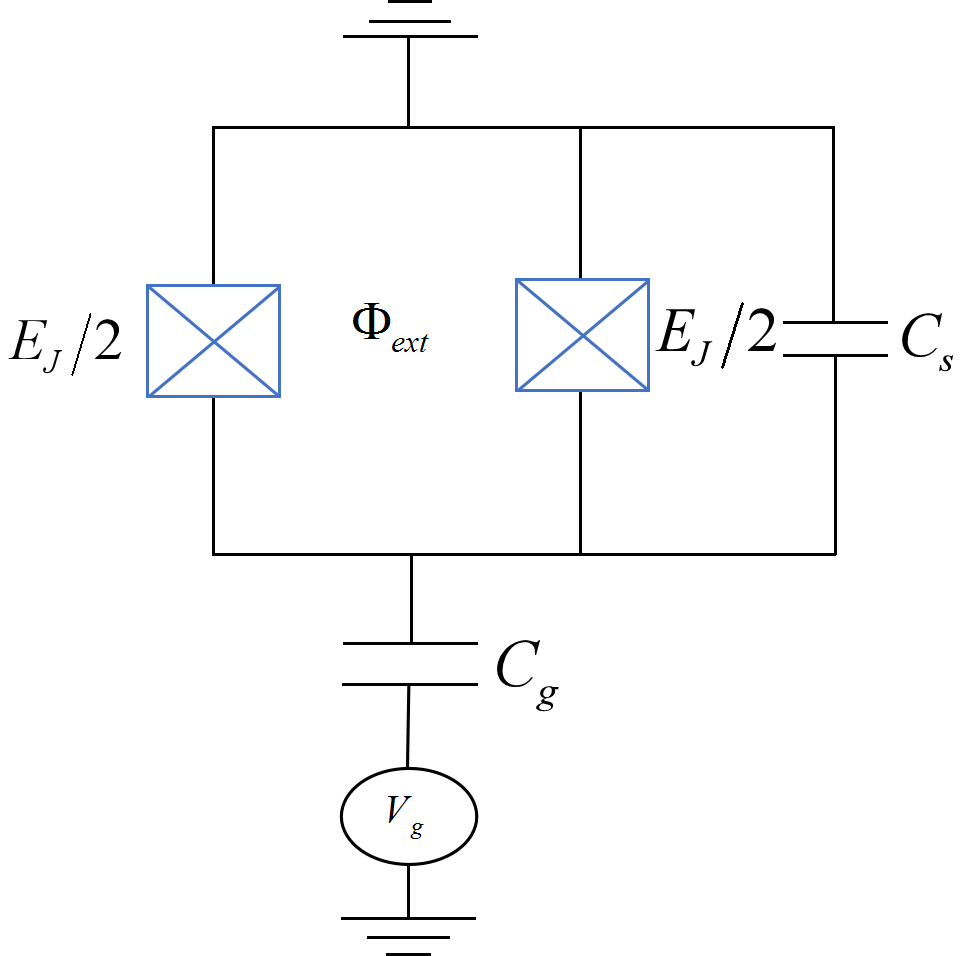}
	 	\caption{ The equivalent circuit of the Transmon qubit, an additional capacitance (Cs) has been added
	 		in parallel to the junction.  }
	 	\label{F27}
	 \end{figure}

\begin{figure}[H]
	 	\centering \includegraphics[scale=0.5]{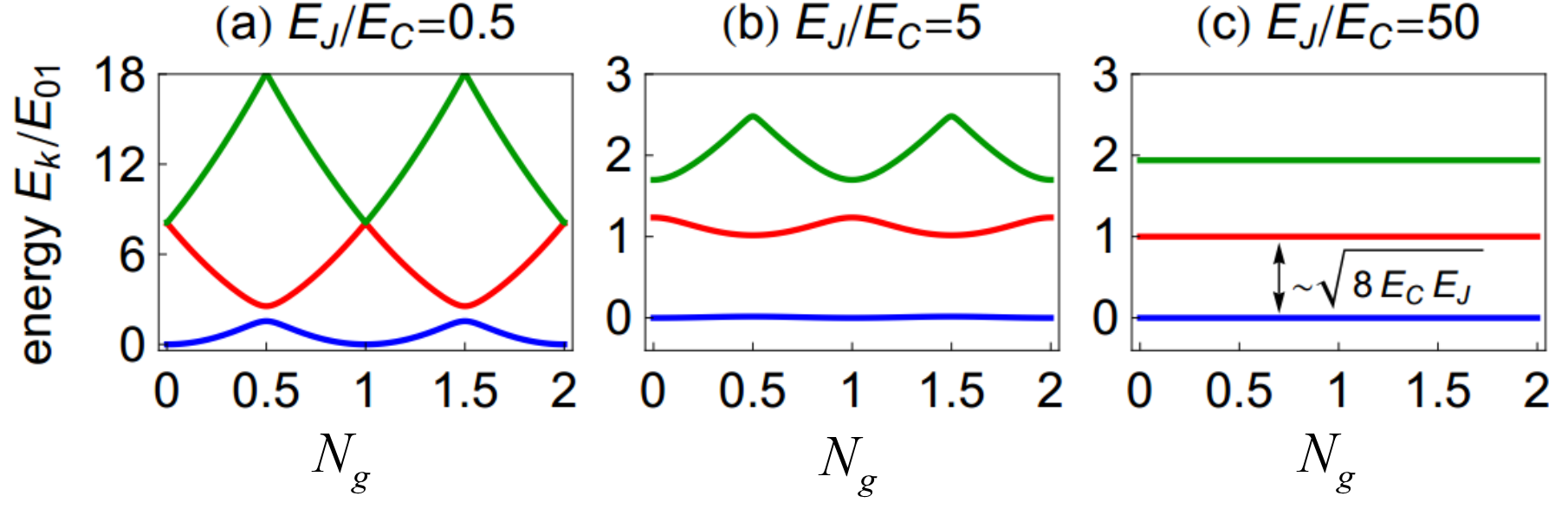}
	 	\caption{ Eigenenergies for three energy levels ($k = 0,1,2$) of the Cooper  pair box Hamiltonian  (\ref{E231})  for different ratios ${{{E_J}} \mathord{\left/
	 				{\vphantom {{{E_J}} {{E_C}}}} \right.
	 				\kern-\nulldelimiterspace} {{E_C}}}$. For very large ${{{E_J}} \mathord{\left/
	 				{\vphantom {{{E_J}} {{E_C}}}} \right.
	 				\kern-\nulldelimiterspace} {{E_C}}}$, the spectrum practically  becomes
	 		that of the harmonic oscillator. The system gets less sensitive to gate charge when   the ${{{E_J}} \mathord{\left/
	 				{\vphantom {{{E_J}} {{E_C}}}} \right.
	 				\kern-\nulldelimiterspace} {{E_C}}}$ is increased from the charging regime with ${{{E_J}} \mathord{\left/
	 				{\vphantom {{{E_J}} {{E_C}}}} \right.
	 				\kern-\nulldelimiterspace} {{E_C}}} = 0.5$ (a) to the transmon regime with  ${{{E_J}} \mathord{\left/
	 				{\vphantom {{{E_J}} {{E_C}}}} \right.
	 				\kern-\nulldelimiterspace} {{E_C}}} = 50$ (c). The figure used from Ref.\cite{65}. }
	 	\label{F28}
	 \end{figure}

	 The difference between the transmon qubit and the charge qubit is that the transmon  qubit is    independent    of  the gate charge by going  to another regime where ${E_J} \gg {E_C}$,  which  is so-called the transmon  regime.  This is done   at the level of the equivalent circuit of the superconducting qubits  by  inroducing  the addition of a large shunting capacitance  placed in parallel with the Josephson junction to increase the overall capacitance as illustrated in figure (\ref{F27}).
	 The increased ${{{E_J}} \mathord{\left/
	 		{\vphantom {{{E_J}} {{E_C}}}} \right.
	 		\kern-\nulldelimiterspace} {{E_C}}}$ ratio makes the Cooper box less sensitive to gate charge and its anharmonicity, leading to the flattening energy bands, whereas the transmon frequency can be tuned by applying the magnetic flux through the loop of SQUID. Such energy bands are illustrated in figure (\ref{F28}), where three eigenenergies ${E_k}$ (k=1,2,3) of the  Cooper pair box  Hamiltonian (\ref{E222}) are shown
	 as a function of gate charge $N_g$ for different values of the  ratio ${{{E_J}} \mathord{\left/
	 		{\vphantom {{{E_J}} {{E_C}}}} \right.
	 		\kern-\nulldelimiterspace} {{E_C}}}$. The      sensitivity  to  charge noise is defined as the charge dispersion,  which is the transition energy between  two   neighboring energy levels
	 \begin{eqnarray}\label{E229}
	 { \in _k} = {E_{k,k + 1}}\left( {{N_g} = 0.5} \right) - {E_{k,k + 1}}\left( {{N_g} = 0} \right).
	 \end{eqnarray}
	 
	 Owing  to  the  charge dispersion  decreases exponentially fast as the ratio  ${{{E_J}} \mathord{\left/
	 		{\vphantom {{{E_J}} {{E_C}}}} \right.
	 		\kern-\nulldelimiterspace} {{E_C}}}$ is increased as shown  in  Ref. \cite{43}
	 \begin{eqnarray}\label{E230}
	 { \in _k}\infty {e^{ - i\sqrt {{{8{E_J}} \mathord{\left/
	 					{\vphantom {{8{E_J}} {{E_C}}}} \right.
	 					\kern-\nulldelimiterspace} {{E_C}}}} }}.
	 \end{eqnarray}
	 
	  The transmon Hamiltonian is the same as that of the hamiltonian (\ref{E218}), with only change being the charging energy ${E_C}$:
	  \begin{eqnarray}\label{E231}
	  H = \frac{{{E_C}}}{2}{\left( {N - {N_g}} \right)^2} - {E_J}\left( {{\Phi _{ext}}} \right)\cos \phi,  
	  \end{eqnarray}
	  where ${E_C} = {{{{\left( {2e} \right)}^2}} \mathord{\left/
	  		{\vphantom {{{{\left( {2e} \right)}^2}} {\left( {{C_J} + {C_g} + {C_s}} \right)}}} \right.
	  		\kern-\nulldelimiterspace} {\left( {{C_J} + {C_g} + {C_s}} \right)}}$ is the charging energy.
	  Since ${{{E_J}} \mathord{\left/
	  		{\vphantom {{{E_J}} {{E_C} \gg 1}}} \right.
	  		\kern-\nulldelimiterspace} {{E_C} \gg 1}}$, we can consider that the fluctuations in $\phi $ in the Hamiltonian (\ref{E231}), are very weak and make a development of the cosine of the tunnel energy term  $- {E_J}\left( {{\Phi _{ext}}} \right)\cos \phi  \simeq  - {E_J}\left( {{\Phi _{ext}}} \right) + \left( {{{{E_J}\left( {{\Phi _{ext}}} \right)} \mathord{\left/
	  			{\vphantom {{{E_J}\left( {{\Phi _{ext}}} \right)} 2}} \right.
	  			\kern-\nulldelimiterspace} 2}} \right){\phi ^2} - \left( {{{{E_J}\left( {{\Phi _{ext}}} \right)} \mathord{\left/
	  			{\vphantom {{{E_J}\left( {{\Phi _{ext}}} \right)} {24}}} \right.
	  			\kern-\nulldelimiterspace} {24}}} \right){\phi ^4}$.  The Hamiltonian (\ref{E231}) becomes
	  \begin{eqnarray}\label{E232}
	  H = \frac{{{E_C}}}{2}{\left( {N - {N_g}} \right)^2} - {E_J}\left( {{\Phi _{ext}}} \right) + \frac{{{E_J}\left( {{\Phi _{ext}}} \right)}}{2}{\phi ^2} - \frac{{{E_J}\left( {{\Phi _{ext}}} \right)}}{{24}}{\phi ^4}. 
	  \end{eqnarray}
	 In figure (\ref{F28}), we observe that in the Transmon system, the energy levels are independent of the charging shift $N_g$. Therefore, we can omit this term from the Hamiltonian. Using the gauge transformation ${U_t} = {e^{i{N_g}\phi }}$, and thus
	 \begin{eqnarray}\label{E233}
	 H \approx {U_t}HU_t^ +  = \frac{{{E_C}}}{2}{N^2} + \frac{{{E_J}}}{2}{\phi ^2} - {E_J} - \frac{{{E_J}}}{{24}}{\phi ^4}, 
	 \end{eqnarray}
	 the first part of this Hamiltonian is clearly that of a harmonic oscillator:
	 \begin{eqnarray}\label{E234}
	 {H_1} = \frac{{\sqrt {8{E_J}{E_C}} }}{2}\left( {{{\left\{ {{{\left( {\frac{{{E_C}}}{{8{E_J}}}} \right)}^{\frac{1}{4}}}N} \right\}}^2} + {{\left\{ {{{\left( {\frac{{{E_J}}}{{8{E_C}}}} \right)}^{\frac{1}{4}}}\phi } \right\}}^2}} \right).
	 \end{eqnarray}
	 
	 We can then introduce the bosonic variables ${b^ + } = \left( {{1 \mathord{\left/
	 			{\vphantom {1 {\sqrt 2 }}} \right.
	 			\kern-\nulldelimiterspace} {\sqrt 2 }}} \right)\left( {i{{\left( {{{8{E_C}} \mathord{\left/
	 						{\vphantom {{8{E_C}} {{E_J}}}} \right.
	 						\kern-\nulldelimiterspace} {{E_J}}}} \right)}^{{1 \mathord{\left/
	 					{\vphantom {1 4}} \right.
	 					\kern-\nulldelimiterspace} 4}}}N + {{\left( {{{{E_J}} \mathord{\left/
	 						{\vphantom {{{E_J}} {8{E_C}}}} \right.
	 						\kern-\nulldelimiterspace} {8{E_C}}}} \right)}^{{1 \mathord{\left/
	 					{\vphantom {1 4}} \right.
	 					\kern-\nulldelimiterspace} 4}}}\phi } \right)$ and $b = \left( {{1 \mathord{\left/
	 			{\vphantom {1 {\sqrt 2 }}} \right.
	 			\kern-\nulldelimiterspace} {\sqrt 2 }}} \right)\left( { - i{{\left( {{{8{E_C}} \mathord{\left/
	 						{\vphantom {{8{E_C}} {{E_J}}}} \right.
	 						\kern-\nulldelimiterspace} {{E_J}}}} \right)}^{{1 \mathord{\left/
	 					{\vphantom {1 4}} \right.
	 					\kern-\nulldelimiterspace} 4}}}N + {{\left( {{{{E_J}} \mathord{\left/
	 						{\vphantom {{{E_J}} {8{E_C}}}} \right.
	 						\kern-\nulldelimiterspace} {8{E_C}}}} \right)}^{{1 \mathord{\left/
	 					{\vphantom {1 4}} \right.
	 					\kern-\nulldelimiterspace} 4}}}\phi } \right)$, and therefore we have $N =  - \left( {{i \mathord{\left/
	 			{\vphantom {i 2}} \right.
	 			\kern-\nulldelimiterspace} 2}} \right){\left( {{{{E_J}} \mathord{\left/
	 				{\vphantom {{{E_J}} {2{E_C}}}} \right.
	 				\kern-\nulldelimiterspace} {2{E_C}}}} \right)^{{1 \mathord{\left/
	 				{\vphantom {1 4}} \right.\\
	 				\kern-\nulldelimiterspace} 4}}}\\\left( {{b^ + } - b} \right)$ and $\phi  = {\left( {{{2{E_C}} \mathord{\left/
	 				{\vphantom {{2{E_C}} {{E_J}}}} \right.
	 				\kern-\nulldelimiterspace} {{E_J}}}} \right)^{{1 \mathord{\left/
	 				{\vphantom {1 4}} \right.
	 				\kern-\nulldelimiterspace} 4}}}\left( {{b^ + } + b} \right)$ where $\left[ {b,{b^ + }} \right] = 1$, ${b^ + } = \sum\limits_J {\sqrt {J + 1} \left| {J + 1} \right\rangle \left\langle J \right|} $. The Hamiltonian  (\ref{E233}) becomes 
	 \begin{eqnarray}\label{E235}
	 H = \sqrt {8{E_J}{E_C}} \left( {{b^ + }b + \frac{1}{2}} \right) - {E_J} - \frac{{{E_C}}}{{12}}{\left( {{b^ + } + b} \right)^4}. 
	 \end{eqnarray}
	 
	 The corresponding eigen-energies are separated by the energy quantum $\sqrt {8{E_J}{E_C}} $.
	 
	 In the transmon regime, the charging energy determines the non-linearity of the oscillator through its anharmonicity $\alpha$
	 \begin{eqnarray}\label{E236}
	 \alpha  = {E_{12}} - {E_{01}} =  - {E_C}. 
	 \end{eqnarray} 
	 
	 Finally, the flat energy bands enable the assignment of a unique frequency to each energy level of the states $\left| J \right\rangle $. So the Hamiltonian is written in the form
	 \begin{eqnarray}\label{E237}
	 H = \hbar \sum\limits_J {{\omega _J}\left| J \right\rangle \left\langle J \right|}. 
	 \end{eqnarray} 
	 
	 For the first two levels, the Hamiltonian in equation (\ref{E237}) becomes
	 \begin{eqnarray}\label{E238}
	 H = \hbar \frac{{{\omega _q}}}{2}{\sigma _z},
	 \end{eqnarray} 
	 where ${\omega _q} = {\omega _2} - {\omega _1}$  is  the qubit transition frequency, ${\sigma _z} = \left| 0 \right\rangle \left\langle 0 \right| - \left| 1 \right\rangle \left\langle 1 \right|$.  
	
\section{Cavity quantum electrodynamics}
   
\begin{figure*}
	\centering \includegraphics[scale=0.5]{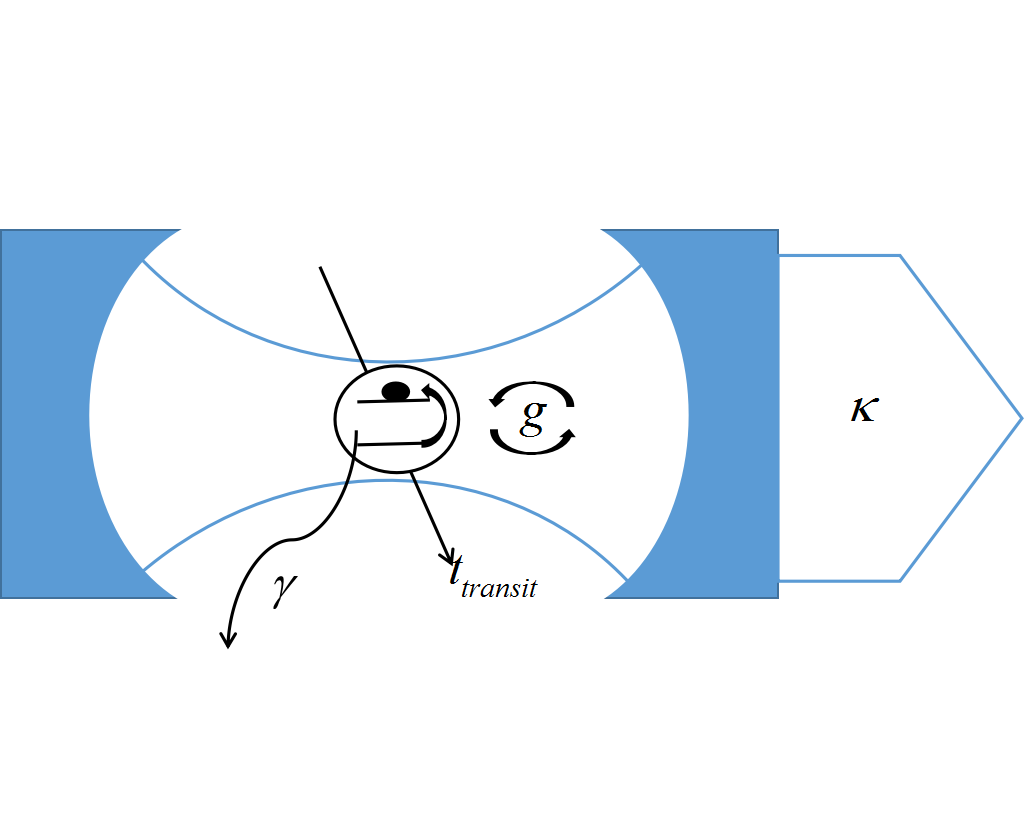}
	\caption{ QED cavity  architecture includes  an  electromagnetic field mode  of decay rate $\kappa $ coupled to a two-level system of  decay rate $\gamma $ and the transition time  ${t_{transit}}$ with a coupling force $g$.  }
	\label{F29}
\end{figure*}

		Cavity quantum electrodynamics is the physical theory describing the interaction between light and matter schematically represented in Fig. (\ref{F29}), which studies the interaction between atoms represented by a
		two-level system and the quantized electromagnetic mode inside a cavity. Cavity QED is interesting in fundamental physics, as well as for applications in quantum information processing.

\subsection{Quantification of the field in a cavity}

We consider a cavity of length L of two perfect mirrors as shown in  Fig. (\ref{F210}). Maxwell's equations:
 \begin{eqnarray}\label{E239}
 \nabla .E = 0, ~~~~~~~~~~\nabla .B = 0, 
 \end{eqnarray} 
 \begin{eqnarray}\label{E240}
 \nabla  \times E =  - \frac{{\partial B}}{{\partial t}},~~~~~~~~~~\nabla  \times B = {\mu _0}{\varepsilon _0}\frac{{\partial B}}{{\partial t}},
 \end{eqnarray}
 where ${\mu _0}$ and ${\varepsilon _0}$ are the vacuum permeability and permittivity. The electric field is polarized along the x axis, $\vec E\left( {r,t} \right) = {E_x}\left( {z,t} \right){\vec e_x}$. The boundary conditions impose the following decomposition ${E_x}\left( {z,t} \right) = {\left[ {{{2{\omega ^2}} \mathord{\left/
 				{\vphantom {{2{\omega ^2}} {\left( {V{\varepsilon _0}} \right)}}} \right.
 				\kern-\nulldelimiterspace} {\left( {V{\varepsilon _0}} \right)}}} \right]^{{1 \mathord{\left/
 				{\vphantom {1 2}} \right.
 				\kern-\nulldelimiterspace} 2}}}q\left( t \right)\sin \left( {kz} \right)$, with  ${k_n} = {{n\pi } \mathord{\left/
 		{\vphantom {{n\pi } L}} \right.
 		\kern-\nulldelimiterspace} L}$($n \succ 0$), ${\omega _n} = c{k_n}$, and   $V$ is the effective volume of the cavity, and $q(t)$ is the amplitude
 of normal mode $n$. From Maxwell's equations (\ref{E239}) and (\ref{E240}), we readily  get the expression of the magnetic field $\vec B\left( {r,t} \right) = {B_y}\left( {z,t} \right){\vec e_y}$, ${B_y}\left( {z,t} \right) = \left( {{{{\mu _0}{\varepsilon _0}} \mathord{\left/
 			{\vphantom {{{\mu _0}{\varepsilon _0}} k}} \right.
 			\kern-\nulldelimiterspace} k}} \right)\left( {{{2{\omega ^2}} \mathord{\left/
 			{\vphantom {{2{\omega ^2}} {V{\varepsilon _0}}}} \right.
 			\kern-\nulldelimiterspace} {V{\varepsilon _0}}}} \right)p\left( t \right)\cos \left( {kz} \right)$, where $p\left( t \right) = \dot q\left( t \right)$ is the canonical momentum operator conjugate to $q$. By using Poynting's theorem, we obtain the energy of the electromagnetic field in the cavity: 
 \begin{figure}[H]
 	\centering \includegraphics[scale=0.5]{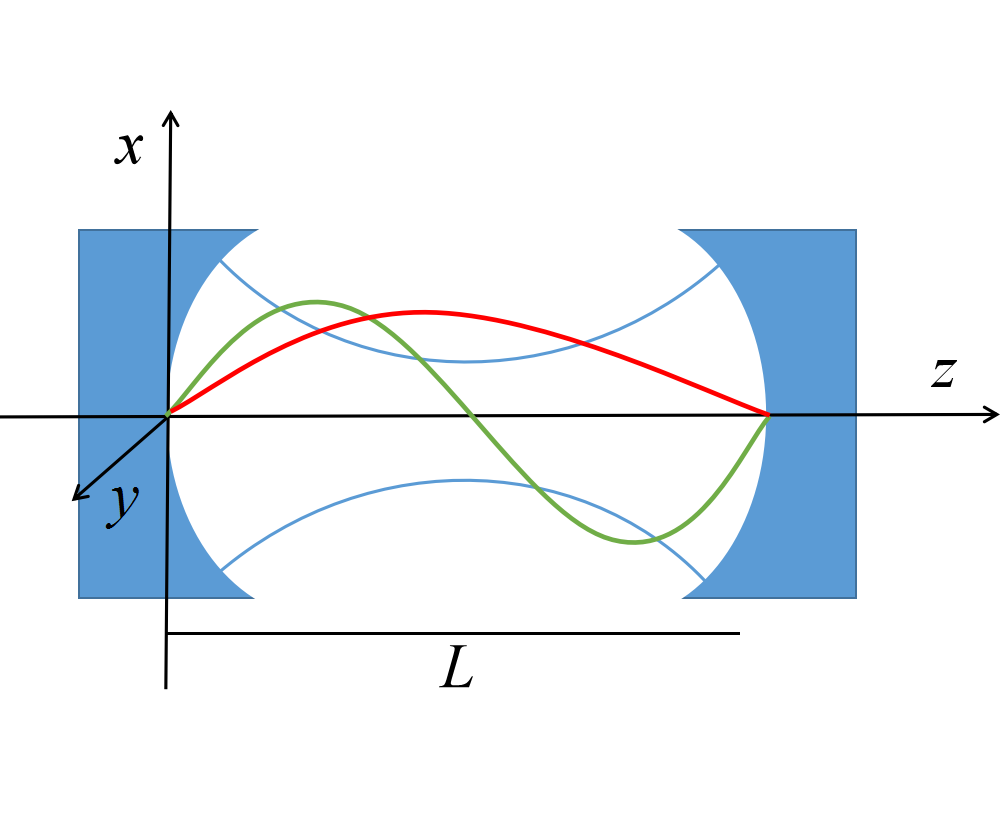}
 	\caption{ An electromagnetic cavity of length $L$, the electric field must therefore be zero at the extremities of the cavity.  }
 	\label{F210}
 \end{figure}
 
\begin{eqnarray}\label{E241}
H = \frac{1}{2}\int {dV\left( {{\varepsilon _0}E_x^2\left( {z,t} \right) + \frac{1}{{{\mu _0}}}B_y^2\left( {z,t} \right)} \right)}. 
\end{eqnarray} 

 We replace the expressions of the electric and magnetic field as mentioned above in this expression, then using the trigonometric relations ${\sin ^2}\left( {kz} \right) = {{\left( {1 - \cos \left( {2kz} \right)} \right)} \mathord{\left/
 		{\vphantom {{\left( {1 - \cos \left( {2kz} \right)} \right)} 2}} \right.
 		\kern-\nulldelimiterspace} 2}$, ${\cos ^2}\left( {kz} \right) = {{\left( {1 + \cos \left( {2kz} \right)} \right)} \mathord{\left/
 		{\vphantom {{\left( {1 + \cos \left( {2kz} \right)} \right)} 2}} \right.
 		\kern-\nulldelimiterspace} 2}$, we finally find the following  expression 
 \begin{eqnarray}\label{E242}
 H = \frac{1}{2}\left( {{p^2} + {\omega ^2}{q^2}} \right).
 \end{eqnarray} 
 
 By moving from the classical variables $p$ and $q$ to their quantum operators, which are given by
 \begin{eqnarray}\label{E243}
 \tilde p = i\sqrt {\frac{\omega }{2}} \left( {{a^ + } - a} \right);~\tilde q = i\sqrt {\frac{1}{{2\omega }}} \left( {{a^ + } + a} \right).
 \end{eqnarray}	
 where $\tilde p$ and $\tilde q$ obey the canonical commutation relation $\left[ {\tilde q,\tilde p} \right] = i\hbar $, the quantum Hamiltonian of the field is written in the following form
 \begin{eqnarray}\label{E244}
 H = \hbar \omega \left( {{a^ + }a + \frac{1}{2}} \right).
 \end{eqnarray}
 
  \subsection{ Jaynes-Cummings model }
\begin{figure}[H]
  	\centering \includegraphics[scale=0.5]{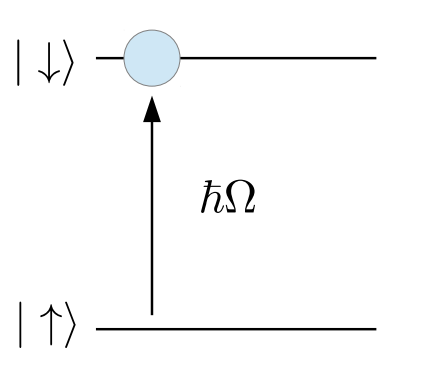}
  	\caption{ Two level atom.  }
  	\label{F211}
  \end{figure}

  We now consider the atom, described as a two-level system whose Hamiltonian, ${H_{atom}} = \left( {{{\hbar \Omega } \mathord{\left/
  			{\vphantom {{\hbar \Omega } 2}} \right.
  			\kern-\nulldelimiterspace} 2}} \right){\sigma _z}$, as indicated  in the diagram(\ref{F211}).
  We can now study the interaction between the two-level atom and the quantized field. The Hamiltonian system is divided into three parts
  \begin{eqnarray}\label{E245}
  H = {H_{atom}} + {H_{field}} + {H_{{\mathop{\rm int}} er}},
  \end{eqnarray}
  where ${H_{atom}}$ is the Hamiltonian of a two-level atom, ${H_{field}} = \hbar {\omega _r}{a^ + }a$ is the Hamiltonian of the quantized field and ${H_{{\mathop{\rm int}} er}}$ is the interaction Hamiltonian between the atom and the quantized field.
  
  \subsection{ Approximation of the dipole moment and the rotating wave }
  The coupling between an atom and an electromagnetic wave is dominated by the coupling of the electric field with the electric dipole. The interaction Hamiltonian  can therefore be written as
  \begin{eqnarray}\label{E246}
  {H_{{\mathop{\rm int}} er}} =  - \tilde d.\tilde E,
  \end{eqnarray}
  where the  dipole operator is written as $\tilde d = d{\sigma _x}$, and from the expression of the electric field and equation  (\ref{E243}). We therefore can write the field operator
  \begin{eqnarray}\label{E247}
  \tilde E = {\left( {\frac{\omega }{{V{\varepsilon _0}}}} \right)^{\frac{1}{2}}}\sin \left( {kz} \right)\left( {{a^ + } + a} \right).
  \end{eqnarray}
  
  Finally, the interaction Hamiltonian (\ref{E246}) becomes
  \begin{eqnarray}\label{E248}
  {H_{{\mathop{\rm int}} er}} = \hbar g\left( {{\sigma ^ + } + \sigma } \right)\left( {a + {a^ + }} \right),
  \end{eqnarray}
  where $g = {{g{\varepsilon _{rms}}} \mathord{\left/
  		{\vphantom {{g{\varepsilon _{rms}}} \hbar }} \right.
  		\kern-\nulldelimiterspace} \hbar }$(${\varepsilon _{rms}} = \sqrt {{\omega  \mathord{\left/
  			{\vphantom {\omega  {V{\varepsilon _0}}}} \right.
  			\kern-\nulldelimiterspace} {V{\varepsilon _0}}}} \sin \left( {kz} \right)$), while  ${\sigma ^ - } = \left|  \uparrow  \right\rangle \left\langle  \downarrow  \right|$ and ${\sigma ^ + } = \left|  \downarrow  \right\rangle \left\langle  \uparrow  \right|$ are Pauli matrices expressed in terms of the ground  $\left| 0 \right\rangle $
  and excited $\left| 1 \right\rangle $ atom  states.
  The Hamiltonian of the system (\ref{E245}) then becomes
  \begin{eqnarray}\label{E249}
  H = \hbar {\omega _r}{a^ + }a + \frac{{\hbar \Omega }}{2}{\sigma _z} + \hbar g\left( {{\sigma ^ + } + {\sigma ^ - }} \right)\left( {{a^ + } + a} \right).
  \end{eqnarray}
  
   The Jaynes-Cummings Hamiltonian can be derived from the Hamiltonian (\ref{E249}) by using the rotating wave approximation (RWA). The RWA is valid in the regime where $g \ll {\omega _r},\Omega $. The terms ${\sigma ^ + }a\left( {{\sigma ^ - }{a^ + }} \right)$ correspond to the processes of excitation (de-excitation) of the atom and a photon is created(annihilated) in the cavity mode. The two other terms corresponding to the  excitation (de-excitation) of both the atom and the cavity field mode do not conserve energy.  In the RWA regime, these two terms can be neglected. Within the framework of this approximation, we  obtain the Hamiltonian corresponding to the Jaynes-Cummings model\cite{112}
   \begin{eqnarray}\label{E250}
   H = \hbar {\omega _r}{a^ + }a + \frac{{\hbar \Omega }}{2}{\sigma _z} + \hbar g\left( {a{\sigma ^ + } + {a^ + }{\sigma ^ - }} \right).
   \end{eqnarray} 
   
   We can transform this expression into a more practical form, by introducing the following excitation number operator $N \simeq {a^ + }a + {{{\sigma _z}} \mathord{\left/
   		{\vphantom {{{\sigma _z}} 2}} \right.
   		\kern-\nulldelimiterspace} 2}$, the Hamiltonian (\ref{E250})   can be rewritten  as  
   \begin{eqnarray}\label{E251}
   H = \hbar {\omega _r}N + \frac{{\hbar \Delta }}{2}{\sigma _z} + \hbar g\left( {a{\sigma ^ + } + {a^ + }{\sigma ^ - }} \right),
   \end{eqnarray} 
   where $\Delta  = \Omega  - {\omega _r}$ is the atom-cavity detuning. We  can easily demonstrate the following  commutation  relation $\left[ {H,N} \right] = 0$. The stationary states of the system without coupling ($g = 0$) are
   \begin{eqnarray}\label{E252}
   \left| { \downarrow ,n} \right\rangle  = \left|  \downarrow  \right\rangle  \otimes \left| n \right\rangle , ~~~~ \left| { \uparrow ,n} \right\rangle  = \left|  \uparrow  \right\rangle  \otimes \left| n \right\rangle. 
   \end{eqnarray}
   
   The  interaction  Hamiltonian $H_{inter}$ couples these two-by-two states $\left| { \uparrow ,n} \right\rangle $ with  $\left| { \downarrow ,n + 1} \right\rangle $. Only the fundamental level $\left| { \uparrow ,0} \right\rangle $ is not coupled to any other state
   \begin{eqnarray}\label{E253}
   {H_{{\mathop{\rm int}} er}}\left| { \uparrow ,0} \right\rangle  = 0, 
   \end{eqnarray}
   \begin{eqnarray}\label{E254}
   {H_{{\mathop{\rm int}} er}}\left| { \downarrow ,0} \right\rangle  = \hbar g\left| { \uparrow ,1} \right\rangle ,~~~{H_{{\mathop{\rm int}} er}}\left| { \uparrow ,1} \right\rangle  = \hbar g\left| { \downarrow ,0} \right\rangle, 
   \end{eqnarray}
   \begin{eqnarray}\label{E255}
   {H_{{\mathop{\rm int}} er}}\left| { \downarrow ,n} \right\rangle  = \hbar g\sqrt {n + 1} \left| { \uparrow ,n + 1} \right\rangle ,{H_{{\mathop{\rm int}} er}}\left| { \uparrow ,n + 1} \right\rangle  = \hbar g\sqrt {n + 1} \left| { \downarrow ,n} \right\rangle.
   \end{eqnarray}
   
    In the double states $\left\{ {\left| { \downarrow ,n} \right\rangle ,\left| { \uparrow ,n + 1} \right\rangle } \right\}$, the Hamiltonian (\ref{E251}) is then written
    \begin{eqnarray}\label{E256}
    H = \hbar {\omega _r}\left( {n + 1} \right)\left( {\begin{array}{*{20}{c}}
    	1&0\\
    	0&1
    	\end{array}} \right) + \frac{\hbar }{2}\left( {\begin{array}{*{20}{c}}
    	{ - \Delta }&{2g\sqrt {n + 1} }\\
    	{2g\sqrt {n + 1} }&\Delta 
    	\end{array}} \right). 
    \end{eqnarray}
    
    The eigen-energies of the Hamiltonian (\ref{E251}) are
    \begin{eqnarray}\label{E257}
    {E_{ + ,n}} = \hbar {\omega _r}\left( {n + 1} \right) + \frac{\hbar }{2}\sqrt {4g^2\left( {n + 1} \right) + {\Delta ^2}} ,{E_{ - ,n}} = \hbar {\omega _r}\left( {n + 1} \right) - \frac{\hbar }{2}\sqrt {4g^2\left( {n + 1} \right) + {\Delta ^2}},  
    \end{eqnarray}
    and the corresponding states are given by
    \begin{eqnarray}\label{E258}
    \left| { + ,n} \right\rangle  = \cos {\theta _n}\left| { \downarrow ,n} \right\rangle  + \sin {\theta _n}\left| { \uparrow ,n + 1} \right\rangle ,\left| { - ,n} \right\rangle  =  - \sin {\theta _n}\left| { \downarrow ,n} \right\rangle  + \cos {\theta _n}\left| { \uparrow ,n + 1} \right\rangle ,  
    \end{eqnarray} 
    while the egen-energy of the ground state $\left| { \uparrow ,0} \right\rangle $ is 
    \begin{eqnarray}\label{E259}
    \frac{{ - \hbar \Delta }}{2},~~and~~~{\theta _n} = \left( {\frac{1}{2}} \right){\tan ^{ - 1}}\left( {\frac{{2g\sqrt {n + 1} }}{2}} \right). 
    \end{eqnarray}
     \subsection{ Resonance regime and dispersive regime }
     \textbf{\underline{ Resonance regime ( $\Delta=0$)}}
     
\begin{figure}[H]
     	\centering \includegraphics[scale=0.6]{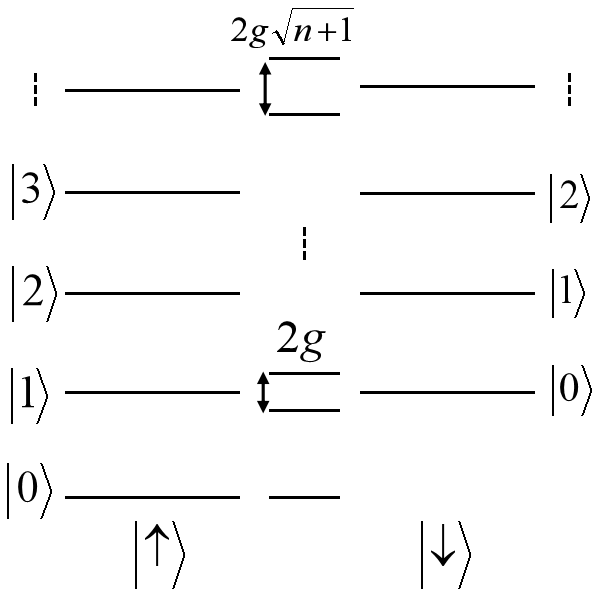}
     	\caption{ Energy diagram in the resonant case.  }
     	\label{F212}
     \end{figure}

     In the resonant case, characterized by $\Delta=0$, the eigenstates are  the combinations of symmetric and anti-symmetric of the  states $\left| { \downarrow ,n} \right\rangle $ and $\left| { \uparrow ,n + 1} \right\rangle $($\left| { \pm ,n} \right\rangle  = \left( {{1 \mathord{\left/
     			{\vphantom {1 {\sqrt 2 }}} \right.
     			\kern-\nulldelimiterspace} {\sqrt 2 }}} \right)\left( {\left| { \uparrow ,1} \right\rangle  \pm \left| { \downarrow ,0} \right\rangle } \right)$), and the interaction term  lifts the degeneracy of these states. This situation is illustrated in 
     Figure (\ref{F212}).
    
     \textbf{\underline{ Dispersive regime (  $\Delta  \gg g$)}}
     
     We now describe the dispersive case  in which the atom  is operated at a frequency for which the detuning $\Delta$ between  atom  and cavity  are large. In this case, no exchanging excitations  between atom and cavity, and  the eigenstates take the form
     \begin{eqnarray}\label{H3}
     \left| { + ,n} \right\rangle  \simeq \left| { \downarrow ,n} \right\rangle  + \frac{{g\sqrt {n + 1} }}{\Delta }\left| { \uparrow ,n + 1} \right\rangle ,\left| { - ,n} \right\rangle  \simeq  - \frac{{g\sqrt {n + 1} }}{\Delta }\left| { \downarrow ,n} \right\rangle  + \left| { \uparrow ,n + 1} \right\rangle.
     \end{eqnarray}
     
\begin{figure}[H]
    	\centering \includegraphics[scale=0.6]{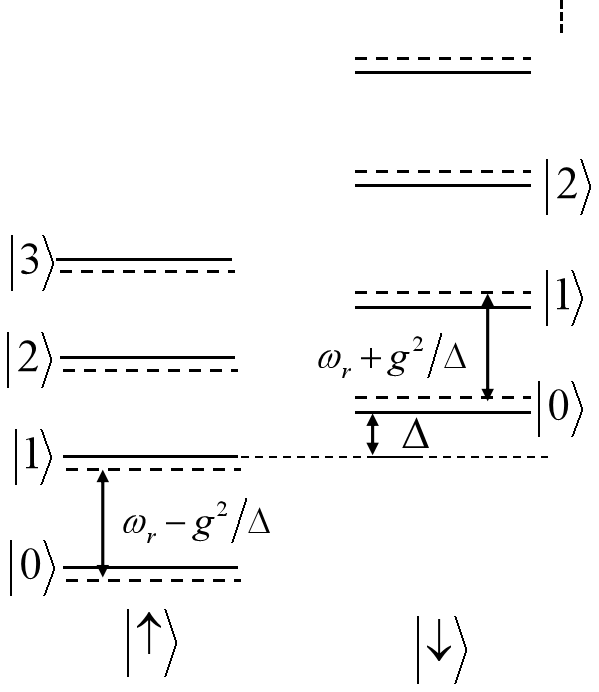}
    	\caption{ Energy diagram in the dispersive regime. }
    	\label{F213}
    \end{figure}

     As shown  in figure (\ref{F213}), the energies of the eigenstates are shifted by the coupling. In this regime, the energy exchange between the qubit and the resonator field is difficult. By applying the unitary transformation, $U = \exp (\lambda \left( {a{\sigma ^ + } - {a^ + }{\sigma ^ - }} \right)$ with $\lambda  = {g \mathord{\left/
     		{\vphantom {g {\Delta  \ll 1}}} \right.
     		\kern-\nulldelimiterspace} {\Delta  \ll 1}}$, it is possible to obtain an approximate effective Hamiltonian
     \begin{eqnarray}\label{E261}
     {H_{eff}} = UH{U^ + } \approx \hbar \left( {{\omega _r} + \frac{{{g^2}}}{\Delta }{\sigma _z}} \right){a^ + }a + \frac{\hbar }{2}\left( {\Omega  + \frac{{{g^2}}}{\Delta }} \right){\sigma _z}.
     \end{eqnarray}
    
	\section{Circuit quantum electrodynamics}
	Circuit quantum  electrodynamics (circuit  QED) is an analog of cavity quantum electrodynamics,  which  is proposed by Blais\cite{44}. Such a circuit  QED is realized in an electrical circuit by coupling capacitively a superconducting qubit to a coplanar resonator (playing the role of the cavity) or the quantum harmonic LC oscillator.
	
	 \subsection{Quantum harmonic LC oscillator}
\begin{figure}[H]
	 	\centering \includegraphics[scale=0.6]{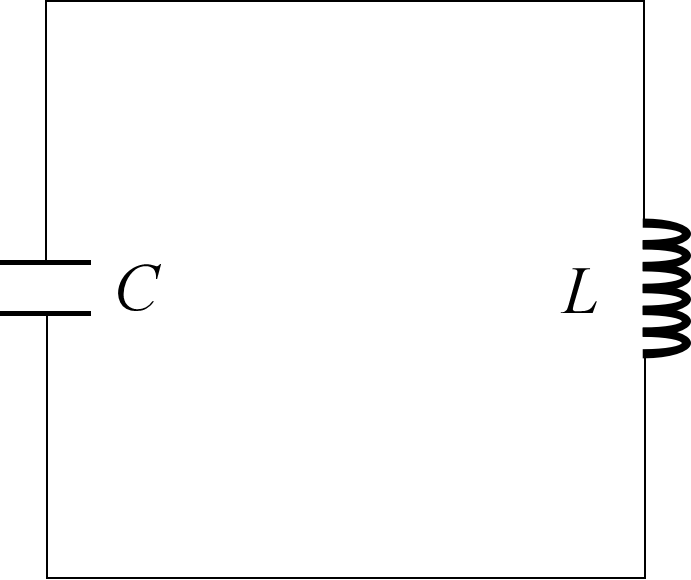}
	 	\caption{ Energy diagram in the dispersive regime. }
	 	\label{F214}
	 \end{figure}

	 The quantum LC oscillator is the simplest quantum circuit as shown in figure (\ref{F214}),  it consists of a capacitor $C$ in series with an inductance $L$. By using  the Kirchhoff's rule for quantum  circuit  shown  in Fig. (\ref{F214}), we find the following equation of motion
	 \begin{eqnarray}\label{E262}
	 L\frac{{d\dot Q}}{{dt}} + \frac{Q}{C} = 0.
	 \end{eqnarray}
	 
	 Using the following Euler-Lagrangian equation ${{d\left( {{{\partial {L_r}} \mathord{\left/
	 					{\vphantom {{\partial {L_r}} {\partial \dot Q}}} \right.
	 					\kern-\nulldelimiterspace} {\partial \dot Q}}} \right)} \mathord{\left/
	 		{\vphantom {{d\left( {{{\partial L} \mathord{\left/
	 								{\vphantom {{\partial {L_r}} {\partial \dot Q}}} \right.
	 								\kern-\nulldelimiterspace} {\partial \dot Q}}} \right)} {dt - {{\partial {L_r}} \mathord{\left/
	 							{\vphantom {{\partial {L_r}} {\partial Q}}} \right.
	 							\kern-\nulldelimiterspace} {\partial Q}} = 0}}} \right.
	 		\kern-\nulldelimiterspace} {dt - {{\partial {L_r}} \mathord{\left/
	 				{\vphantom {{\partial L} {\partial Q}}} \right.
	 				\kern-\nulldelimiterspace} {\partial Q}} = 0}}$,  we can determine the Lagrangian of the system
	 \begin{eqnarray}\label{E263}
	 {L_r} = \frac{L}{2}{\dot Q^2} - \frac{{{Q^2}}}{{2C}}.
	 \end{eqnarray}
	 
	 The  corresponding  conjugate moment $p = {{\partial {L_r}} \mathord{\left/
	 		{\vphantom {{\partial {L_r}} {\partial \dot Q}}} \right.
	 		\kern-\nulldelimiterspace} {\partial \dot Q}} = L\dot Q$. Therefore, the Hamiltonian is written in the following form
	\begin{eqnarray}\label{E264}
	 H = \frac{\Phi }{{2L}} + \frac{Q}{{2C}}.
	 \end{eqnarray}
	 
	The flux $\Phi $ in the inductor can be chosen as a position variable, whereas the charge $Q$ of the capacitor can be seen as its conjugate variable, $\left[ {\Phi ,Q} \right] = i\hbar $. We introduce ${a^ + }$ and $a$ to diagonalize the hamiltonian with the quantum harmonic oscillator. $Q$ and  $\Phi $ can thus be written as $a =  - i{\Phi  \mathord{\left/
			{\vphantom {\Phi  {\sqrt {2\hbar {\omega _r}L}  + {Q \mathord{\left/
								{\vphantom {Q {\sqrt {2\hbar {\omega _r}C} }}} \right.
								\kern-\nulldelimiterspace} {\sqrt {2\hbar {\omega _r}C} }}}}} \right.
			\kern-\nulldelimiterspace} {\sqrt {2\hbar {\omega _r}L}  + {Q \mathord{\left/
					{\vphantom {Q {\sqrt {2\hbar {\omega _r}C} }}} \right.
					\kern-\nulldelimiterspace} {\sqrt {2\hbar {\omega _r}C} }}}}$ and  ${a^ + } = i{\Phi  \mathord{\left/
			{\vphantom {\Phi  {\sqrt {2\hbar {\omega _r}L}  + {Q \mathord{\left/
								{\vphantom {Q {\sqrt {2\hbar {\omega _r}C} }}} \right.
								\kern-\nulldelimiterspace} {\sqrt {2\hbar {\omega _r}C} }}}}} \right.
			\kern-\nulldelimiterspace} {\sqrt {2\hbar {\omega _r}L}  + {Q \mathord{\left/
					{\vphantom {Q {\sqrt {2\hbar {\omega _r}C} }}} \right.
					\kern-\nulldelimiterspace} {\sqrt {2\hbar {\omega _r}C} }}}}$. We  then plug these expressions in equation  (\ref{E264})  we get
	\begin{eqnarray}\label{E265}
	H = \hbar {\omega _r}\left( {{a^ + }a + \frac{1}{2}} \right),
	\end{eqnarray}
	where ${\omega _r} = {1 \mathord{\left/
			{\vphantom {1 {\sqrt {LC} }}} \right.
			\kern-\nulldelimiterspace} {\sqrt {LC} }}$.
	 \subsection{Transmission line (Coplanar waveguide)}
\begin{figure}[H]
		\centering \includegraphics[scale=0.7]{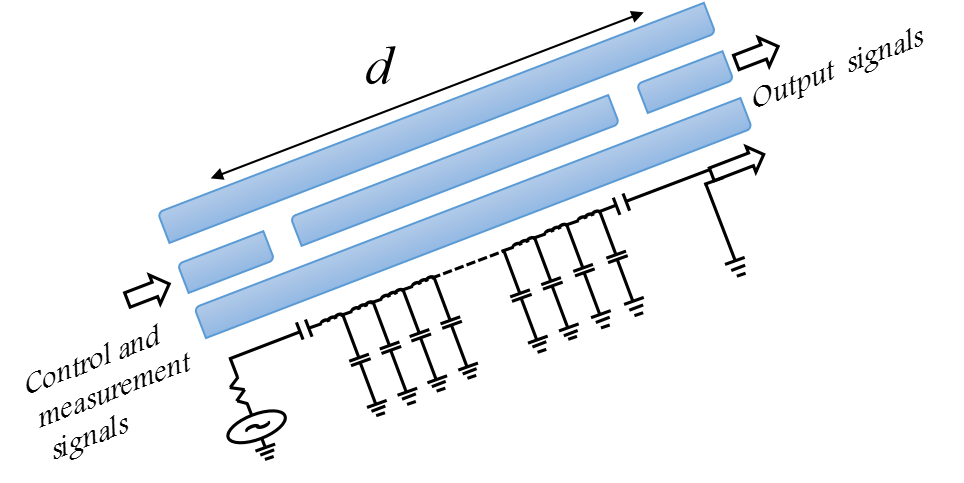}
		\caption{ Representation of the transmission line by a set of inductors in parallel with capacitors. }
		\label{F215}
	\end{figure}

	 The aim of this section is to obtain a quantum description of the electromagnetic field inside a transmission line. We use the standard approach of modeling the line as a set of inductors in parallel with capacitors as shown (\ref{F215}).
	 
\begin{figure}[H]
	 	\centering \includegraphics[scale=0.7]{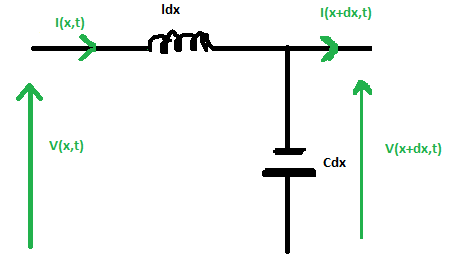}
	 	\caption{ Representation of an inductor in parallel with the capacitor. }
	 	\label{F216}
	 \end{figure}

	 We assume inductance $l$ and capacitance $c$ per unit the length of the  line. Each segment of the line of the length $dx$ has an inductance $ldx$ in parallel with a capacitance $cdx$, as shown in  figure(\ref{F216}). We apply Kirchhoff's rule to this  figure, and we find the following equation $V\left( {x,t} \right) = ldx{{\partial I\left( {x,t} \right)} \mathord{\left/
	 		{\vphantom {{\partial I\left( {x,t} \right)} {\partial t + V\left( {x + dx,t} \right)}}} \right.
	 		\kern-\nulldelimiterspace} {\partial t + V\left( {x + dx,t} \right)}}$ (i.e ${{V\left( {x + dx,t} \right)} \mathord{\left/
	 		{\vphantom {{V\left( {x + dx,t} \right)} {dx - {{V\left( {x,t} \right)} \mathord{\left/
	 							{\vphantom {{V\left( {x,t} \right)} {dx = l{{\partial I\left( {x,t} \right)} \mathord{\left/
	 												{\vphantom {{\partial I\left( {x,t} \right)} {\partial t}}} \right.
	 												\kern-\nulldelimiterspace} {\partial t}}}}} \right.
	 							\kern-\nulldelimiterspace} {dx = l{{\partial I\left( {x,t} \right)} \mathord{\left/
	 									{\vphantom {{\partial I\left( {x,t} \right)} {\partial t}}} \right.
	 									\kern-\nulldelimiterspace} {\partial t}}}}}}} \right.
	 		\kern-\nulldelimiterspace} {dx - {{V\left( {x,t} \right)} \mathord{\left/
	 				{\vphantom {{V\left( {x,t} \right)} {dx = l{{\partial I\left( {x,t} \right)} \mathord{\left/
	 									{\vphantom {{\partial I\left( {x,t} \right)} {\partial t}}} \right.
	 									\kern-\nulldelimiterspace} {\partial t}}}}} \right.
	 				\kern-\nulldelimiterspace} {dx = l{{\partial I\left( {x,t} \right)} \mathord{\left/
	 						{\vphantom {{\partial I\left( {x,t} \right)} {\partial t}}} \right.
	 						\kern-\nulldelimiterspace} {\partial t}}}}}}$). Finally, we find the following constitutive relation 
	 \begin{eqnarray}\label{E266}
	 l{\partial _t}I\left( {x,t} \right) + {\partial _x}V\left( {x,t} \right) = 0.
	 \end{eqnarray}
We choose the generalized flux as the degree of freedom instead of choosing current or voltage, the  generalized flux is defined as $\Phi \left( {x,t} \right) = \int {V\left( {x,t} \right)dx} $. By replacing  the expression of V (x, t) in equation (\ref{E266}), we find the expression of the current  $I\left( {x,t} \right) = \left( {{1 \mathord{\left/
			{\vphantom {1 l}} \right.
			\kern-\nulldelimiterspace} l}} \right){\partial _x}\Phi \left( {x,t} \right)$. The Lagrangian of the system is given by
\begin{eqnarray}\label{E267}
{L_g} = \int_{\frac{{ - d}}{2}}^{\frac{d}{2}} {dxL\left( {x,t} \right) = \int_{\frac{{ - d}}{2}}^{\frac{d}{2}} {dx\left[ {\frac{c}{2}{{\left( {{\partial _t}\Phi } \right)}^2} - \frac{1}{{2l}}{{\left( {{\partial _x}\Phi } \right)}^2}} \right]} }. 
\end{eqnarray}

The associated Euler-Lagrange equations are those of a wave propagating at the speed $\nu  = {1 \mathord{\left/
		{\vphantom {1 {\sqrt {lc} }}} \right.
		\kern-\nulldelimiterspace} {\sqrt {lc} }}$. By  using  boundary conditions ($\Phi \left( {{d \mathord{\left/
			{\vphantom {d 2}} \right.
			\kern-\nulldelimiterspace} 2},t} \right) = \Phi \left( { - {d \mathord{\left/
			{\vphantom {d 2}} \right.
			\kern-\nulldelimiterspace} 2},t} \right) = 0$), $\Phi \left( {x,t} \right)$ can  be written under
the form
\begin{eqnarray}\label{E268}
\Phi \left( {x,t} \right) = \sqrt {\frac{2}{d}} \sum\limits_{{k_0} = 1} {{\phi _{{k_0}}}\left( t \right)} \cos \left( {\frac{{{k_0}\pi x}}{d}} \right) + \sqrt {\frac{2}{d}} \sum\limits_{{k_e} = 2} {{\phi _{{k_e}}}\left( t \right)} \sin \left( {\frac{{{k_e}\pi x}}{d}} \right).
\end{eqnarray}
We replace (\ref{E227}) in (\ref{E266}), the Lagrangian $L_g$ is then written in terms of the new variables ${\phi _k}\left( t \right)\left( {k = {k_0},{k_e}} \right)$
\begin{eqnarray}\label{E269}
{L_g} = \sum\limits_k {\frac{c}{2}} {\left( {{{\dot \phi }_k}} \right)^2} - \frac{1}{{2l}}{\left( {\frac{{k\pi }}{d}} \right)^2}{\left( {{\phi _k}} \right)^2}.
\end{eqnarray}

It is the Lagrangian of the sum of decoupled harmonic oscillators. The calculation of the moments ${p_k}$ combined with the variables ${{\phi _k}}$: ${p_k} = c{\dot \phi _k}$ by the Legendre transformation, then the
commutation relation is  $\left[ {{a_k},{a^ + }_{k'}} \right] = {\delta _{k,k'}}$. The Hamiltonian of the resonator is then written
\begin{eqnarray}\label{E270}
H = \sum\limits_k {{\omega _k}\left( {a_k^ + {a_k} + \frac{1}{2}} \right)},
\end{eqnarray}
where ${\omega _k} = \left( {{{k\pi } \mathord{\left/
			{\vphantom {{k\pi } d}} \right.
			\kern-\nulldelimiterspace} d}} \right)\nu $, ${a_k} =  - i\sqrt {\left( {{{k\pi } \mathord{\left/
				{\vphantom {{k\pi } {2\hbar }}} \right.
				\kern-\nulldelimiterspace} {2\hbar }}} \right)\sqrt {{c \mathord{\left/
				{\vphantom {c l}} \right.
				\kern-\nulldelimiterspace} l}} } {\tilde \phi _k} + \sqrt {\left( {{d \mathord{\left/
				{\vphantom {d {2\hbar k\pi }}} \right.
				\kern-\nulldelimiterspace} {2\hbar k\pi }}} \right)\sqrt {{l \mathord{\left/
				{\vphantom {l c}} \right.
				\kern-\nulldelimiterspace} c}} } {\tilde p_k}$ and $a_k^ +  = i\sqrt {\left( {{{k\pi } \mathord{\left/
				{\vphantom {{k\pi } {2\hbar }}} \right.
				\kern-\nulldelimiterspace} {2\hbar }}} \right)\sqrt {{c \mathord{\left/
				{\vphantom {c l}} \right.
				\kern-\nulldelimiterspace} l}} } {\tilde \phi _k} + \sqrt {\left( {{d \mathord{\left/
				{\vphantom {L {2\hbar k\pi }}} \right.
				\kern-\nulldelimiterspace} {2\hbar k\pi }}} \right)\sqrt {{l \mathord{\left/
				{\vphantom {l c}} \right.
				\kern-\nulldelimiterspace} c}} } {\tilde p_k}$.
\subsection{Coupling a transmon-type superconducting qubit to a coplanar waveguide resonator}
\begin{figure*}
	\centering \includegraphics[scale=0.5]{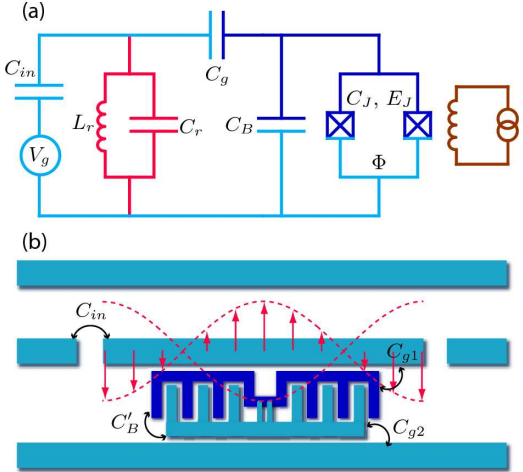}
	\caption{ Circuit QED for the transmon qubit: (a)  The transmon qubit is coupled to the $LC$ resonator through a gate capacitance $C_d$. The two Josephson junctions, with capacitance $C_J$ and Josephson energy $E_J$, are in parallel with an additional large capacitance $C_B$. (b) The transmon device consists of a transmon qubit coupled to a transmission line. The figure used from Ref.\cite{43}. }
	\label{F217}
\end{figure*}

The coupling  between  the transmon-type superconducting qubit  and the coplanar   waveguide resonator  is an electrostatic capacitive interaction. Such a transmon is located at the center of a transmission line resonator, as shown in Fig.(\ref{F217}b). As seen in Fig. (\ref{F217}a),   the transmon qubit is capacitively coupled to the resonator in the equivalent circuit of the transmon-resonator system (figure \ref{F217}b), it is composed of two Josephson junctions, which are hijacked by two large capacitances $C_g$ and $C_B$.  These large capacities lead to low charge energy, ${E_c} = {{{{\left( {2e} \right)}^2}} \mathord{\left/
		{\vphantom {{{{\left( {2e} \right)}^2}} {{C_t}}}} \right.
		\kern-\nulldelimiterspace} {{C_t}}}$ with $C_t$ = $C_B$ + $C_g$ + $C_J$  which distinguishes the   transmon from the charge qubit. 

The coupling between the transmon and the resonator is demonstrated by following an approach similar to that used for dissipation in the cavity. The standing wave in the resonator
(Fig. \ref{F217}(b)) induces tension between the two islands. We consider a capacitive coupling between these two devices, the interaction Hamiltonian takes the following form  
\begin{eqnarray}\label{E271}
{H_{{\mathop{\rm int}} er}} = 2\frac{{{C_g}}}{{{C_t}}}eV_{rms}^0\tilde N\left( {a + {a^ + }} \right),
\end{eqnarray}
where ${\omega _r} = {1 \mathord{\left/
		{\vphantom {1 {\sqrt {{L_r}{C_r}} }}} \right.
		\kern-\nulldelimiterspace} {\sqrt {{L_r}{C_r}} }}$ is the  resonator  frequency , and $V_{rms}^0 = \sqrt {{{\hbar {\omega _r}} \mathord{\left/
			{\vphantom {{\hbar {\omega _r}} {{C_r}}}} \right.
			\kern-\nulldelimiterspace} {{C_r}}}} $. The total Hamiltonian of the system is then written
\begin{eqnarray}\label{E272}
H = \frac{{{E_c}}}{2}{\left( {N - {N_g}} \right)^2} - {E_J}\cos \phi  + \hbar {\omega _r}{a^ + }a + 2\beta eV_{rms}^0\tilde N\left( {a + {a^ + }} \right),
\end{eqnarray}
where $\beta  = {{{C_g}} \mathord{\left/
		{\vphantom {{{C_g}} {{C_t}}}} \right.
		\kern-\nulldelimiterspace} {{C_t}}}$. We can rewrite the Hamiltonian in terms of the eigenstates of the transmon  $\left| i \right\rangle $, we then  obtain the generalized Jaynes-Cummings Hamiltonian 
\begin{eqnarray}\label{E273}
H = \sum\limits_j {{\omega _j}\left| j \right\rangle \left\langle j \right| + \hbar {\omega _r}{a^ + }a + \hbar \sum\limits_{ij} {{g_{ij}}\left| i \right\rangle } } \left\langle j \right|\left( {a + {a^ + }} \right),
\end{eqnarray}
where  $\hbar {g_{ij}} = 2\beta eV_{rms}^0\left\langle i \right|\tilde N\left| j \right\rangle $ are   nearest-neighbor energy levels of the coupling, with $\left| {\left\langle {j + 1} \right|\tilde N\left| j \right\rangle } \right| \approx \sqrt {{{\left( {j + 1} \right)} \mathord{\left/
			{\vphantom {{\left( {j + 1} \right)} 2}} \right.
			\kern-\nulldelimiterspace} 2}} {\left( {{{{E_J}} \mathord{\left/
				{\vphantom {{{E_J}} {8{E_c}}}} \right.
				\kern-\nulldelimiterspace} {8{E_c}}}} \right)^{{1 \mathord{\left/
				{\vphantom {1 2}} \right.
				\kern-\nulldelimiterspace} 2}}}$.
    
We omit the terms corresponding to a double excitation and a double desexcitation, we obtain the generalized Hamiltonian of Jaynes-Cummings
\begin{eqnarray}\label{E274}
H = \hbar \sum\limits_j {{\omega _j}\left| j \right\rangle \left\langle j \right| + \hbar {\omega _r}{a^ + }a + \hbar \sum\limits_i {{g_{i,i + 1}}\left| i \right\rangle \left\langle {i + 1} \right| + h.c} }.
\end{eqnarray}
With sufficient anharmonicity for a transmon system, we can keep only  the first two levels, the Hamiltonian (\ref{E274}) becomes Jaynes-Cummings in equation (\ref{E249}).

Similar to cavity quantum electrodynamics, we will be able to describe a dispersive regime in circuit quantum electrodynamics. This is done when the detuning $\Delta  = {\omega _q} - {\omega _r}$ between the transmon and resonator is large. We diagonalize the Hamiltonian (\ref{E274}), by applying the unitary transformation $D = {e^{S - {S^ + }}}$, with $S = \sum\limits_i {{\beta _i}a\left| {i + 1} \right\rangle \left\langle i \right|} $ (${\beta _i} = {{{g_{i,i + 1}}} \mathord{\left/
		{\vphantom {{{g_{i,i + 1}}} {\left( {{\omega _{i,i + 1}} - {\omega _r}} \right)}}} \right.
		\kern-\nulldelimiterspace} {\left( {{\omega _{i,i + 1}} - {\omega _r}} \right)}}$). We use the "Baker-Campbell-Hausdorff\cite{Hall}" relation and keeping the terms up to ${{{g^2}_{i,i + 1}} \mathord{\left/
		{\vphantom {{{g^2}_{i,i + 1}} {\Delta _i^2}}} \right.
		\kern-\nulldelimiterspace} {\Delta _i^2}}$ (${\Delta _i} = {\omega _{i,i + 1}} - {\omega _r}$), we get
\begin{eqnarray}\label{E275}
{H_{eff}} = DH{D^ + } &=& \hbar \sum\limits_j {{\omega _j}\left| i \right\rangle \left\langle i \right| + \hbar {\omega _r}{a^ + }a + \hbar \sum\limits_i {{\chi _{i,i + 1}}\left| {i + 1} \right\rangle \left\langle {i + 1} \right|} }  - \hbar {\chi _{01}}\left| 0 \right\rangle \left\langle 0 \right|\nonumber\\
&+& \hbar \sum\limits_{i = 1} {\left( {{\chi _{i - 1,i}} - {\chi _{i,i + 1}}} \right){a^ + }a\left| i \right\rangle \left\langle i \right| + \hbar \sum\limits_i {{\eta _i}aa\left| {i + 2} \right\rangle } } \left\langle i \right| + h.c,
\end{eqnarray}
where ${\chi _{i,i + 1}} = {{g_{i,i + 1}^2} \mathord{\left/
		{\vphantom {{g_{i,i + 1}^2} {\left( {{\omega _{i,i + 1}} - {\omega _r}} \right)}}} \right.
		\kern-\nulldelimiterspace} {\left( {{\omega _{i,i + 1}} - {\omega _r}} \right)}}$ and ${\eta _i} = {{{g_{i,i + 1}}{g_{i + 1,i + 2}}} \mathord{\left/
		{\vphantom {{{g_{i,i + 1}}{g_{i + 1,i + 2}}} {\left[ {2\left( {{\omega _{i + 1}} - {\omega _i} - {\omega _r}} \right)\left( {{\omega _{i + 2}} - {\omega _{i + 1}} - {\omega _r}} \right)} \right]}}} \right.
		\kern-\nulldelimiterspace} {\left[ {2\left( {{\omega _{i + 1}} - {\omega _i} - {\omega _r}} \right)\left( {{\omega _{i + 2}} - {\omega _{i + 1}} - {\omega _r}} \right)} \right]}}$.
 
The last term ${\eta _i}$ corresponding to two-photon transitions can be neglected from ${\chi _{i,i + 1}}$. Then, equation (\ref{E275}) becomes
\begin{eqnarray}\label{E276}
{H_{eff}} = \frac{\hbar }{2}\underbrace {\left( {{\omega _{01}} + {\chi _{01}}} \right)}_{{{\omega '}_{01}}}{\sigma _z} + \hbar \underbrace {\left( {{\omega _r} - \frac{1}{2}{\chi _{12}}} \right)}_{{{\omega '}_r}}{a^ + }a + \hbar \underbrace {\left( {{\chi _{01}} - \frac{{{\chi _{12}}}}{2}} \right)}_\chi {\sigma _z}{a^ + }a.
\end{eqnarray}

Finally, the dispersive Hamiltonian can be written as
\begin{eqnarray}\label{E277}
{H_{eff}} = \hbar \frac{{{{\omega '}_{01}}}}{2}{\sigma _z} + \hbar \left( {{{\omega '}_r} + \chi {\sigma _z}} \right){a^ + }a.
\end{eqnarray}

\section{Quantum  gates  with  superconducting device }

\subsection{ Single-qubit  gates }

\subsubsection*{ $X$-rotation gate for one qubit }

Applying  the classical field polarized according to ${e_x}$, $\vec E\left( t \right) = \left( {{E_0}{e^{ - i\left( {\omega t + \theta } \right)}} + E_0^ * {e^{i\left( {\omega t + \theta } \right)}}} \right){\vec e_x}$(where ${E_0}$ and $\theta $ are the complex amplitude of the field and the frequency) introduces a dipole interaction between the transmon-type superconducting qubit and the microwave field. Within the framework of the dipole approximation, we get
\begin{eqnarray}\label{E278}
{H_{dis}} =  - \vec d.\vec E\left( {r,t} \right),
\end{eqnarray}
$d$ is the dipole moment for the two-level system. Now, introducing the following relation ($\left| e \right\rangle \left\langle e \right| + \left| g \right\rangle \left\langle g \right| = 1$) then equation (\ref{E278}) becomes ${H_{dis}} =  - \left( {{\rho _{ab}}{\sigma _ + } + {\rho _{ba}}{\sigma _ - }} \right)E\left( t \right)$, where ${\rho _{ab}} = \left\langle a \right|{d_x}\left| b \right\rangle $ are the matrix elements of the electric dipole moment.  On the other hand, we can rewrite the Hamiltonian expression as follows
\begin{eqnarray}\label{E279}
{H_{disp}} =  - \hbar {\Omega _R}\left( {{\sigma _ + } + {\sigma _ - }} \right)\left( {{e^{ - i\left( {w  t + \phi } \right)}} + {e^{i\left( {w  t + \phi } \right)}}} \right),
\end{eqnarray}
with ${\Omega _R} = {{\left| {{\rho _{ab}}} \right|\left| {{E_0}} \right|} \mathord{\left/
		{\vphantom {{\left| {{\rho _{ab}}} \right|\left| {{E_0}} \right|} \hbar }} \right.
		\kern-\nulldelimiterspace} \hbar }$ is the classical Rabi frequency which characterizes the exchange of energy between the qubit and the field mode. By  applying    the  unitary transformation $U\left( t \right) = {e^{ - i\left( {{{{H_0}} \mathord{\left/
					{\vphantom {{{H_0}} \hbar }} \right.
					\kern-\nulldelimiterspace} \hbar }} \right)t}}$, where ${H_0} = \left( {{{\hbar {\omega _0}} \mathord{\left/
			{\vphantom {{\hbar {\omega _0}} 2}} \right.
			\kern-\nulldelimiterspace} 2}} \right){\sigma _z}$. The interaction Hamiltonian becomes
\begin{eqnarray}\label{E280}
{H_I}\left( t \right) =  - \hbar {\Omega _R}\left( {{\sigma _ + }{e^{i{\omega _0}t}} + {\sigma _ - }{e^{ - i{\omega _0}t}}} \right)\left( {{e^{ - i\left( {\nu t + \phi } \right)}} + {e^{i\left( {\nu t + \phi } \right)}}} \right).
\end{eqnarray}

We choose $\phi  = 0$ and when  $\nu  = {\omega _0}$, we obtain the Hamiltonian corresponding to a rotation around the $x$-axis of the following Bloch sphere
\begin{eqnarray}\label{E281}
{H_{{\sigma _x}}} = \hbar {\Omega _R}\left( {{\sigma _ + } + {\sigma _ - }} \right) \equiv \hbar {\Omega _R}{\sigma _x}.
\end{eqnarray} 

The corresponding evolution operator is $U\left( t \right) = {e^{ - i\left( {{{{H_{{\sigma _x}}}} \mathord{\left/
					{\vphantom {{{H_{{\sigma _x}}}} \hbar }} \right.
					\kern-\nulldelimiterspace} \hbar }} \right)t}}$. By using  the Taylor expansion,  the matrix form is given by
\begin{eqnarray}\label{E282}
{U_{{\sigma _x}}}\left( t \right) = \left( {\begin{array}{*{20}{c}}
	{\cos \left( {{\Omega _R}t} \right)}&{ - i\sin \left( {{\Omega _R}t} \right)}\\
	{ - i\sin \left( {{\Omega _R}t} \right)}&{\cos \left( {{\Omega _R}t} \right)}
	\end{array}} \right),
\end{eqnarray}
therefore
\begin{eqnarray}\label{E283}
{R_x}\left( \theta  \right) = \left( {\begin{array}{*{20}{c}}
	{\cos \left( {\frac{\theta }{2}} \right)}&{ - i\sin \left( {\frac{\theta }{2}} \right)}\\
	{ - i\sin \left( {\frac{\theta }{2}} \right)}&{\cos \left( {\frac{\theta }{2}} \right)}
	\end{array}} \right),
\end{eqnarray}
where $\theta  = 2{\Omega _R}t$ is the angle of rotation.
\subsubsection*{ $Y$-rotation gate for one qubit }
The rotation around the $y$-axis in the Bloch sphere corresponds to the $X$-rotation with the phase difference of ${\pi  \mathord{\left/
		{\vphantom {\pi  2}} \right.
		\kern-\nulldelimiterspace} 2}$. Therefore, we have chosen ${{\phi  = \pi } \mathord{\left/
		{\vphantom {{\phi  = \pi } 2}} \right.
		\kern-\nulldelimiterspace} 2}$ in equation (\ref{E279}), and we obtain the following interaction Hamiltonian
\begin{eqnarray}\label{E284}
H = \hbar {\Omega _R}\left( {i{\sigma _ - } - i{\sigma _ + }} \right) = \hbar {\Omega _R}{\sigma _y}.
\end{eqnarray}
The corresponding evolution operator is given by ${U_{{\sigma _y}}}\left( t \right) = {e^{ - i{\Omega _R}{\sigma _y}t}}$. Therfore, the matrix form is given by
\begin{eqnarray}\label{E285}
{U_{{\sigma _y}}}\left( t \right) = \left( {\begin{array}{*{20}{c}}
	{\cos \left( {{\Omega _R}t} \right)}&{ - \sin \left( {{\Omega _R}t} \right)}\\
	{\sin \left( {{\Omega _R}t} \right)}&{\cos \left( {{\Omega _R}t} \right)}
	\end{array}} \right),
\end{eqnarray}
which implies
\begin{eqnarray}\label{E286}
{R_y} = \left( {\begin{array}{*{20}{c}}
	{\cos \left( {\frac{\theta }{2}} \right)}&{ - \sin \left( {\frac{\theta }{2}} \right)}\\
	{\sin \left( {\frac{\theta }{2}} \right)}&{\cos \left( {\frac{\theta }{2}} \right)}
	\end{array}} \right).
\end{eqnarray}
\subsubsection*{ $Z$-rotation gate for one qubit }
With the change of the phase, it is not possible to obtain the matrix ${\sigma _z}$ in the equation (\ref{E279}), it is necessary to use the dispersive regime $\Delta  \gg {\Omega _R}$. The interaction Hamiltonian is then written
\begin{eqnarray}\label{E287}
{H_{{\sigma _z}}} = \hbar \frac{{\Omega _R^2}}{\Delta }{\sigma _z}.
\end{eqnarray}
In the same way as before, we find the corresponding matrix form of the evolution operator (${U_{{\sigma _z}}}\left( t \right) = {e^{ - \left( {{i \mathord{\left/
					{\vphantom {i \hbar }} \right.
					\kern-\nulldelimiterspace} \hbar }} \right){H_{{\sigma _z}}}t}}$)
\begin{eqnarray}\label{E288}
{U_{{\sigma _z}}}\left( t \right) = \left( {\begin{array}{*{20}{c}}
	{{e^{ - i\frac{{\Omega _R^2}}{\Delta }t}}}&0\\
	0&{{e^{i\frac{{\Omega _R^2}}{\Delta }t}}}
	\end{array}} \right).
\end{eqnarray}
Finally, we thus find the rotation around  the $z$-axis with ${\theta  \mathord{\left/
		{\vphantom {\theta  2}} \right.
		\kern-\nulldelimiterspace} 2} = \left( {{{\Omega _R^2} \mathord{\left/
			{\vphantom {{\Omega _R^2} \Delta }} \right.
			\kern-\nulldelimiterspace} \Delta }} \right)t$
\begin{eqnarray}\label{E289}
{R_z}\left( \theta  \right) = \left( {\begin{array}{*{20}{c}}
	{{e^{ - i\frac{\theta }{2}}}}&0\\
	0&{{e^{i\frac{\theta }{2}}}}
	\end{array}} \right).
\end{eqnarray}
\subsection{Two-qubit  gates}

\subsubsection*{$\sqrt {iSWAP}$ gate}
The following Hamiltonian describes the capacitive coupling  of  the two  superconducting  qubits\cite{62}
\begin{eqnarray}\label{E290}
{H_{2q}} =  - \frac{{{\omega _1}}}{2}\sigma _z^1 - \frac{{{\omega _2}}}{2}\sigma _z^2 + g_{qq}\left( {\sigma _ + ^1\sigma _ - ^2 + \sigma _ - ^1\sigma _ + ^2} \right),
\end{eqnarray}
where ${\omega _j}\left( {j = 1,2} \right)$ are transition frequencies  between the two lowest energy states  ${\left| 0 \right\rangle _j}$ and ${\left| 1 \right\rangle _j}$, and $g_{qq}$ is the coupling frequency\cite{62}. On resonance (${\omega _1} = {\omega _2}$), the frequencies will keep on resonance during an operation time. In the two-qubit basis $\left\{ {\left| {00} \right\rangle ,\left| {01} \right\rangle ,\left| {10} \right\rangle ,\left| {11} \right\rangle } \right\}$,  the evolution  operator ${U_{{\mathop{\rm int}} }}\left( t \right)$ corresponding to ${H_{{\mathop{\rm int}} }} = g_{qq}\left( {\sigma _ + ^1\sigma _ - ^2 + \sigma _ - ^1\sigma _ + ^2} \right)$ can be  expressed as
\begin{eqnarray}\label{E291}
{U_{{\mathop{\rm int}} }}\left( t \right) = \left( {\begin{array}{*{20}{c}}
	1&0&0&0\\
	0&{\cos \left( {g_{qq}t} \right)}&{ - i\sin \left( {g_{qq}t} \right)}&0\\
	0&{ - i\sin \left( {g_{qq}t} \right)}&{\cos \left( {g_{qq}t} \right)}&0\\
	0&0&0&1
	\end{array}} \right).
\end{eqnarray} 
This swapping evolution allows the implementation of a universal two-qubit gate that allows the preparation of maximally  entangled two-qubit states,  when   switching on the interaction  time $t = {\pi  \mathord{\left/
		{\vphantom {\pi  {\left( {4G} \right)}}} \right.
		\kern-\nulldelimiterspace} {\left( {4g_{qq}} \right)}}$, one realizes the $\sqrt {iSWAP} $ gate, represented by the following  matrix    
\begin{eqnarray}\label{E292}
{U_{{\mathop{\rm int}} }}\left( {\frac{\pi }{{4g_{qq}}}} \right) = \left( {\begin{array}{*{20}{c}}
	1&0&0&0\\
	0&{{1 \mathord{\left/
				{\vphantom {1 {\sqrt 2 }}} \right.
				\kern-\nulldelimiterspace} {\sqrt 2 }}}&{{i \mathord{\left/
				{\vphantom {i {\sqrt 2 }}} \right.
				\kern-\nulldelimiterspace} {\sqrt 2 }}}&0\\
	0&{{i \mathord{\left/
				{\vphantom {i {\sqrt 2 }}} \right.
				\kern-\nulldelimiterspace} {\sqrt 2 }}}&{{1 \mathord{\left/
				{\vphantom {1 {\sqrt 2 }}} \right.
				\kern-\nulldelimiterspace} {\sqrt 2 }}}&0\\
	0&0&0&1
	\end{array}} \right) = \sqrt {iSWAP},
\end{eqnarray}
which forms universal set of gates together with single qubit gates.  More details about  the implementation with the superconducting  device and characterization  by QPT of this gate can be found in  Ref.\cite{62}.   In addition,  the combination  of more than one  $\sqrt {iSWAP} $ gate with  single qubit  gates allows generating a three-qubit  GHZ-state\cite{49} and implementing  two-qubit  Grover's algorithm\cite{113}  with the superconducting  devices, as well  as a 10-qubit  GHZ-state with a transmon device\cite{52}.
\subsubsection*{Two-qubit C-Phase  gate}
Another coupling is the ZZ- interaction, which arises in many experimental approach such as flux\cite{40} and charge\cite{42} qubits, where the coupling between qubit 1 and 2 is described by  the interaction  Hamiltonian
\begin{eqnarray}\label{E293}
{H_{ZZ}} = \frac{{E_{1,2}^{ZZ}}}{4}\sigma _z^1 \otimes \sigma _z^2,
\end{eqnarray}
where ${E_{1,2}^{ZZ}}$ is the coupling  energies. The corresponding   unitary time-evolution of this interaction in the two-qubit basis  is given by ${U_{ZZ}} = \exp \left( { - i{H_{ZZ}}t} \right)$. For  $t = {\pi  \mathord{\left/
		{\vphantom {\pi  {E_{1,2}^{ZZ}}}} \right.
		\kern-\nulldelimiterspace} {E_{1,2}^{ZZ}}}$ ,  the matrix form  of ${U_{ZZ}}$  can be expressed by 
\begin{eqnarray}\label{E294}
{U_{ZZ}} = {e^{i\frac{\pi }{4}}}\left( {\begin{array}{*{20}{c}}
	1&0&0&0\\
	0&{ - i}&0&0\\
	0&0&{ - i}&0\\
	0&0&0&1
	\end{array}} \right).
\end{eqnarray}

By  combining this  unitary operation  with rotations around z of
each qubit, $R_z^1\left( { - {\pi  \mathord{\left/
			{\vphantom {\pi  2}} \right.
			\kern-\nulldelimiterspace} 2}} \right)$ and $R_z^2\left( {{\pi  \mathord{\left/
			{\vphantom {\pi  2}} \right.
			\kern-\nulldelimiterspace} 2}} \right)$, leading  to  the conditional-phase( c-Phase) gate\cite{17},
\begin{eqnarray}\label{E295}
c{P_{11}} = \left[ {R_z^1\left( {{{ - \pi } \mathord{\left/
				{\vphantom {{ - \pi } 2}} \right.
				\kern-\nulldelimiterspace} 2}} \right) \otimes R_z^2\left( {{\pi  \mathord{\left/
				{\vphantom {\pi  2}} \right.
				\kern-\nulldelimiterspace} 2}} \right)} \right]{U_{ZZ}} = {e^{i\frac{\pi }{2}}}\left( {\begin{array}{*{20}{c}}
	1&0&0&0\\
	0&1&0&0\\
	0&0&1&0\\
	0&0&0&{ - 1}
	\end{array}} \right).
\end{eqnarray}

Later, this gate was experimentally  realized using a transmon-resonator system  in  the dispersive regime. It has also been  implemented to  run the  Grover and Deatsch-Josza algorithms and  create Bell  states\cite{16}. Furthermore, the combination of more than one c-Phase gate with single rotation gates has been reported to generate maximally entangled GHZ states in experimental studies with transmon qubit devices  in Refs.\cite{ 48, 50}.
\section{Chapter summary}
In this chapter, we have reviewed some superconducting qubits such  as  charge and transmon qubits and discussed the coupling of transmon-type to a microwave resonator in circuit QED.  We have also introduced the difference between transom and charge regimes and some of the basic regimes of the QED circuit by making an association with its similar which is the well-known cavity QED. We have associated the quantum information processing protocols of chapter \ref{Ch. 1} with the transmon-cavity system discussed here by realizing the $X$-, $Y$-, and  $Z$-rotations gates for one qubit. we have also seen  some two-qubit operations, such as the c-Phase and $\sqrt {iSWAP} $  gates,  which are arisen from two-qubit
interaction Hamiltonian. The remaining chapters of the thesis will detail our    proposal for the realization  of the $X$-rotation gate for two- and multi-qubit and its implementation in  the Grover search algorithm, as well as the entangling  gates  with the transmon-cavity  device, using a one-step approach.

	\chapter{ Realizing two- and multi-qubit quantum gates in open quantum  systems via  QED crcuit for superconducting qubits}\label{Ch. 3}
	The realization of two- and multi-qubit quantum gates or algorithms typically decomposes into single and two-qubit gates that  have a very important practical role in quantum information processing. Such an approach has been experimentally demonstrated using  superconducting  devices\cite{16,26,48,49,50,52,63}. However,  these approaches  become difficult to build  if only  basic gates are available because the number of single- and two-qubit operations drastically  increases as the number of qubits increases\cite{82,114,115}.   Therefore, the direct implementation of quantum gates and algorithms acting on more than two qubits is  useful and efficient method, because it requires a shorter execution time and can be executed with high fidelity than the equivalent ones. In this framework, a one-step operation for generating a GHZ-entangled state and realizing multi-qubit gates have been previously proposed in systems of superconducting qubits in Refs.\cite{91,92,93,94,95,96,97,98,99,100,101,102}.
	
	In this chapter, we will suggest an efficient and fast method to realize the $X$-rotation and entangling gate for two and serval qubits using a one-step approach based on the transmon-resonator system. Such a system comprises transmon-type superconducting  qubits capacitively coupled to a resonator, driven by a strong microwave field. This chapter will begin with a discussion of our transmon -resonator device and its physical description. Next, we suggest a fast scheme to achieve the $X$-rotation gate for two- and multi-qubit requiring  only  one-step operation. Next,  we  will realize the single-shot  entangling  gate for two and several qubits by exploiting the amount of existing entanglement between the transmon-type superconducting  
	qubits.   Finally, we end this chapter with a conclusion.
	
	\section{Multi-qubit  device description}
	\subsection{Multi-qubit  device}
\begin{figure}[H]
		\centering \includegraphics[scale=0.4]{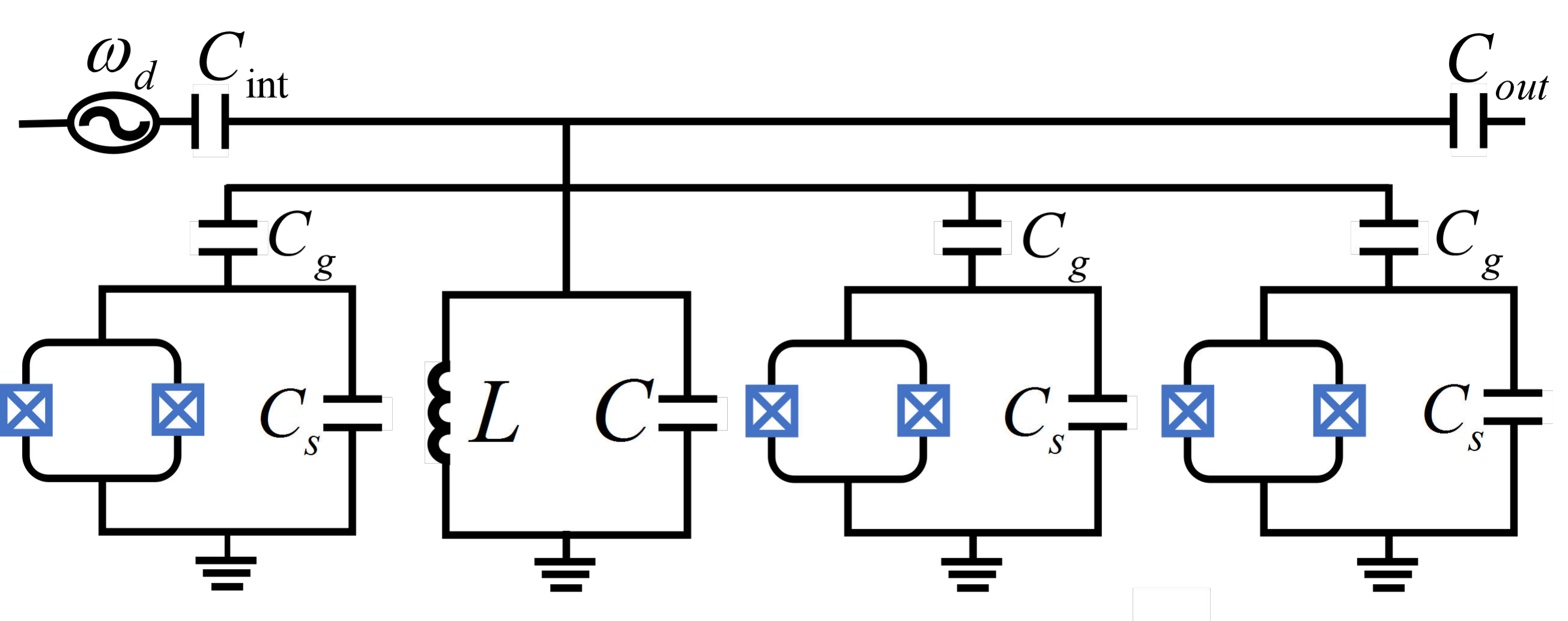}
		\caption{ Electrical circuit diagram of a three-qubit device\cite{sakhouf}. }
		\label{F31}
	\end{figure}

	We here describe our quantum device used during the two last chapters of this thesis, it is comprised of transmon-type superconducting qubits capacitively coupled with a superconducting resonator with the assistance of a classical driving field. Moreover,  we consider the three-qubit device, where three transmon-type superconducting qubits (${Q_j}$ for $j=1,2,3$)  are  capacitively coupled to  a microwave transmission line resonator, as shown  in figure(\ref{F31}). Similar architectures have been studied in previous references \cite{63,116,117}. As shown in figure(\ref{F31}),  the resonator is driven by an external classical field  with the frequency ${\omega _{d}}$  and it can be described by the Hamiltonian\cite{117}
	\begin{eqnarray}\label{E31}
	{H_d} = \varepsilon \left( t \right)\left( {{a^ + }{e^{ - i{\omega _d}t}} + a{e^{i{\omega _d}t}}} \right),
	\end{eqnarray}
	where $\varepsilon \left( t \right)$ is the slowly-varying amplitude of the microwave
	field.
	 The schematic diagram in figure(\ref{F31}) consists of three  transmon qubits with the transition frequency between the two  adjacent  states, this frequency is given by ${\omega _q} = \sqrt {8{E_J}{E_c}} $ (which  is   typically between 3 and 10 GHz) and  can be controlled through the ﬂux dependent of Josephson energy ${E_J} = {E_{J,\max }}\cos \left( {\pi {{{\Phi _e}} \mathord{\left/
	 			{\vphantom {{{\Phi _e}} {{\Phi _0}}}} \right.
	 			\kern-\nulldelimiterspace} {{\Phi _0}}}} \right)$  by means
	 of an external magnetic ﬂux ${{\Phi _e}}$, where ${\Phi _0} = {h \mathord{\left/
	 		{\vphantom {h {2e}}} \right.
	 		\kern-\nulldelimiterspace} {2e}}$ is the magnetic ﬂux
	 quantum\cite{43}.  These qubits  are coupled to the superconducting resonator via identical capacitors ${C_g}$  of frequency ${\omega _r} = {1 \mathord{\left/
	 		{\vphantom {1 {\sqrt {LC} }}} \right.
	 		\kern-\nulldelimiterspace} {\sqrt {LC} }}$ (typically lies between $5$ and $10$ GHz) which  is connected to input and output with capacitors ${C_{{\mathop{\rm int}} }}$
	 and ${C_{{\mathop{\rm out}} }}$, respectively, and a
	 strong microwave field  is applied to the input wire of the
	 resonator.
	 \begin{tcolorbox}
	 	\textbf{Important remark}: During the two last chapters of this thesis and without losing any generality, we consider the transmon-type superconducting qubits identical.  This is valid for tunable transmon when its frequency can be tuned using magnetic flux. In this case,  we will drop off the subscript $j$ for the system parameters.
	 \end{tcolorbox}
	 \subsection{Hamiltonian description and the corresponding time evolution operator}
	 
	 \section*{ Jaynes-Cummings model  and Beyond}
	 The coupled system depicted in figure (\ref{F31}) can be described by  the following Hamiltonian 
	 \begin{eqnarray}\label{E32}
	 \tilde H = {H_{JC}} + {H_D},
	 \end{eqnarray}
	 where ${H_D}$    in  equation (\ref{E31}), while $H_{JC}$ denotes  the Jaynes–Cummings Hamiltonian and is given by  
	 \begin{eqnarray}\label{E33}
	 {H_{JC}} = {\omega _r}{a^ + }a + \frac{1}{2}\sum\limits_{j = 1}^3 {{\omega _{q,j}}\sigma _z^j + \sum\limits_{j = 1}^3 {{g_j}} } \left( {{a^ + }\sigma _ - ^j + a\sigma _ + ^j} \right),
	 \end{eqnarray}
	 where ${a^ + }$ and $(a)$ are  the creation and annihilation operator of the resonator, respectively. ${\omega _{q,j}}$ are the
	 frequencies of transmon-type superconducting  qubits and ${g_j}$ denotes  the qubit–resonator
	 coupling strengths, and the $j$ subscript is used to distinguish the different elements and their parameters. The operators $\sigma _z^j = \left| {{1_j}} \right\rangle \left\langle {{1_j}} \right| - \left| {{0_j}} \right\rangle \left\langle {{0_j}} \right|$, $\sigma _ + ^j = \left| {{1_j}} \right\rangle \left\langle {{0_j}} \right|$ and $\sigma _ - ^j = \left| {{0_j}} \right\rangle \left\langle {{1_j}} \right|$ are Pauli matrices.
	 
	 For a large amplitude driving field $\varepsilon \left( t \right)$, quantum ﬂuctuations in the drive can be neglected as well as being modeled as a classical field. In this case, we can   displace  the resonator field operators by using the time-dependent displacement operator which will be  more convenient
	 \begin{eqnarray}\label{E34}
	 D\left( \alpha  \right) = \exp \left( {\alpha {a^ + } - {\alpha ^*}a} \right).
	 \end{eqnarray}
	 
	 Under this transformation, the operator $a$ becomes   $a + \alpha $   where $ \alpha $ is a complex number that denotes  the classical part of the field. Then,  the displaced Hamiltonian reads\cite{117}
	 \begin{eqnarray}\label{E35}
	 H &=& {D^ + }\left( \alpha  \right)\tilde HD\left( \alpha  \right) - i{D^ + }\left( \alpha  \right)\dot D\left( \alpha  \right)\\ 
	 \nonumber
	 &=& {\omega _r}{a^ + }a + \frac{1}{2}\sum\limits_{j = 1}^3 {{\omega _q}\sigma _z^j + \sum\limits_{j = 1}^3 g \left( {{a^ + }\sigma _ - ^j + a\sigma _ + ^j} \right) - \sum\limits_{j = 1}^3 {g\left( {{\alpha ^*}\sigma _ - ^j + \alpha \sigma _ + ^j} \right)} }, 
	 \end{eqnarray}
	  where  we chose  $\alpha \left( t \right)$ to satisfy
	  \begin{eqnarray}\label{E36}
	  \dot \alpha  =  - i{\omega _r}\alpha  - i\varepsilon \left( t \right){e^{ - i{\omega _d}t}}.
	  \end{eqnarray}
	  
	  The reason  for this $ \alpha $ choice is to eliminate the direct drive on the resonator equation(\ref{E32}). When the  amplitude $\varepsilon $ is   independent of
	  time,  we easily find $\alpha  =  - \left( {{\varepsilon  \mathord{\left/
	  			{\vphantom {\varepsilon  {{\Delta _r}}}} \right.
	  			\kern-\nulldelimiterspace} {{\Delta _r}}}} \right){e^{ - i{\omega _d}t}}$ with ${\Delta _r} = {\omega _r} - {\omega _d}$, and equation (\ref{E35}) becomes\cite{92,119,120,121}
	  \begin{eqnarray}\label{E37}
	  H = {H_1} + {H_2} + {H_3},
	  \end{eqnarray}
	  where
	  \begin{eqnarray}\label{E38}
	  {H_1} = {\omega _r}{a^ + }a + \frac{1}{2}\sum\limits_{j = 1}^3 {{\omega _q}\sigma _z^j},
	  \end{eqnarray}
	  \begin{eqnarray}\label{E39}
	  {H_2} = \sum\limits_{j = 1}^3 {g\left( {{a^ + }\sigma _ - ^j + a\sigma _ + ^j} \right)} ,
	  \end{eqnarray}
	  \begin{eqnarray}\label{E310}
	  {H_3} = \sum\limits_{j = 1}^3 {{\Omega _R}\left( {\sigma _ - ^j{e^{i{\omega _d}t}} + \sigma _ + ^j{e^{ - i{\omega _d}t}}} \right)},
	  \end{eqnarray}
	  where $H_1$ is the free Hamiltonian of the transmon systems  and  the resonator, $H_2$ is the interaction Hamiltonian between  
	  the transmon systems and the resonator and  $H_3$ is the interaction Hamiltonian between the transmon systems and the microwave field(with ${\Omega _R} = {{\varepsilon g} \mathord{\left/
	  		{\vphantom {{\varepsilon g} {{\Delta _r}}}} \right.
	  		\kern-\nulldelimiterspace} {{\Delta _r}}}$ is the Rabi frequency).\\
	  
	  We  choose  the frequencies ${\omega _q}$ and ${\omega _d}$ to satisfy ${\omega _d} = {\omega _q}$ we get ${\Delta _r} = {\omega _r} - {\omega _q}$. Then the interaction Hamiltonian  in the interaction picture with respect to $H_1$ (${H_I} = {e^{i{H_1}t}}H{e^{ - i{H_1}t}}$),  is given by  
	  \begin{eqnarray}\label{E311}
	  {H_I} = {H_0} + {H_i},
	  \end{eqnarray} 
	  where 
	  \begin{eqnarray}\label{E312}
	  {H_0} = 2{\Omega _R}{S_x},~~~with ~~~ {S_x} = \frac{1}{2}\sum\limits_{j = 1}^3 {\left( {\sigma _ - ^j + \sigma _ + ^j} \right)} ,
	  \end{eqnarray} 
	  and 
	  \begin{eqnarray}\label{E313}
	  {H_i} = g\sum\limits_{j = 1}^3 {\left( {{a^ + }\sigma _ - ^j{e^{i{\Delta _r}t}} + a\sigma _ + ^j{e^{ - i{\Delta _r}t}}} \right)}.
	  \end{eqnarray} 
	  
	  In the interaction picture with respect to ${H_0}$, the  Hamiltonian ${H_i}$ yields (${H'_i} = {e^{i{H_0}t}}{H_i}{e^{ - i{H_0}t}}$)
	  \begin{eqnarray}\label{E314}
	  {H'_i} &=&g\left\{ {{e^{ - i{\Delta _r}t}}{a^ + }\left[ {{S_x} + \frac{1}{2}\sum\limits_{j = }^3 {\left( {\sigma _z^j - \sigma _ - ^j + \sigma _ + ^j} \right){e^{ - 2i{\Omega _R}t}} - \frac{1}{2}\sum\limits_{j = }^3 {\left( {\sigma _z^j + \sigma _ - ^j - \sigma _ + ^j} \right){e^{2i{\Omega _R}t}} } } } \right]} \right\}\nonumber \\ 
	  &+& H.c.
	  \end{eqnarray}
	  
	  In the strong driving regime   when  $2{\Omega _R} \gg g$ and $2{\Omega _R} \gg {\Delta _r}$, we can neglect from equation (\ref{E314}) the fast-oscillating terms with 
	  high frequencies. Then the Hamiltonian H can be reduced to \cite{92,119}
	  \begin{eqnarray}\label{E315}
	  {H'_i} = g\left( {{a^ + }{e^{ - i{\Delta _r}t}} + a{e^{i{\Delta _r}t}}} \right){S_x}.
	  \end{eqnarray}
	  The corresponding   evolution operator  to  the Hamiltonian (\ref{E315}) can be written in the form\cite{92,122}
	  \begin{eqnarray}\label{E316}
	  {U'_i}\left( t \right) = {e^{ - iA\left( t \right)S_x^2}}{e^{ - iB\left( t \right){S_x}a}}{e^{ - i{B^*}\left( t \right){S_x}{a^ + }}},
	  \end{eqnarray}
	  where 
	  \begin{eqnarray}\label{E317}
	  A\left( t \right) = \frac{{{g^2}\left[ {t + {{\left( {{e^{ - i{\Delta _r}t}} - 1} \right)} \mathord{\left/
	  					{\vphantom {{\left( {{e^{ - i{\Delta _r}t}} - 1} \right)} {i{\Delta _r}}}} \right.
	  					\kern-\nulldelimiterspace} {i{\Delta _r}}}} \right]}}{{{\Delta _r}}},~~~ and~~~ B\left( t \right) = \frac{{g\left( {{e^{i{\Delta _r}t}} - 1} \right)}}{{i{\Delta _r}}}.
	  \end{eqnarray}
	  In the following, if we choose the evolution time, $t$, to satisfy  ${\Delta _r}t = 2\pi $, we can  obtain  $B (t) = 0$ and $A\left( t \right) = {{{g^2}t} \mathord{\left/
	  		{\vphantom {{{g^2}t} {{\Delta _r}}}} \right.
	  		\kern-\nulldelimiterspace} {{\Delta _r}}}$. The corresponding evolution operator to the whole system becomes independent  of the resonator mode and is given  by  
	  \begin{eqnarray}\label{E318}
	  {U_I}\left( t \right) = {e^{ - i{H_0}t}}{U'_i}\left( t \right) = {e^{ - 2i{\Omega _R}{S_x}t}}{e^{ - 2i\lambda S_x^2t}},
	  \end{eqnarray}
	  where $\lambda  = {{{g^2}} \mathord{\left/
	  		{\vphantom {{{g^2}} {2{\Delta _r}}}} \right.
	  		\kern-\nulldelimiterspace} {2{\Delta _r}}}$ is the effective qubit–qubit coupling
	  strength\cite{120}. We note that the evolution operator described by Eq. (\ref{E318}) does not include the photon operator $a$ or ${a^ + }$ of the resonator mode. Therefore, the cavity can be  initially in an arbitrary state (e.g., in a vacuum state, a coherent state,  a Fock state, or even a thermal state).
	  \begin{tcolorbox}
	  	\textbf{Important remark}: We note that the deriving evolution operator of our transmon-resonator does not necessarily need the condition  ${\Delta _r} \gg g$, indicating that the schemes existing in the two last chapters do not necessarily work in the dispersive regime.  In addition, the preceding schemes work in the dispersive regime, leading to a rather slow energy exchange between transmon systems. Therefore, our schemes do not limit  any of the dispersive approximation conditions being valid. 
	  \end{tcolorbox}
	 \section{Open  system  dynamics }
	 An open quantum system is a quantum system that is coupled to its environment.  Describing the evolution of the open quantum system dynamics using a Schrödinger equation approach alone will make it impossible to achieve.  In this subsection, we, therefore, employ the master  equation, which is a more powerful approach that can describe the full dynamics of open quantum systems. 
	\subsection{Master  equation}
	We start with the general formalism of the master equation and then apply it to the case of our transmon-resonator system discussed above. In general, the time evolution is a mixed state that can be described by the density matrix $\rho $, by taking into account the system dissipation. This description  is governed   by  the master equation, which  can be generally written  in the Lindblad form as (more details on the master equation derivation can be found in Red.\cite{123})
	\begin{eqnarray}\label{E319}
	\frac{{d\rho }}{{dt}} = \frac{{ - i}}{\hbar }\left[ {{H_{sys}},\rho } \right] + \pounds\left[ \Lambda  \right]\rho ,
	\end{eqnarray}
	where  $\pounds\left[ \Lambda  \right]\rho= \frac{1}{2}\sum\limits_i {\left( {{L_i}\rho L_i^ +  - L_i^ + {L_i}\rho  - \rho L_i^ + {L_i}} \right)} $ and ${L_i} = \sqrt {{\gamma _n}} {\Lambda _i}$ are collapse operators, with ${\Lambda _i}$ the Lindblad operators, describing the non-unitary dynamics under  a Hamiltonian of the system  ${H_{sys}}$  due to  the coupling of the system to its environment, and ${{\gamma _n}}$ are the corresponding rates.
	
	 As an example,  we consider the transmon-resonator system consisting of the transmon qubits coupled to a resonator with the assistance of a classical field.   Coupling with additional uncontrollable degrees of freedom leads to dissipation in the system.  In the Born-Markov approximation, this dissipation can be characterized by the resonator dissipation which is described by a photon leakage rate $\kappa  \simeq {{{\omega _r}} \mathord{\left/
	 		{\vphantom {{{\omega _r}} Q}} \right.
	 		\kern-\nulldelimiterspace} Q}$,  the relaxation rate ${\gamma _{1,j}} = {1 \mathord{\left/
	 		{\vphantom {1 {{T_1}}}} \right.
	 		\kern-\nulldelimiterspace} {{T_1}}}$, and the dephasing rate ${\gamma _{\varphi ,j}} = {1 \mathord{\left/
	 		{\vphantom {1 {{T_1}}}} \right.
	 		\kern-\nulldelimiterspace} {{T_1}}} - {1 \mathord{\left/
	 		{\vphantom {1 {2{T_2}}}} \right.
	 		\kern-\nulldelimiterspace} {2{T_2}}}$ for each transmon qubit. In these expressions, $Q$ is the quality factor of the resonator, ${{T_1}}$ is the decoherence time, and ${{T_2}}$ is the dephasing time.  In the presence of these
	 processes, the master equation in Eq. (\ref{E319}) can be rewritten  as  
	 \begin{eqnarray}\label{E320}
	 \frac{{d\rho }}{{dt}} =  - \frac{i}{\hbar }\left[ {H',\rho } \right] + \kappa \pounds\left[ a \right] + \sum\limits_j {{\gamma _{1,j}}\pounds\left[ {\sigma _ - ^j} \right] + \sum\limits_j {{\gamma _{\varphi ,j}}\pounds\left[ {\sigma _z^j} \right]} } ,
	 \end{eqnarray}
	 where $H'$ is the system  Hamiltonian, which  can be either $H$ in Eq. (\ref{E37}) or  ${H_I}$ in Eq. (\ref{E311}). While $\pounds\left[ \Lambda  \right] = \Lambda \rho {\Lambda ^ + } - {{{\Lambda ^ + }\Lambda \rho } \mathord{\left/
	 		{\vphantom {{{\Lambda ^ + }\Lambda \rho } {2 - {{\rho {\Lambda ^ + }\Lambda } \mathord{\left/
	 							{\vphantom {{\rho {\Lambda ^ + }\Lambda } 2}} \right.
	 							\kern-\nulldelimiterspace} 2}}}} \right.
	 		\kern-\nulldelimiterspace} {2 - {{\rho {\Lambda ^ + }\Lambda } \mathord{\left/
	 				{\vphantom {{\rho {\Lambda ^ + }\Lambda } 2}} \right.
	 				\kern-\nulldelimiterspace} 2}}}$  with  $\Lambda  = a,\sigma _ - ^j,~~ or ~\sigma _z^z$.
	
	\subsection{Fidelity  under master equation }
	In order to characterize the performance of the quantum states or operations, it is necessary to compute the fidelity. The fidelity presents the measurement  of the distance between two  mixed states  $\rho $ and $\sigma $, and it  is given  by 
	\begin{eqnarray}\label{E321}
	F = {\left( {Tr\left[ {\sqrt {\sqrt \sigma  \rho \sqrt \sigma  } } \right]} \right)^2}.
	\end{eqnarray}
	
	In the case where one of the two  states  chosen  pure, e.g.  $\sigma  = \left| \psi  \right\rangle \left\langle \psi  \right|$, equation (\ref{E321}) can be readily simplifled to
	\begin{eqnarray}\label{E322}
	F &=& Tr\left[ {\rho \left| \psi  \right\rangle \left\langle \psi  \right|} \right]\nonumber\\ 
	&=& \left\langle \psi  \right|\rho \left| \psi  \right\rangle ,
	\end{eqnarray}
	where $\psi $ is the pure output state expected from the ideal unitary
	operation and $\rho $ is the  density operator of the system at the end
	of the quantum  operation, obtained by
	numerically solving the master equation(\ref{E320}), using the
	QuTiP software \cite{124,125}, which is an open-source software for simulating the  open quantum systems dynamics.
	\section{ Realizing $X$-rotation gate for multi-qubit in the open quantum systems}
	The $X$-rotation gate plays a similar role to Hadamard gate for creating  quantum  superposition and it is an important tool in quantum computation\cite{1}. Therefore, the direct realization of a multi-qubit $X$-rotation gate can be expected to provide fast and practical ways in quantum computation in comparison with the ones provided by single qubit $X$-rotation gates. In addition, its  realization plays a critical role in Grover's search algorithm as we  will  show in the next  chapter. Here, we present a scheme to realize the X-rotation gate for two and  three qubits with an angle ${{ - \pi } \mathord{\left/
			{\vphantom {{ - \pi } 2}} \right.
			\kern-\nulldelimiterspace} 2}$ using only a single-step operation based on the evolution operation (\ref{E318}) in the open quatum  systems.
	\subsection{$X$-rotation gate for two  transmon systems }
	
	\subsubsection{Two-qubit  $X$-rotation gate generation}
	In the two-qubit basis $\left\{ {\left| {00} \right\rangle ,\left| {01} \right\rangle ,\left| {10} \right\rangle ,\left| {11} \right\rangle } \right\}$, we define the two-qubit $X$-rotation gate with an angle  ${{ - \pi } \mathord{\left/
			{\vphantom {{ - \pi } 2}} \right.
			\kern-\nulldelimiterspace} 2}$
	\begin{eqnarray}\label{E323}
	R_x^{ \otimes 2}\left( { - \frac{\pi }{2}} \right) = \left( {\begin{array}{*{20}{c}}
		1&i&i&{ - 1}\\
		i&1&{ - 1}&i\\
		i&{ - 1}&1&i\\
		{ - 1}&i&i&1
		\end{array}} \right),
	\end{eqnarray}
	where single qubit gate $R_x^{ \otimes 2}\left( { - {\pi  \mathord{\left/
				{\vphantom {\pi  2}} \right.
				\kern-\nulldelimiterspace} 2}} \right)$ transform states $\left| 0 \right\rangle $ and $\left| 1 \right\rangle $ to the superposition states $\left( {{1 \mathord{\left/
				{\vphantom {1 {\sqrt 2 }}} \right.
				\kern-\nulldelimiterspace} {\sqrt 2 }}} \right)\left( {\left| 0 \right\rangle  + i\left| 1 \right\rangle } \right)$ and $\left( {{i \mathord{\left/
				{\vphantom {i {\sqrt 2 }}} \right.
				\kern-\nulldelimiterspace} {\sqrt 2 }}} \right)\left( {\left| 0 \right\rangle  - i\left| 1 \right\rangle } \right)$, respectively.
    
	The evolution  operator ${U_I}\left( t \right)$ of the two  transmon  system (Eq. (\ref{E318})) can be expressed on the   same basis as\cite{119,126}
	\begin{eqnarray}\label{E324}
	{U_I}\left( t \right) &=& {e^{ - 2i{\Omega _R}{S_x}t}}{e^{ - 2i\lambda S_x^2t}}\nonumber\\ 
	&=&  \left( {\begin{array}{*{20}{c}}
		{\frac{1}{2} + \frac{1}{2}\cos \left( {2{\Omega _R}t} \right){e^{ - 2i\lambda t}}}&{ - \frac{i}{2}\sin \left( {2{\Omega _R}t} \right){e^{ - 2i\lambda t}}}&{ - \frac{i}{2}\sin \left( {2{\Omega _R}t} \right){e^{ - 2i\lambda t}}}\\
		{ - \frac{i}{2}\sin \left( {2{\Omega _R}t} \right){e^{ - 2i\lambda t}}}&{\frac{1}{2} + \frac{1}{2}\cos \left( {2{\Omega _R}t} \right){e^{ - 2i\lambda t}}}&{ - \frac{1}{2} + \frac{1}{2}\cos \left( {2{\Omega _R}t} \right){e^{ - 2i\lambda t}}}\\
		{ - \frac{i}{2}\sin \left( {2{\Omega _R}t} \right){e^{ - 2i\lambda t}}}&{ - \frac{1}{2} + \frac{1}{2}\cos \left( {2{\Omega _R}t} \right){e^{ - 2i\lambda t}}}&{\frac{1}{2} + \frac{1}{2}\cos \left( {2{\Omega _R}t} \right){e^{ - 2i\lambda t}}}\\
		{ - \frac{1}{2} + \frac{1}{2}\cos \left( {2{\Omega _R}t} \right){e^{ - 2i\lambda t}}}&{ - \frac{i}{2}\sin \left( {2{\Omega _R}t} \right){e^{ - 2i\lambda t}}}&{ - \frac{i}{2}\sin \left( {2{\Omega _R}t} \right){e^{ - 2i\lambda t}}}
		\end{array}} \right.\nonumber \\
	&&\left. {\begin{array}{*{20}{c}}
		{ - \frac{1}{2} + \frac{1}{2}\cos \left( {2{\Omega _R}t} \right){e^{ - 2i\lambda t}}}\\
		{ - \frac{i}{2}\sin \left( {2{\Omega _R}t} \right){e^{ - 2i\lambda t}}}\\
		{ - \frac{i}{2}\sin \left( {2{\Omega _R}t} \right){e^{ - 2i\lambda t}}}\\
		{\frac{1}{2} + \frac{1}{2}\cos \left( {2{\Omega _R}t} \right){e^{ - 2i\lambda t}}}
		\end{array}} \right),
	\end{eqnarray} 
	where  ${S_x} = \left( {{1 \mathord{\left/
				{\vphantom {1 2}} \right.
				\kern-\nulldelimiterspace} 2}} \right)\sum\limits_{j = 1}^2 {\left( {\sigma _ + ^j + \sigma _ - ^j} \right)} $. In the situation where $2\lambda t = \pi $ and $2{\Omega _R}t = {{\left( {2n + 1} \right)\pi } \mathord{\left/
			{\vphantom {{\left( {2n + 1} \right)\pi } 2}} \right.
			\kern-\nulldelimiterspace} 2}$ (n is an interger), i.e.,  the operation time can be chosen to  be  $t = {\pi  \mathord{\left/
			{\vphantom {\pi  {2\lambda }}} \right.
			\kern-\nulldelimiterspace} {2\lambda }}$ and  the Rabi frequency satisfying ${{{\Omega _R}} \mathord{\left/
			{\vphantom {{{\Omega _R}} \lambda }} \right.
			\kern-\nulldelimiterspace} \lambda } = {{\left( {2n + 1} \right)} \mathord{\left/
			{\vphantom {{\left( {2n + 1} \right)} 2}} \right.
			\kern-\nulldelimiterspace} 2}$. Therefore,  the above evolution operator (Eq. (\ref{E322}))  leads to the two-qubit  X-rotation gate  with
	the angle $- {\pi  \mathord{\left/
			{\vphantom {\pi  2}} \right.
			\kern-\nulldelimiterspace} 2}$ , writes\cite{126}
	\begin{eqnarray}\label{E326}
	{U_I}\left( t \right) = \left( {\begin{array}{*{20}{c}}
		1&i&i&{ - 1}\\
		i&1&{ - 1}&i\\
		i&{ - 1}&1&i\\
		i&i&i&1
		\end{array}} \right) = R_x^{ \otimes 2}\left( { - \frac{\pi }{2}} \right).
	\end{eqnarray}
	
	This result depicts  that the coupling  between
	the two transmon qubits mediated by a resonator
	with the assistance of a strong classical  field can easily
	achieve the two-qubit $X$-rotation gate. This gate is efficient as it can replace the need for two individual $X$-rotation gates applied to two separate qubits. Combining this gate with other two-qubit gates is crucial in forming a set of universal gates for quantum computation \cite{1}. 
	\subsubsection{Fidelity  and discussion}
	In order to examine the robustness with respect to the dissipation of our transmon-resonator system for realizing the two-qubit $X$-rotation gate (Eq. (\ref{E326})), it is necessary to estimate the fidelity. The gate  fidelity  is given in equation (\ref{E322}), where  $\left| \psi  \right\rangle $ is the ideal output state without any dissipation after applying a two-qubit $X$-rotation gate performed on the transmon system initially in the state $\left| {00} \right\rangle $ and the resonator mode initially in the
	state ${\left| 0 \right\rangle _r}$. It is given by
	\begin{eqnarray}\label{E327}
	\left| \psi  \right\rangle  = \frac{1}{2}\left( {\left| {00} \right\rangle  + i\left| {01} \right\rangle  + i\left| {10} \right\rangle  - \left| {11} \right\rangle } \right),
	\end{eqnarray}
	while $\rho $ is the final density operator of the system at the end of the two-qubit $X$-rotation  gate, obtained by solving the master equation (\ref{E320}) numerically  for $j$=1,2, where  $\pounds\left[ \Lambda  \right] = \Lambda \rho {\Lambda ^ + } - {{{\Lambda ^ + }\Lambda \rho } \mathord{\left/
			{\vphantom {{{\Lambda ^ + }\Lambda \rho } {2 - {{\rho {\Lambda ^ + }\Lambda } \mathord{\left/
								{\vphantom {{\rho {\Lambda ^ + }\Lambda } 2}} \right.
								\kern-\nulldelimiterspace} 2}}}} \right.
			\kern-\nulldelimiterspace} {2 - {{\rho {\Lambda ^ + }\Lambda } \mathord{\left/
					{\vphantom {{\rho {\Lambda ^ + }\Lambda } 2}} \right.
					\kern-\nulldelimiterspace} 2}}}$  with $\Lambda  = a$, $\sigma _ - ^1$, $\sigma _ - ^2$, $ \sigma _z^1$, or $\sigma _z^2$. In addition,  we assume ${\gamma _{1,1}} = {\gamma _{1,2}} = {\gamma _1}$ and  ${\gamma _{\varphi ,1}} = {\gamma _{\varphi ,2}} = {\gamma _\varphi }$ for simplicity.
	
\begin{figure}[H]
	 	\subfloat[\label{b1}]{%
	 		\includegraphics[width=0.5\columnwidth]{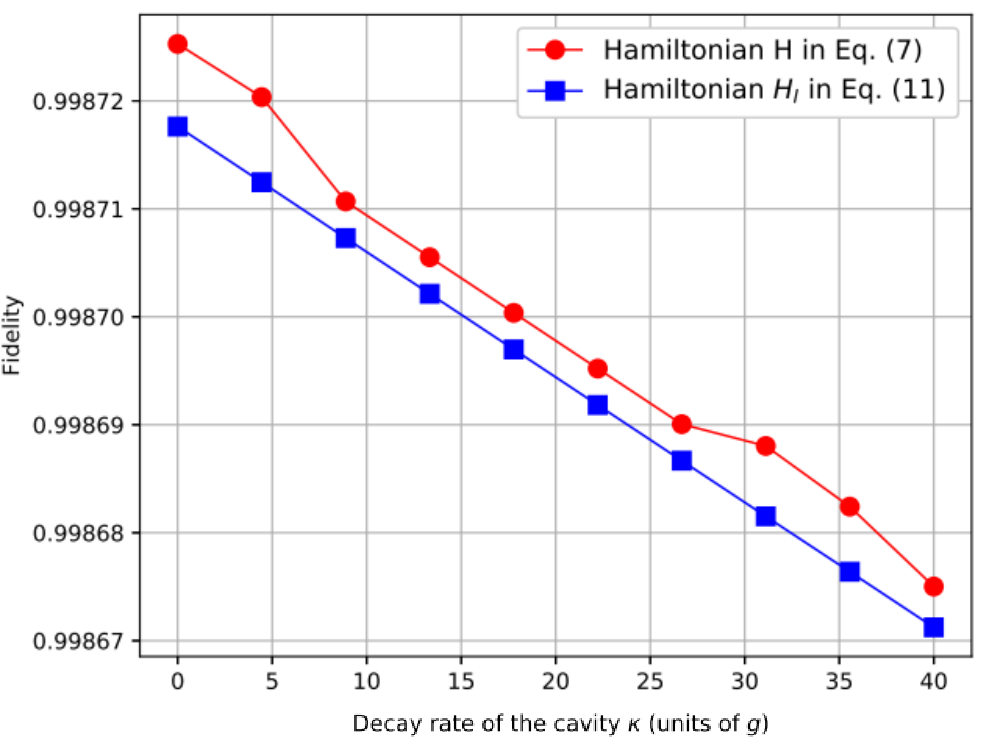}%
	 	}\hfill
	 	\subfloat[\label{b2} ]{%
	 		\includegraphics[width=0.5\columnwidth]{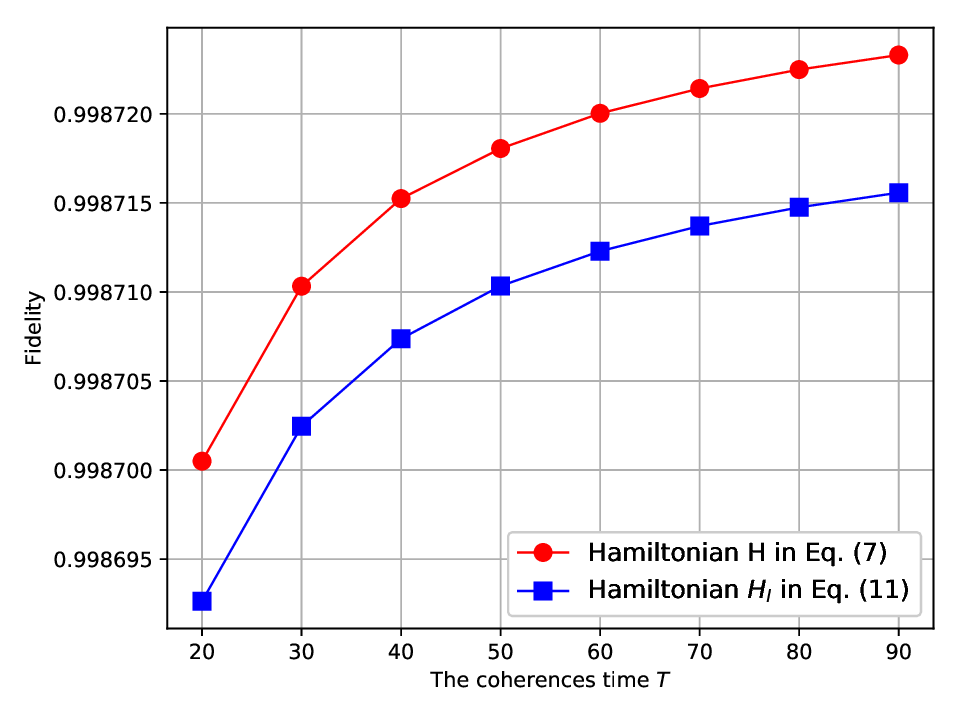}%
	 	}\hfill
	 	\caption{ (Color online) (a) Fidelity of the two-qubit $X$-rotation gate as a function of the decay rate   of resonator  $\kappa $ (a) and as a function of the coherence time $T$ (b).  The blue solid line means that our calculation was performed by using the Hamiltonian (\ref{E37}) for $j$=1,2,  while the calculation referred to by the solid line was performed using the Hamiltonian  (\ref{E311}) for $j$=1,2.   The system parameters used in the numerical simulation are mentioned in the text.}
	 	\label{F32}

	 \end{figure}

	We have numerically simulated the fidelity of the two-qubit  $X$-rotation gate in equation (\ref{E326}) under both Hamiltonians (\ref{E37}) and (\ref{E311}) versus $\kappa $ and $T$ as depicted in figure (\ref{F32}), where the resonator mode is in the vacuum state. The fidelities of the two-qubit $X$-rotation gate corresponding to both  Hamiltonians (\ref{E37}) and (\ref{E311}) are the almost same. From figure (\ref{F32}a), one can see that the fidelity is achieved at a high $ \sim 0.998$ even though the resonator decay $\kappa $ reaching to large values, indicating that the gate fidelity is insensitive to the resonator decay. For the numerical simulations of  this figure, we  have chosen  the system  parameters as ${{{\omega _r}} \mathord{\left/
			{\vphantom {{{\omega _r}} {2\pi  = 6.4}}} \right.
			\kern-\nulldelimiterspace} {2\pi  = 6.4}}$ GHz and ${{{\omega _q}} \mathord{\left/
			{\vphantom {{{\omega _q}} {2\pi  = 4.8}}} \right.
			\kern-\nulldelimiterspace} {2\pi  = 4.8}}$GHz,  leading to ${{{\Delta _r}} \mathord{\left/
			{\vphantom {{{\Delta _r}} {2\pi  = 0.896}}} \right.
			\kern-\nulldelimiterspace} {2\pi  = 0.896}}$GHz. In addition, we choose  ${g \mathord{\left/
			{\vphantom {g {2\pi  = 60}}} \right.
			\kern-\nulldelimiterspace} {2\pi  = 60}}$ MHz and ${{{\Omega _R}} \mathord{\left/
			{\vphantom {{{\Omega _R}} {2\pi  = 200Mh}}} \right.
			\kern-\nulldelimiterspace} {2\pi  = 200}}$MHz, which are readily available as ${g \mathord{\left/
			{\vphantom {g {2\pi  = 10 - 200}}} \right.
			\kern-\nulldelimiterspace} {2\pi  = 10 - 200}}$ MHz and ${{{\Omega _R}} \mathord{\left/
			{\vphantom {{{\Omega _R}} {2\pi  = 300}}} \right.
			\kern-\nulldelimiterspace} {2\pi  = 300}}$ MHz which have been reported in references\cite{48,63,116,117,118, 127}.  We note that these values satisfy the conditions for realizing the two-qubit $X$-rotation gate chosen above.
	In addition,  the measuring of the  decoherence time ${T_1}$ and dephasing time ${T_2}$ for transmon devices between  20 and 95 $\mu s$ have been reported in references\cite{45,128,129}.  Therefore,  we here choose ${T_1} = 95\mu s$ and ${T_2} = 70\mu s$, leading to a relaxation rate ${{{\gamma _1}} \mathord{\left/
			{\vphantom {{{\gamma _1}} {2\pi  = 1.68}}} \right.
			\kern-\nulldelimiterspace} {2\pi  = 1.68}}$kHz and dephasing rate  ${{{\gamma _\varphi }} \mathord{\left/
			{\vphantom {{{\gamma _\varphi }} {2\pi  = 1.68}}} \right.
			\kern-\nulldelimiterspace} {2\pi  = 1.68}}$kHz.  Furthermore, we  have chosen  the  operation time based  on the gate conditions above  for $n=0$ to  be  $t = {\pi  \mathord{\left/
			{\vphantom {\pi  {\left( {4{\Omega _R}} \right)}}} \right.
			\kern-\nulldelimiterspace} {\left( {4{\Omega _R}} \right)}} \sim 0.62ns$ (where  $\lambda  = 3{\Omega _R}$), which is much shorter than the coherence time of the transmon system and the cavity decay time. While the effect of the coherences time $\sim 0.998695$ on the performance of the two-qubit $X$-roation  gate has been explored in figure (\ref{F32}b). As shown in Fig. (\ref{F32}b),   the effect of the decoherence becomes less dominant, and the fidelity of a quantum gate has achieved a high fidelity (even for a  small value of the coherences time, whereas, it is clear that the gate fidelity slowly increases as the coherences time increases and it reaches to  $0.99872$  high value. Here, we have chosen  $\kappa  = 2.5$MHz and the other parameters are the same as in panel (a).
	
\subsection{$X$-rotation gate for three  transmon systems  }
\subsubsection{Three-qubit  $X$-rotation gate generation}
In the three-qubit  basis $\left\{ {\left| {000} \right\rangle ,\left| {001} \right\rangle ,\left| {010} \right\rangle ,\left| {011} \right\rangle ,\left| {100} \right\rangle ,\left| {101} \right\rangle ,\left| {110} \right\rangle ,\left| {111} \right\rangle } \right\}$,   we define
the  $X$-rotation gate for three  qubits  with an angle $- {\pi  \mathord{\left/
		{\vphantom {\pi  2}} \right.
		\kern-\nulldelimiterspace} 2}$ 
 \begin{eqnarray}\label{E328}
 {W_3} &=& R_x^{ \otimes 3}\left( { - \frac{\pi }{2}} \right)\sigma _x^{ \otimes 3}\nonumber \\ 
 &=& \frac{1}{{2\sqrt 2 }}\left( {\begin{array}{*{20}{c}}
 	{ - i}&{ - 1}&{ - 1}&i&{ - 1}&i&i&1\\
 	{ - 1}&{ - i}&i&{ - 1}&i&{ - 1}&1&i\\
 	{ - 1}&i&{ - i}&{ - 1}&i&1&{ - 1}&i\\
 	i&{ - 1}&{ - 1}&{ - i}&1&i&i&{ - 1}\\
 	{ - 1}&i&i&1&{ - i}&{ - 1}&{ - 1}&i\\
 	i&{ - 1}&1&i&{ - 1}&{ - i}&i&{ - 1}\\
 	i&1&{ - 1}&i&{ - 1}&i&{ - i}&{ - 1}\\
 	1&i&i&{ - 1}&i&{ - 1}&{ - 1}&{ - i}
 	\end{array}} \right),
 \end{eqnarray}
 where ${\sigma _x}$ denotes the $NOT$ gate operation for single qubit,  while $\sigma _x^{ \otimes 3} = {\sigma _x} \otimes {\sigma _x} \otimes {\sigma _x}$. The evolution operator  of equation (\ref{E318}) can be expressed in the same basis as\cite{sakhouf}
 \begin{eqnarray}\label{E329}
 {U_I}\left( t \right) &=& {e^{ - ibh{S_x}}}{e^{ - ibS_x^2}}\nonumber\\
 &=& \left( {\begin{array}{*{20}{c}}
 	A'&C'&C'&D'&C'&D'&D'&B'\\
 	C'&A'&D'&C'&D'&C'&B'&D'\\
 	C'&D'&A'&C'&D'&B'&C'&D'\\
 	D'&C'&C'&A'&B'&D'&D'&C'\\
 	C'&D'&D'&B'&A'&C'&C'&D'\\
 	D'&C'&B'&D'&C'&A'&D'&C'\\
 	D'&B'&C'&D'&C'&D'&A'&C'\\
 	B'&D'&D'&C'&D'&C'&C'&A'
 	\end{array}} \right) ,
 \end{eqnarray}
 where $A'$, $B'$, $C'$ and $D'$ are the elements of the evolution operator matrix ${U_I}\left( t \right)$ (Eq. (\ref{E328})) given by
 \begin{eqnarray}\label{EB317}
 A' &=& \frac{1}{4}{e^{ - i\frac{{9b}}{4}}}\left( {\cos \left( {\frac{{3bh}}{2}} \right) + 3\cos \left( {\frac{{bh}}{2}} \right){e^{2ib}}} \right),\\
 B' &=& \frac{{ - i}}{4}{e^{ - i\frac{{9b}}{4}}}\left( {\sin \left( {\frac{{3bh}}{2}} \right) - 3\sin \left( {\frac{{bh}}{2}} \right){e^{2ib}}} \right),\\
 C' &=& \frac{{ - i}}{4}{e^{ - i\frac{{9b}}{4}}}\left( {\sin \left( {\frac{{3bh}}{2}} \right) + \sin \left( {\frac{{bh}}{2}} \right){e^{2ib}}} \right),\\
 D' &=& \frac{1}{4}{e^{ - i\frac{{9b}}{4}}}\left( {\cos \left( {\frac{{3bh}}{2}} \right) - \cos \left( {\frac{{bh}}{2}} \right){e^{2ib}}} \right),
 \end{eqnarray}
  with $b = 2\lambda t$ and $h = {{{\Omega _R}} \mathord{\left/
  		{\vphantom {{{\Omega _R}} \lambda }} \right.
  		\kern-\nulldelimiterspace} \lambda }$. Taking the condition
  chosen above ${\Delta _r}t = 2\pi $, and setting $bh = {\pi  \mathord{\left/
  		{\vphantom {\pi  2}} \right.
  		\kern-\nulldelimiterspace} 2}$ and
  $b = 8\left( {m + 1} \right)\pi $ ($m$ is an integer), we can choose the
  interaction time $t = {t_R} = {\pi  \mathord{\left/
  		{\vphantom {\pi  {4{\Omega _R}}}} \right.
  		\kern-\nulldelimiterspace} {4{\Omega _R}}}$ and ${{{\Delta _r}} \mathord{\left/
  		{\vphantom {{{\Delta _r}} {{\Omega _R} = 8}}} \right.
  		\kern-\nulldelimiterspace} {{\Omega _R} = 8}}$. Then the evolution operator (Eq. \ref{E329}) leads to
  \begin{eqnarray}\label{E334}
  {U_I}\left( {{t_R}} \right) = \exp \left( {{{i\pi } \mathord{\left/
  			{\vphantom {{i\pi } 2}} \right.
  			\kern-\nulldelimiterspace} 2}} \right){W_3},
  \end{eqnarray}
  where $\exp \left( {{{i\pi } \mathord{\left/
  			{\vphantom {{i\pi } 2}} \right.
  			\kern-\nulldelimiterspace} 2}} \right)$ is an overall phase and ${W_3}$ is the single three-qubit $X$-rotation gate  defined in equation (\ref{E328}). This gate
  together with single two-qubit rotation gates is very important to form a set of universal gates for quantum computation\cite{1}.
 \subsubsection{Fidelity  and discussion}
 In order to evaluate the validity of our proposal, we calculate the fidelity of the three-qubit $X$-rotation gate in Eq.  (\ref{E334}). It is given in equation (\ref{E322}), where $\left| \psi  \right\rangle $ is the  output state without dissipation system after a joint three-qubit $X$-rotation operation is performed on the initial state $\left| {000} \right\rangle {\left| 0 \right\rangle _r}$ and it is given by 
 \begin{eqnarray}\label{Eq}
 \left| {{\psi }} \right\rangle &=& \left( {{1 \mathord{\left/
 			{\vphantom {1 {2\sqrt 2 }}} \right.
 			\kern-\nulldelimiterspace} {2\sqrt 2 }}} \right)\left( { - i} \right.\left| {{0_1}{0_2}{0_3}} \right\rangle  - \left| {{0_1}{0_2}{1_3}} \right\rangle
 - \left| {{0_1}{1_2}{0_3}} \right\rangle  + i\left| {{0_1}{1_2}{1_3}} \right\rangle  - \left| {{1_1}{0_2}{0_3}} \right\rangle\nonumber\\
 &+& \left. {i\left| {{1_1}{0_2}{1_3}} \right\rangle  + i\left| {{1_1}{1_2}{0_3}} \right\rangle  + \left| {{1_1}{1_2}{1_3}} \right\rangle } \right) \otimes {\left| n \right\rangle _r}  , 
 \end{eqnarray}
 while $\rho $ is the final  density  of the whole system obtained by solving the master equation (\ref{E320}) (for $j$= 1,2,3) using the numerical simulation, where  $\Lambda  = a,\sigma _ - ^1,\sigma _ - ^2,\sigma _ - ^3,\sigma _z^1,\sigma _z^2$, or $\sigma _z^3$ and for simplicity, we assume ${\gamma _{1,1}} = {\gamma _{1,2}} = {\gamma _{1,3}} = {\gamma _1}$ and ${\gamma _{\varphi ,1}} = {\gamma _{\varphi ,2}} = {\gamma _{\varphi ,3}} = {\gamma _\varphi }$. 
 
\begin{figure}[H]
	\subfloat[\label{b1}]{%
		\includegraphics[width=0.5\columnwidth]{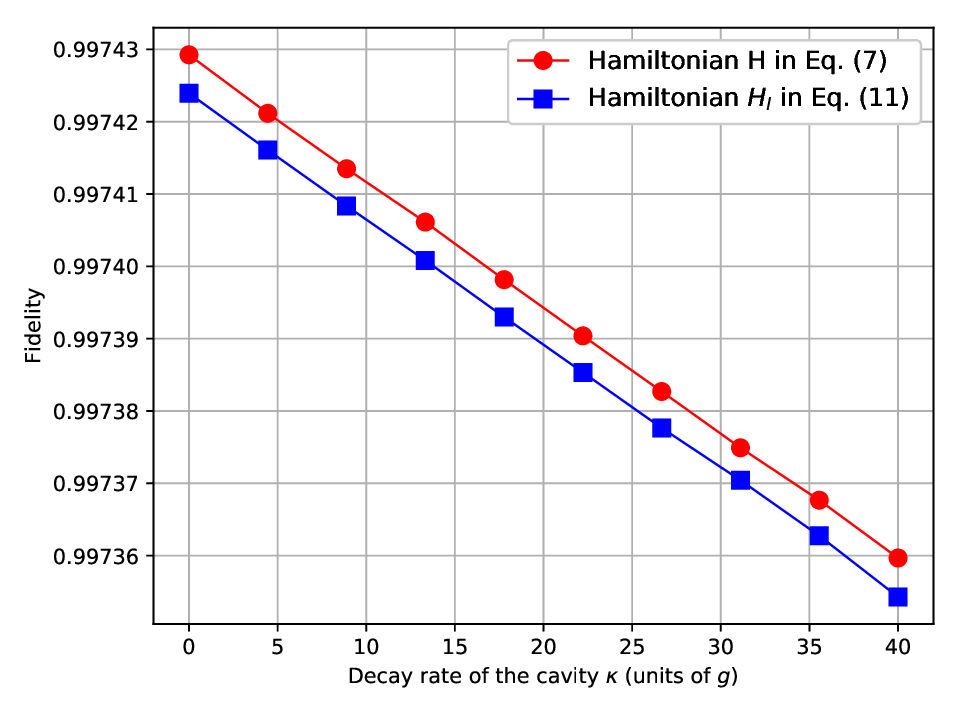}%
	}\hfill
	\subfloat[\label{b2} ]{%
		\includegraphics[width=0.5\columnwidth]{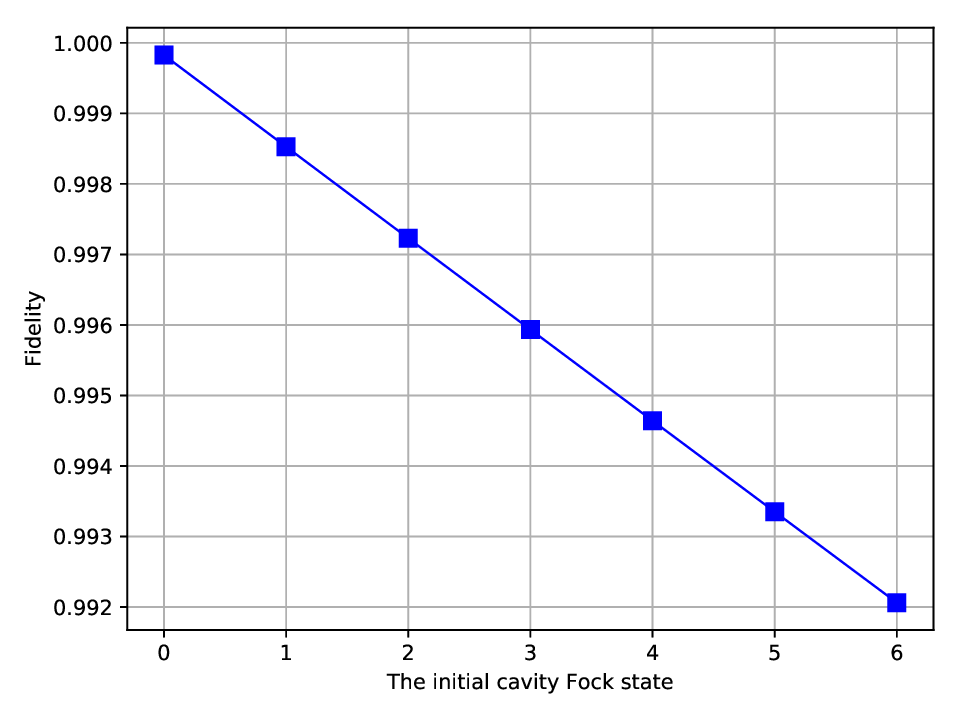}%
	}\hfill
	\caption{ (Color online) (a) Fidelity of the three-qubit $X$-rotation gate  for different  values of the cavity decay rate $\kappa $ (a) and initial cavity Fock states $\left| n \right\rangle $ (b).  The blue solid line means that our calculation was performed by using the Hamiltonian (\ref{E37}) for $j$=1,2,3,  while the calculation referred by the solid line was performed using the Hamiltonian  (\ref{E311}) for $j$=1,2,3.   The system parameters used in the numerical simulation are the same as a figure (\ref{F32}) while the others are mentioned in the text.}
	\label{F33}

\end{figure}

 We numerically calculated the fidelity of the single three-qubit  $X$-rotation gate in equation (\ref{E334})  for different initial states of the transmon system. Due to the similarity, we only demonstrate the fidelity corresponding to the initial state $\left| {000} \right\rangle $  in figure (\ref{F33}) as an example. Figure (\ref{F33}a) shows the fidelity versus the cavity decay rate $\kappa $ for the operation gate defined in Eq. (\ref{E334}), when the resonator is initially in the ground state and it will remain in this state throughout the procedure. Our simulations show that the fidelities corresponding to Hamiltonians (\ref{E37}) and (\ref{E311}) are almost the same, and the gate fidelity decreases while increasing the cavity 
 decay rate $\kappa $ as expected.
 However,  one can see the gate operational fidelity nearly keeps $0.997$ for a large range of cavity decay. This result means that cavity decay does not influence the gate fidelity of our proposal. Additionally, the fidelity of this scheme can be further increased as the qubit–resonator coupling strength $g$ is increased. This improvement will be clearly shown in figure (\ref{F33}b) for $n = 0$ compared with figure (\ref{F33}a).  The system parameters used here are the same of the figure (\ref{F32}). For the resonator frequency chosen in the previous subsection and for the decay rate ${\kappa  \mathord{\left/
 		{\vphantom {\kappa  {2\pi }}} \right.
 		\kern-\nulldelimiterspace} {2\pi }} = 0.5$ MHz used in the numerical simulation, the required quality factor  for the
 resonator is $Q \sim 1.28 \times {10^4}$. The cavity quality factor
 here is achievable in experiments because the transmission
 line resonator (TLR) with a quality factor $Q \sim {10^3}$ has been demonstrated in experiments\cite{48,118}.
 
 Furthermore, the fidelity of an operation gate for different initial cavity Fock state $\left| n \right\rangle $ is plotted in Fig. (\ref{F33}b), it is clear that the fidelity slightly decreases with the increase of the photon number. In contrast, we find that the fidelity can still be over $0.99$ even for $n =$ 6, which means the cavity state almost does not influence the gate fidelity. Finally, our presented analysis is given throughout the two subsections implying that  high-fidelity implementation of a two- and  multi-qubit $X$-rotation gates are  feasible with the present circuit QED technology. 	
\section{Single-shot entangling gate in the open quantum systems}
Due to its high fidelity, a single-shot entangling gate plays a key role in quantum information processing. This operation gate quickly creates a maximally entangled state and forms a universal entrance set for quantum computing. Currently, the preparation and demonstration of multi-qubit entangling gates are  achieved based on the decomposition  of single- and two-qubit operations,
yielding lower fidelity and requiring longer execution time. Here, we propose the two- and three-qubit entangling gates to generate a maximally entangled state based on our transmon-resonator system requiring only a one-step operation, as well as we numerically investigate the system dynamics of these operation gates under realistic assumptions about the system parameters.
\subsection{ Entangling gate for two  transmon systems}

\subsubsection{Two-qubit  entangling gate generation}
In the following, if we choose the Rabi  frequency satisfying  ${\Omega _R}t = n\pi $ in the evolution operator of two transmon systems ${U_I}\left( t \right)$ (Eq.(\ref{E324})), we have the two-qubit entangling  gate 
\begin{eqnarray}\label{E335}
{U_I}\left( t \right) = {e^{ - i\lambda t}}\left( {\begin{array}{*{20}{c}}
	{\cos \left( {\lambda t} \right)}&0&0&{ - i\sin \left( {\lambda t} \right)}\\
	0&{\cos \left( {\lambda t} \right)}&{ - i\sin \left( {\lambda t} \right)}&0\\
	0&{ - i\sin \left( {\lambda t} \right)}&{\cos \left( {\lambda t} \right)}&0\\
	{ - i\sin \left( {\lambda t} \right)}&0&0&{\cos \left( {\lambda t} \right)}
	\end{array}} \right), 
\end{eqnarray}
where the factor ${e^{ - i\lambda t}}$ is an overall phase. Remarkably,  where the transmon systems   are initially in the state  $\left| {00} \right\rangle $ and the operation time $t = {\pi  \mathord{\left/
		{\vphantom {\pi  {4\lambda }}} \right.
		\kern-\nulldelimiterspace} {4\lambda }}$, the
gate described by equation (\ref{E335}) can fastly generate a two-qubit
maximally entangled state ${U_I}\left( {{\pi  \mathord{\left/
			{\vphantom {\pi  {4\lambda }}} \right.
			\kern-\nulldelimiterspace} {4\lambda }}} \right)\left| {00} \right\rangle  = \left( {{1 \mathord{\left/
			{\vphantom {1 {\sqrt 2 }}} \right.
			\kern-\nulldelimiterspace} {\sqrt 2 }}} \right)\left( {\left| {00} \right\rangle  - i\left| {11} \right\rangle } \right)$, which was realized
experimentally with a superconducting transmon
device using sequences of single- and two-qubit gates\cite{16,62}.
\subsubsection{Two-qubit entangling gate characterization}
\subsubsection*{Time evolution}
\begin{figure}[H]
		\subfloat[\label{b1}]{%
			\includegraphics[width=0.5\columnwidth]{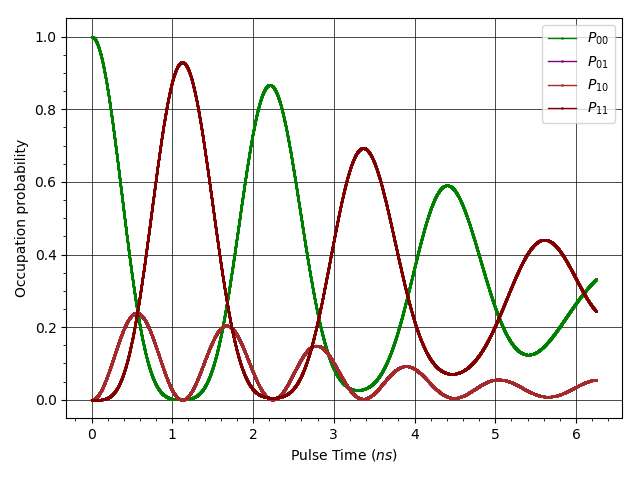}%
		}\hfill
		\subfloat[\label{b2} ]{%
			\includegraphics[width=0.5\columnwidth]{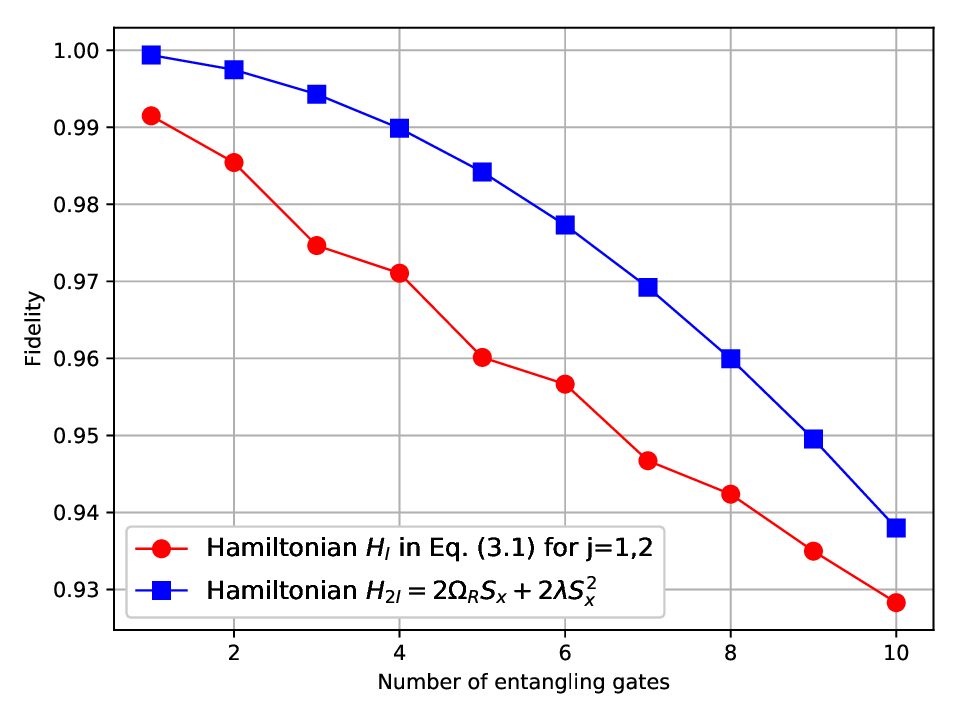}%
		}\hfill
		\caption{ (Color online) Simulated measurements of the single-shot entangling gate (Eq.\ref{E335}) (a) The
			occupation probabilities versus the interaction time. The system parameters used in the numerical simulation are ${g \mathord{\left/
					{\vphantom {g {2\pi }}} \right.
					\kern-\nulldelimiterspace} {2\pi }} = 40$ MHz, $2{\Omega _R} = 9g$, ${T_c} = 20\mu s$, and ${\kappa  \mathord{\left/
					{\vphantom {\kappa  {2\pi  = 1.5}}} \right.
					\kern-\nulldelimiterspace} {2\pi  = 1.5}}$ MHz. (b) Fidelity versus the number of the two-qubit entangling gates with environment decoherence under the Hamiltonian in equation (\ref{E311}) and without  environment decoherence under the following  Hamiltonian ${H_{2I}} = 2{\Omega _R}{S_x} + 2\lambda S_x^2$.   The data parameters used are the
			same as in panel (b).}
		\label{F34}

	\end{figure}

The result which  is shown  in figure (\ref{F34})  presents the time evolution of the state occupation
probabilities ${P_{ij}}$ as a function of the interaction time corresponding
to applying a single entangling gate,  assuming    the transmon
systems initially prepared in state $\left| {00} \right\rangle $ and the resonator
in the vacuum state (where ${P_{ij}}$ are the probabilities of
finding $\left| {ij} \right\rangle $ with $ij$ = 00; 01; 10;  or 11). These results were obtained by numerically solving the Lindblad master equation in equation (\ref{E320}) (for j=1,2),  we here assume that ${T_c} = {T_1} = {T_2}$, which is valid for tunable transmons\cite{87}.	
 As shown in Fig. (\ref{F34}b), the swap oscillation between  $\left| {00} \right\rangle $ and $\left| {11} \right\rangle $  is nearly perfect. However, due to the environmental decoherence of our transmon-resonator system, these  oscillations decrease with the increasing operation time. On the other hand,  the probabilities ${P_{00}}$ and ${P_{11}}$ should approach  50$\%$  for an ideal entangled state. Still, in our results, these probabilities are reduced to less than 50$\%$ in equal superposition crossing points due to the effects of environmental noise. Comparing the simulated with ideal states using the equation ( \ref{E322}), we find the fidelities corresponding to applying 10-entangling  gates are shown in Fig. (\ref{F34}b).  It is clear that  from  this  figure the fidelities decrease as the growing number of gates, however,   the Hamiltonian in Eq. (\ref{E311}) decreases more rapidly   caused by the effect of the system decoherence and the resonator decay.  In addition,  we note that the fidelity for one run of the entangling gate is  99.15$\%$,  which means that our scheme is efficient and faster compared to one of the maximally entangled states created from sequences of single- and two-qubit gates\cite{16,62}.
 \subsection{Entangling gate for three  transmon systems }
 \subsubsection{Three-qubit  entangling gate generation}
 Following the evolution operator in equation (\ref{E328}),  in which if setting  ${\Delta _r}t = 2\pi $ and ${\Delta _r} \approx g$, we
 can make the interaction time  $t$ and Rabi frequency ${\Omega _R}$
 satisfy $t = {T} = {{3\pi } \mathord{\left/
 		{\vphantom {{3\pi } {4{\Omega _R}}}} \right.
 		\kern-\nulldelimiterspace} {4{\Omega _R}}}$ and ${\Omega _R} = 3\lambda $. Leading  to  the
 three-qubit entangling gate\cite{sakhouf} 
 \begin{eqnarray}\label{E336}
 M\left( {{T}} \right) = {e^{ - i\theta }}\left( {\begin{array}{*{20}{c}}
 	{\tilde A'}&0&0&0&0&0&0&{\tilde B'}\\
 	0&{\tilde A'}&0&0&0&0&{\tilde B'}&0\\
 	0&0&{\tilde A'}&0&0&{\tilde B'}&0&0\\
 	0&0&0&{\tilde A'}&{\tilde B'}&0&0&0\\
 	0&0&0&{\tilde B'}&{\tilde A'}&0&0&0\\
 	0&0&{\tilde B'}&0&0&{\tilde A'}&0&0\\
 	0&{\tilde B'}&0&0&0&0&{\tilde A'}&0\\
 	{\tilde B'}&0&0&0&0&0&0&{\tilde A'}
 	\end{array}} \right),
 \end{eqnarray}
 where ${e^{ - i\theta }}$ is an overall phase (with $\theta  = {{27\pi } \mathord{\left/
 		{\vphantom {{9\pi } 8}} \right.
 		\kern-\nulldelimiterspace} 8}$), $\tilde A' =\cos \left( {{{{\Omega _R}{\rm T}} \mathord{\left/
 			{\vphantom {{{\Omega _R}{\rm T}} 3}} \right.
 			\kern-\nulldelimiterspace} 3}} \right)$ and $\tilde B' =  - i\sin \left( {{{{\Omega _R}{\rm T}} \mathord{\left/
 			{\vphantom {{{\Omega _R}{\rm T}} 3}} \right.
 			\kern-\nulldelimiterspace} 3}} \right)$. It is worth noticing that if  the transmon  systems are initially prepared  in the state $\left| {000} \right\rangle $, the operator $M\left( {{T}} \right)$ can generate a three-qubit  $GHZ$-state ${\rm M}\left( {{{3\pi } \mathord{\left/
 			{\vphantom {{3\pi } {4{\Omega _R}}}} \right.
 			\kern-\nulldelimiterspace} {4{\Omega _R}}}} \right)\left| {{0}{0}{0}} \right\rangle  = \left( {{{{e^{i\theta }}} \mathord{\left/
 			{\vphantom {{{e^{i\theta }}} {\sqrt 2 }}} \right.
 			\kern-\nulldelimiterspace} {\sqrt 2 }}} \right)\left( {\left| {{0}{0}{0}} \right\rangle  - i\left| {{1}{1}{1}} \right\rangle } \right)$. This maximally entangled state for three transmon qubits was experimentally generated with a superconducting transmon and phase  device using one- and two-qubit gates\cite{48}. 
\subsubsection{Three-qubit entangling gate characterization}

\begin{figure}[H]
	\subfloat[\label{b1}]{%
		\includegraphics[width=0.49\columnwidth]{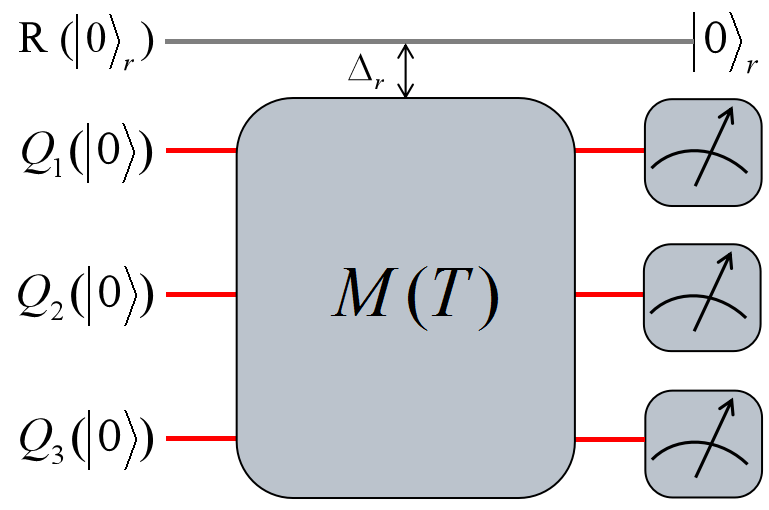}%
	}\hfill
	\subfloat[\label{b2} ]{%
		\includegraphics[width=0.5\columnwidth]{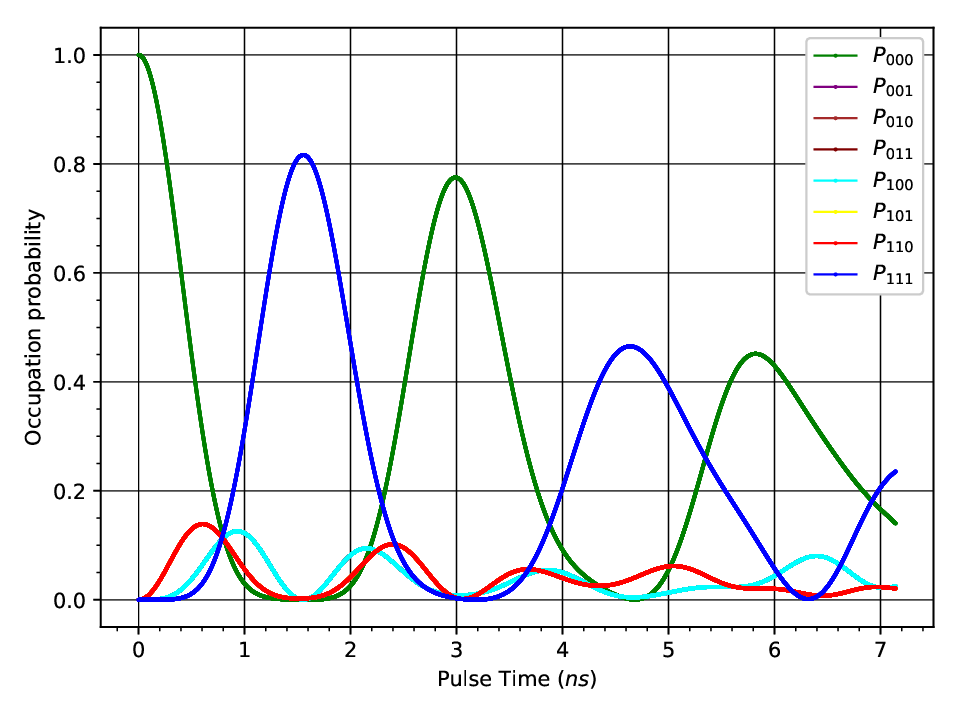}%
	}\hfill
	\caption{ (Color online) (a) Quantum circuit for generating $GHZ$-state using the single-shot entangling gate(Eq.\ref{E335}) (b)  The occupation probabilities as a function of the operation time ${T_m}$\cite{130}. The system  parameters used here are ${g \mathord{\left/
				{\vphantom {g {2\pi }}} \right.
				\kern-\nulldelimiterspace} {2\pi }} = 20$ MHz, $2{\Omega _R} = 9g$, ${T_c} = 20\mu s$, and ${\kappa  \mathord{\left/
				{\vphantom {\kappa  {2\pi  = 1.5}}} \right.
				\kern-\nulldelimiterspace} {2\pi  = 1.5}}$ MHz.}
	\label{F35}
\end{figure}

To generate a $GHZ$-state based on three transmon systems when the resonator mode in the state ${\left| 0 \right\rangle _r}$. We initialize all three transmon systems in their ground state $\left| {000} \right\rangle $, whereas assuming that the resonator remains in the vacuum  state  throughout the operation[See figure (\ref{F35}a)]. This assumption is valid for transmon-resonator systems\cite{52}. We depict  in Fig. (\ref{F35}b), the time evolution of 
probabilities ${P_{ijk}}$ during the GHZ-state protocol versus   the interaction time corresponding to applying a one-entangling gate(where ${P_{ijk}}$ are the probabilities to find the state $\left| {ijk} \right\rangle $ with  $ijk = 000,001,010,011,100,101,110$, or $ 111$ ). We get these results by solving the Lindblad master equation in equation (\ref{E320}) (for $j$=1,2,3) for the transmon-cavity  density matrix $\rho $  using numerical simulation,  we  have assumed  here   the coherence times to  be equal  ${T_c} = {T_1} = {T_2}$.

As shown  in Fig (\ref{F35}b), the swap oscillation between the states 
$\left| {000} \right\rangle $ and $\left| {111} \right\rangle $ are almost perfect, however, we  can
see these oscillations decrease with the increase of the
pulse time, and also the other  occupation probabilities
are almost visible owing to the environmental decoherence of both transmon qubits and resonator, which have been taken into account throughout our simulations.  In addition,  unlike an ideal  $ GHZ$ state in which  the probabilities ${P_{000}}$ and ${P_{111}}$  can reach  50$\%$, from the figure (\ref{F35}b)   we  see these probabilities are reduced to less than 50$\%$ in crossing points due to the environmental noise. \\

\begin{figure}[H]
	\subfloat[\label{b1}]{%
		\includegraphics[width=0.49\columnwidth]{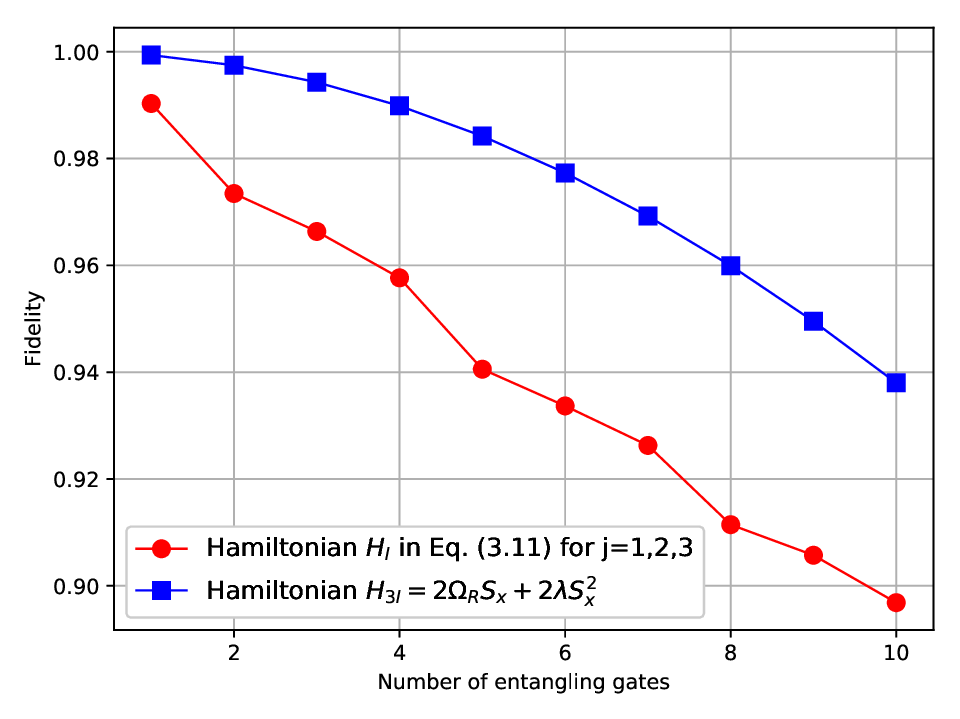}%
	}\hfill
	\subfloat[\label{b2} ]{%
		\includegraphics[width=0.5\columnwidth]{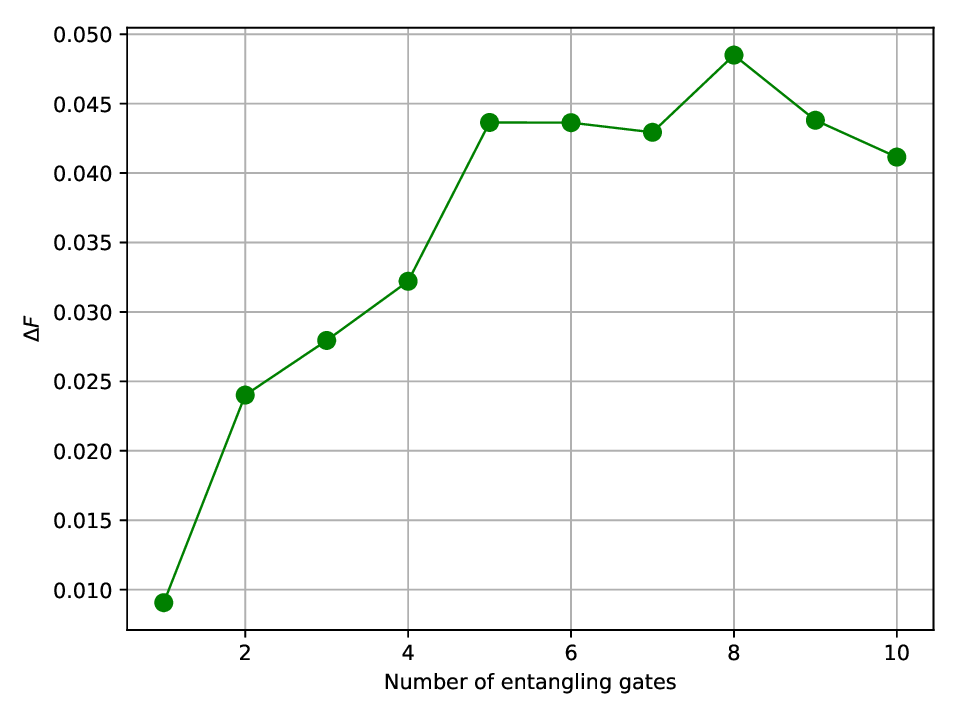}%
	}\hfill
	\caption{ (Color online)  (a) Numerical simulation of the fidelity as a function of the number of entangling gates under both Hamiltonians in equation (\ref{E311}) and ${H_{3I}}$. (b) A number of entangling operations of the error are caused by decoherence. $\Delta F$ denotes the difference of the fidelity for generating  $GHZ$ state with and without dissipation\cite{130}.    The data parameters used are the
		same as in figure (\ref{F35}a).}
	\label{F36}
\end{figure}

We compare the simulated and ideal states by estimating  the gate fidelity of the three-qubit entangling gate using the equation (\ref{E322}), where $\left| \psi  \right\rangle $ is the output state of the  system without considering the dissipation and dephasing performed on the transmon  system initially prepared  in the state $\left| {000} \right\rangle $ and the resonator mode initially in the state  ${\left| 0 \right\rangle _r}$, while $\rho $ is the simulated matrix obtained by numerically solving the equation (\ref{E320}).  Figure (\ref{F36}a) shows the gate fidelity of the entangling gate with and without taking into account the decoherence of the transmon system and the cavity decay under the hamiltonians   (\ref{E311}) and ${H_{3I}}$, respectively. We find the fidelity of both Hamiltonians corresponding to the applied 10-gate by estimating decreases as the number of the three-qubit entangling gates increases, whereas, due to the environmental  noise of both transmon  system  and resonator decay   the fidelity corresponding to   Hamiltonian in Eq. (\ref{E311}) decreases more rapidly as expected. Higher  fidelity is achieved when a single entangling gate is applied ($ \sim 99.03\%$).
For more understanding of the significance of the
estimated fidelities, we compare our result to the fidelity
obtained from a measurement of GHZ protocols based
on the sequence of one- and two-qubit gates. Three-qubit
entanglement using superconducting qubits for maximally GHZ-entangled states such as phase qubits\cite{49}, containing one- and two-C-Phase-qubit operations, measured to be ${F_{GHZ}}$=87$\%$
and transmon qubits\cite{48}, containing one- and two-iSWAP-qubit operations, measured to be ${F_{GHZ}}$= 62$\%$. The lower fidelities of these protocols compared to our  protocol are caused by a longer total period of operations, resulting in decomposing into sequences of one and two iSWAP or C-Phase gates, which leads to destroying the coherence of the system by environmental noise. Our protocol proposal is fast and more efficient especially for complicated quantum circuits because it contains only entangling gate implementation requiring a one-step operation. 

Finally, we evaluate the influence of environmental decoherence in generating a GHZ state by  comparing the fidelity with and without
dissipation. Such a result is shown in Figure (\ref{F36}b), in which 
the difference between the two fidelities is plotted versus the number of gates. As expected, the error due to the decoherence increases with the growing number of entangling gates. The error caused by the effects of decoherence corresponding to the one entangling gate is around 0.95$\%$.
\begin{tcolorbox}
	\textbf{Important remark}: It is important to note that the less sensitive to charge noise for the transmon system the bigger the price to pay which is a reduction in anharmonicity.  For this reason,  we need a sufficiently large anharmonicity to suppress the transitions to higher transmon levels outside of the two levels  $\left| 0 \right\rangle $ and  $\left| 1 \right\rangle $. We thus  can discuss    a rough estimate of the influence of the
	second excited state $\left| 2 \right\rangle $. In the transmon regime (${E_J} \gg {E_C}$), an approximate estimate of the relative and absolute anharmonicity are ${\alpha _r} =  - {\left( {8{{{E_J}} \mathord{\left/
					{\vphantom {{{E_J}} {{E_C}}}} \right.
					\kern-\nulldelimiterspace} {{E_C}}}} \right)^{ - {1 \mathord{\left/
					{\vphantom {1 2}} \right.
					\kern-\nulldelimiterspace} 2}}}$ and $\alpha  = {E_{01}}{\alpha _r}$, respectively\cite{131}, with  ${E_{10}} = \sqrt {8{E_J}{E_C}}  - {E_C} $  denoted by the energy separation between the levels $\left| 0 \right\rangle $ and $\left| 1 \right\rangle $. We now choose ${{{E_J}} \mathord{\left/
			{\vphantom {{{E_J}} {{E_C}}}} \right.
			\kern-\nulldelimiterspace} {{E_C}}} = 50$,  ${E_C} = 2\pi  \times 2$ GHz, and therfore the absolute anharmonicity is given by $\alpha  =  - 2\pi  \times 1.9$ GHz.  Assuming the neighbor energy levels are coupled by a microwave driven by  Rabi frequency ${\Omega _R} = 2\pi  \times 330$MHz, the occupation
	probability  of  $\left| 2 \right\rangle $  is given by  ${P_2} \approx {2 \mathord{\left/
			{\vphantom {2 {\left( {4 + {{\left( {{\alpha  \mathord{\left/
												{\vphantom {\alpha  {{\Omega _R}}}} \right.
												\kern-\nulldelimiterspace} {{\Omega _R}}}} \right)}^2}} \right)}}} \right.
			\kern-\nulldelimiterspace} {\left( {4 + {{\left( {{\alpha  \mathord{\left/
									{\vphantom {\alpha  {{\Omega _R}}}} \right.
									\kern-\nulldelimiterspace} {{\Omega _R}}}} \right)}^2}} \right)}}$\cite{92,132}, which is  less than 8$\%$. We thus can treat the transmon-type superconducting qubit as a two-level system even in the strong driving regime. 
\end{tcolorbox}

\section{Chapter summary }
	
	In this chapter, based on the multi-qubit device constructed as  two and more transmon-type superconducting qubits that are capacitively coupled  to a resonator ($LC$) driven by a strong classical microwave field, we have proposed a faster and more efficient approach showing that the quantum gates or algorithms are realized using only  one-step operation. Using this device and approach,  we have achieved the two- and three-qubit quantum gates such as the $X$-rotation gate for creating superposition  states and another one called the  single-shot entangling gate for generating maximally   entangled state. In addition,  we suggested the dynamics in open quantum systems of these schemes by estimating the gate fidelities and especially the time evolution of the single-shot entangling gates. High fidelities are achieved to  implement  our schemes using  realistic assumptions about system parameters and the gate time is  significantly shorter than the lifetimes of superconducting systems  and than most current implementations.
	
	\chapter{Implementing $X$-rotation gates in Grover's algorithm and quantum process tomography of the entangling gates}\label{Ch. 4}

	Searching through large and unsorted databases is a crucial problem with broad applications. The Grover search algorithm is famous and one of the most popular quantum algorithms providing a robust method for quantum computers to accomplish a database search and quadratically overcome  nowadays methods in terms of time due to laws of quantum mechanics such as quantum superposition, entanglement, and interference.  Additionally,  Grover’s algorithm is important as it is easy to achieve and is therefore considered as a good algorithm which  demonstrates the power of quantum computation.
	
	Many efforts have been devoted to  theoretically and experimentally implementing this algorithm in many  physical  systems,  for instance,  Nuclear Magnetic Resonance (NMR)\cite{29,30}, trapped ions\cite{31,33,34,134}, cavity QED\cite{35,36,133,135}, as well as   an interesting  system about a  quantum  device consisting of  two   transmon  qubits coupled to a resonator\cite{16,113}. Recently, Figatt et al \cite{39}  experimentally implemented three-qubit Grover’s
	search algorithm  using a decomposition  of a one-qubit rotation  and two-qubit gates with a scalable trapped atomic ion system.  The execution of this algorithm for one iteration was done using 
	two methods of the quantum oracle, the Boolean oracle and the  phase oracle , which  are achieved with a high accuracy of   70.5$\%$ fdelity. In the past decades, W L Yang et al\cite{37} used the cavity QED system  for implementing Grover’s algorithm for three qubits, and subsequently, M Waseem et al \cite{136} proposed a scheme for implementing  of same algorithm  using four-level SQUID  in cavity QED. 
	
	In  this chapter,  we  discuss that  the Grover's search  algorithm  can  be performed using  the $X$-roataion  gate,  which  is similar to  Hadamard gate as discussed in  chapter \ref{Ch. 1}. This method
	is  able to find a specific object  in an unsorted database of size $N$ using only  ${\rm O}\left( {\sqrt N } \right)$ operations. We then  implement   this algorithm  using the realization of the $X$- rotation gate based on the     transmon-resonator  system for two  and three qubits, this gate   was realized in the previous chapter.  In  addition,  we fully characterize the performance of the single-shot entangling  gate  for two and three tranmon-type superconducting qubits using   quantum process tomography(QPT). The QPT is a useful method for experimentally obtaining a complete implementation of the quantum gate or algorithm,  it has been used with superconducting systems  for the entangling gate to characterize a square root i-SWAP
	gate with  two  qubits\cite{113,137,138} and for multi-qubit operations to characterize a Toffoli and C-Phase gates\cite{26,63}, as well as with trapped ions  to obtain  a complete implementation of  two- and three-qubit operations\cite{80,139,140}.   
		\section{  Grover's algorithm with X-rotation gates   }
		In this section, we first discuss Grover's search algorithm for three qubits using $X$-rotation gates which are similar to the Hadamard gate with a probability higher than $0.94$ to find the correct state. Then we generalize the realization of this algorithm for several qubits.
		\subsection{ Three-Qubit Grover’s Search Algorithm  }
		Two quantum gates can be used to prepare the states into a quantum superposition, they play a similar role to the Hadamard gate. The first logic gate is the $X$-rotation gate defined in equation (\ref{E12}).  While the second gate is the $NOT$  gate defined in  Eq.(\ref{E6}).So,  in the special  case  when the rotation   $\theta  =  - {\pi  \mathord{\left/
				{\vphantom {\pi  2}} \right.
				\kern-\nulldelimiterspace} 2}$  is in Eq. \ref{E12},  we get
		\begin{eqnarray}\label{E41}
			{R_x}\left( { - \frac{\pi }{2}} \right) = \frac{1}{{\sqrt 2 }}\left( {\begin{array}{*{20}{c}}
					1&i\\
					i&1
				\end{array}} \right).
			\end{eqnarray}
			
			Then,  we can  generate the superposition  states in the case of three qubits by  applying  the following operation 
			\begin{eqnarray}\label{E42}
			{W_3} &=& R_x^{ \otimes 3}\left( { - \frac{\pi }{2}} \right)\sigma _x^{ \otimes 3}.
			\end{eqnarray}
			
\begin{figure}[H]
					\centering \includegraphics[scale=0.5]{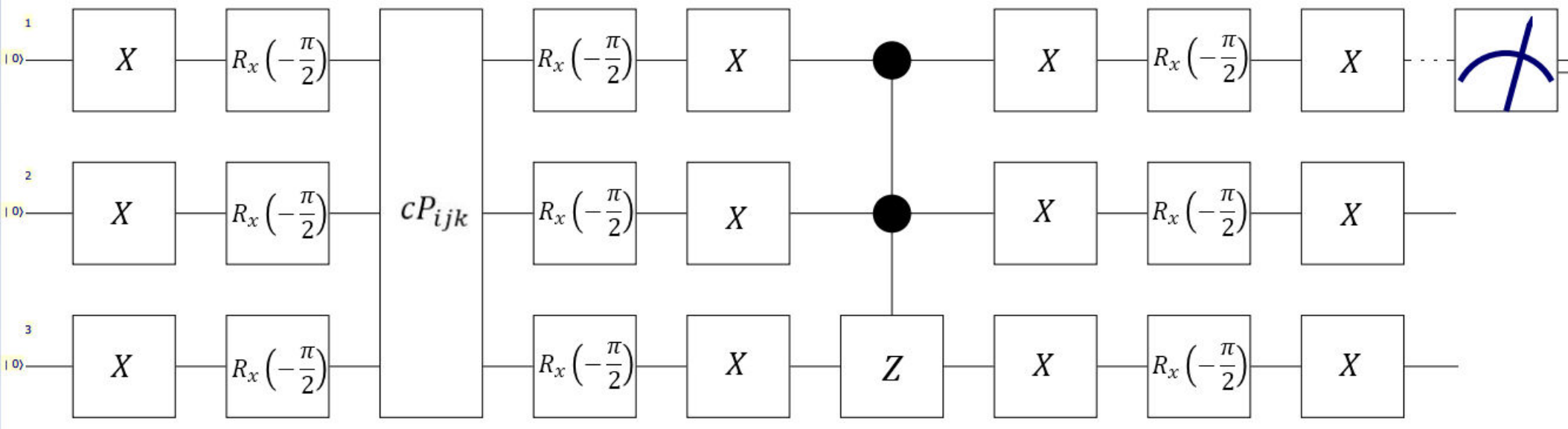}
					\caption{ Schematic circuit of a three-qubit Grover's algorithm\cite{126}. }
					\label{F41}
				\end{figure}

\begin{figure*} 
					\subfloat[\label{b1}]{%
						\includegraphics[width=0.5\columnwidth]{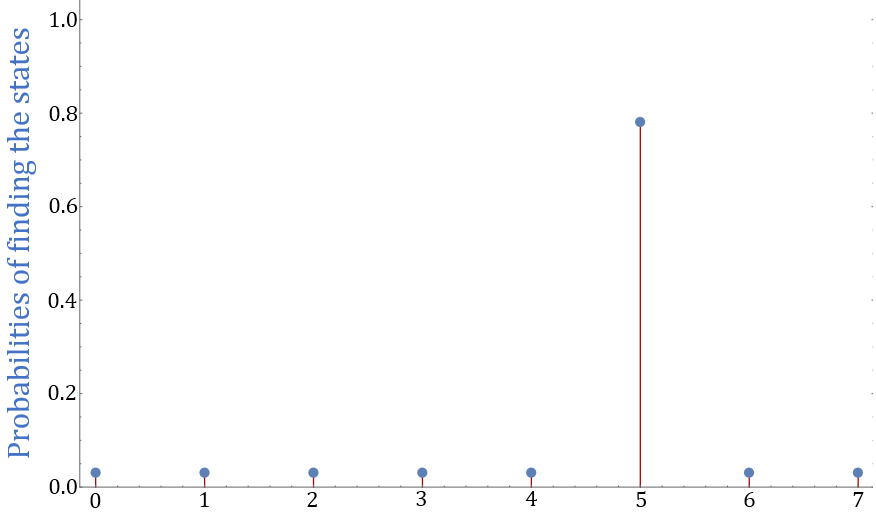}%
					}\hfill
					\subfloat[\label{b2} ]{%
						\includegraphics[width=0.5\columnwidth]{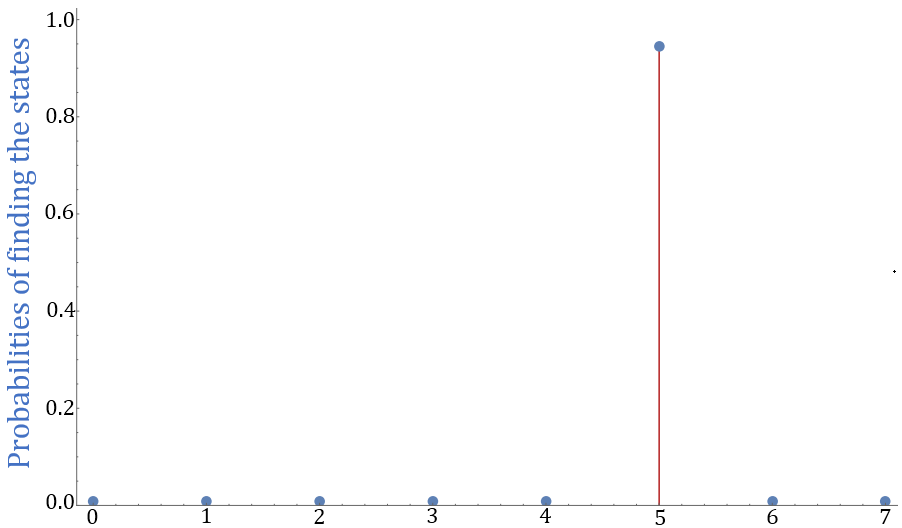}%
					}\hfill
					\caption{ (Color online)  The probability of finding the state after the first (a) and second (b)  Grover’s iteration.}
					\label{F42}
					\end{figure*}

			According to \cite{6}, and also the search algorithm discussed in the  chapter \ref{Ch. 1}, the implementation of Grover’s algorithm in the second step  an oracle in which we can flip the sign of the desired state $\left| x \right\rangle $ is nedeed. This can be expressed as
			\begin{eqnarray}\label{E43}
			\left| x \right\rangle \mathop  \to \limits^O {\left( { - 1} \right)^{f\left( x \right)}}\left| x \right\rangle ,
			\end{eqnarray}
			where $f\left( x \right) = 1$ if $x$ is the desired state and  $f\left( x \right) = 0$ otherwise.  For the three qubits case  the quantum  oracle writes\cite{126}
			\begin{eqnarray}\label{E44}
			O = c{P_{ijk}} = {I_.} - \left| {ijk} \right\rangle \left\langle {ijk} \right|,
			\end{eqnarray}
			where $ijk = \left\{ {000,001,010,011,100,101,110,111} \right\}$. As shown in Fig. \ref{F41}, with three qubits initially prepared in the state $\left| {000} \right\rangle $,  then these can be prepared in a superposition state by applying the $W_3$ gate given by equation (\ref{E42}). This operation is followed by the algorithm iteration and must be applied twice. This includes the phase oracle $O = c{P_{ijk}}$  (Eq.  (\ref{E43})) which allows the inversion of the amplitude of the desired state and the diffusion operator.
				In figure (\ref{F42}a),  it is clear that   the probability of finding the desired state by  applying the first  iteration is ${{25} \mathord{\left/
						{\vphantom {{25} {32}}} \right.
						\kern-\nulldelimiterspace} {32}}$, whereas the probability of finding the desired state for the second iteration which achieves ${{121} \mathord{\left/
						{\vphantom {{121} {128}}} \right.
						\kern-\nulldelimiterspace} {128}}$  as dipected in  Fig. (\ref{F42}b). We note that similar results were obtained for the other states in the two Grover iterations.
				\subsection{  Grover’s Search Algorithm for $n$-qubit }
\begin{figure}[H]
					\centering \includegraphics[scale=0.5]{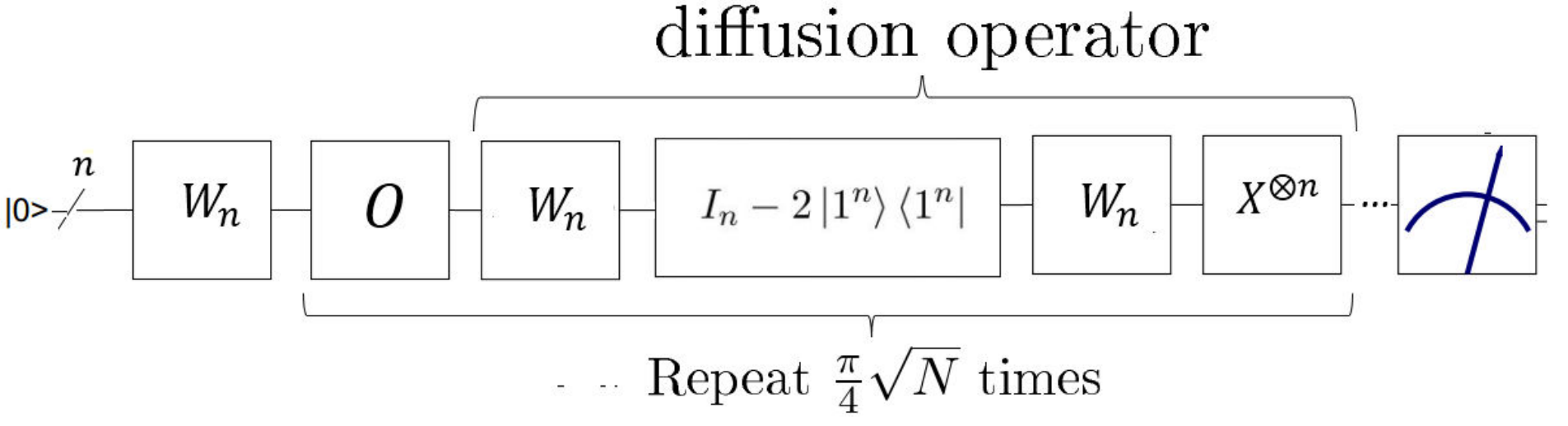}
					\caption{ Schematic circuit   of  Grover's   algorithm\cite{126}. }
					\label{F43}
				\end{figure}

				In the generalized case, searching in an unordered database of $N$  elements (where $n$ is the number of qubits) requires the following steps of Grover's algorithm (see Fig. \ref{F43}):
				
				1.  Starting  with  the register  ${\left| 0 \right\rangle ^{ \otimes n}}$.
				
				2.  Applying  the ${W_n}$ gate,  this operation  can   map an initial state into a superposition  state  
				
				\begin{eqnarray}\label{E45}
				{W_n} = R_x^{ \otimes n}\left( { - \frac{\pi }{2}} \right)\sigma _x^{ \otimes n}.
				\end{eqnarray}
				3.  Grover's  algorithm iteration can be repeated nearly  $\left( {{\pi  \mathord{\left/
							{\vphantom {\pi  4}} \right.
							\kern-\nulldelimiterspace} 4}} \right)\sqrt N $ times:\\
				a). Applying  the quantum  oracle   as discussed  in  chapter  \ref{Ch. 1}.\\ 
				b). Applying  the diffusion operator. 
				\begin{eqnarray}\label{E46}
				{D_n} = \sigma _x^{ \otimes n}R_x^{ \otimes n}\left( { - \frac{\pi }{2}} \right)\sigma _x^{ \otimes n}c{P_{\underbrace {111...1}_n}}\sigma _x^{ \otimes n}R_x^{ \otimes n}\left( { - \frac{\pi }{2}} \right),
				\end{eqnarray}
				with  $c{P_{\underbrace {111...1}_n}} = {I_n} - 2{\left| 1 \right\rangle ^{ \otimes n}}{\left\langle 1 \right|^{ \otimes n}}$. \\
				4. Measurement  of te final  state.
\section{Implementing the $X$-rotation gate in Grover's search algorithm }
				  			In this section,  we implement the Grover algorithm with a transmon-resonator system based on two and three-qubit $X$-rotaion  gates which were realized in the last chapter using the transmon-resonator system and  requiring one-step operations.
				  			\subsection{Grover's algorithm for a two-qubit $X$-rotation gate}
\begin{figure}[H]
	\centering \includegraphics[scale=0.5]{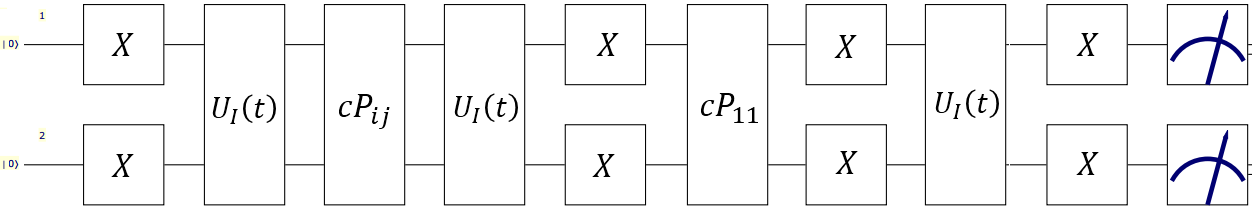}
	\caption{ Schematic circuit   of the two-qubit  Grover  algorithm\cite{126}. }
	\label{F44}
\end{figure}

				  			We propose a simple scheme of two-qubit Grover's algorithm,  shown  in figure (\ref{F44}),  based on the two-qubit $X$-rotation  gate realized in chapter \ref{Ch. 3} (Eq. \ref{E326}) and the two-qubit controlled phase gate.	
				  			\begin{eqnarray}\label{E47}
				  			c{P_{11}} = V_2^{ - 1}{U_{11}}{V_2},
				  			\end{eqnarray}
				  			where ${V_2} = {I_2} \otimes {W_1}$ and  ${U_{11}}$ and ${U_{11}}$ plays a similar role  as a controlled-NOT gate and is given  by  
				  			
				  			\begin{eqnarray}\label{E48}
				  			{U_{11}} = \left( {\begin{array}{*{20}{c}}
				  				1&0&0&0\\
				  				0&1&0&0\\
				  				0&0&0&{ - i}\\
				  				0&0&i&0
				  				\end{array}} \right),
				  			\end{eqnarray}
				  			while in order to implement the other  oracle operators  $c{P_{ij}} = {V_2}{U_{ij}}V_2^{ - 1}$ (with  $ij = \left\{ {01,10,00} \right\}$), we  can  also define other operators   in terms of the following
				  			two-qubit operators  
				  			
				  			\begin{eqnarray}\label{E49}
				  			{U_{01}} = \left( {\begin{array}{*{20}{c}}
				  				1&0&0&0\\
				  				0&1&0&0\\
				  				0&0&0&i\\
				  				0&0&{ - i}&0
				  				\end{array}} \right),{U_{10}} = \left( {\begin{array}{*{20}{c}}
				  				0&{ - i}&0&0\\
				  				i&0&0&0\\
				  				0&0&1&0\\
				  				0&0&0&1
				  				\end{array}} \right),{U_{00}} = \left( {\begin{array}{*{20}{c}}
				  				0&i&0&0\\
				  				{ - i}&0&0&0\\
				  				0&0&1&0\\
				  				0&0&0&1
				  				\end{array}} \right).
				  			\end{eqnarray}
				  			
				  			To carry out the scheme as shown in Fig. \ref{F44}, we consider the two transmon systems that are initially prepared in state $\left| {00} \right\rangle $. Then,  we apply the $\sigma _x^{ \otimes 2}$ and $U\left( {{\pi  \mathord{\left/
				  						{\vphantom {\pi  {2\lambda }}} \right.
				  						\kern-\nulldelimiterspace} {2\lambda }}} \right)$ [Eq.\ref{E326}] operations which prepare the initial states into a superposition state. The obtained state evolves the operations of figure (\ref{F44})  from left to right looking for the final state. We note that in our scheme only the $c{P_{ij}}$  operator can change. 
\subsection*{Fidelity  and discussion} 
\begin{figure}[H]
	\centering \includegraphics[scale=0.5]{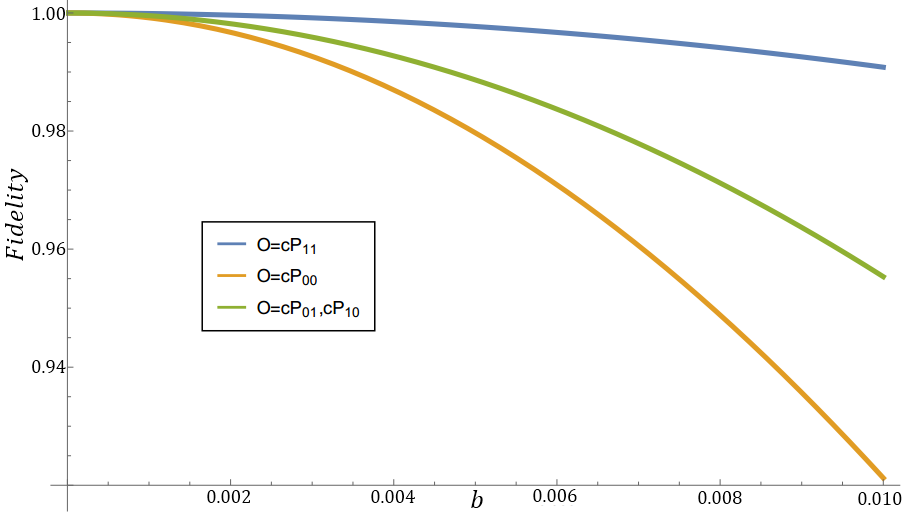}
	\caption{ Numerical results for the fidelities of two-qubit in Grover algorithm for the four possible oracles as the function $b = 2\lambda t$.  These plots   are drawn for ${{{\Omega _R}} \mathord{\left/
				{\vphantom {{{\Omega _R}} \lambda }} \right.
				\kern-\nulldelimiterspace} \lambda } = 13.5$\cite{126}. }
	\label{F45}
\end{figure}

In order to check the validity of our algorithm for different cases of the quantum oracle shown in Fig. (\ref{F44}), it is necessary to compute the fidelity of finding the target state $\left| {00} \right\rangle $ using the equation (\ref{E322}),   where  $\left| \psi  \right\rangle $  represents the target state and $\rho $  is the final density matrix obtained without considering the dissipation of our system. The algorithm fidelities for different possible oracles are plotted in Fig.  (\ref{F45}).  The obtained numerical analysis shows that the algorithm fidelities decrease slightly with increasing values of $b = 2\lambda t$ and are still larger than 90$\%$.  
Now, we discuss the experimental realization of the implementation scheme algorithm in Fig. \ref{F44}  for two transmon-type superconducting qubits capacitively coupled to a resonator using the $R_x^{ \otimes 2}\left( { - {\pi  \mathord{\left/
			{\vphantom {\pi  2}} \right.
			\kern-\nulldelimiterspace} 2}} \right)$  gate. The conditional phase gate of the transmon device has been already experimentally realized  in  Refs.\cite{16}.  The
lifetime of the transmon device was measured in references\cite{128,129}. There,   the decoherence time ${T_1} = 95\mu s$ and the dephasing time ${T_2} = 70\mu s$ were reported. We
assume that  strong qubit-resonator coupling of   ${g \mathord{\left/
		{\vphantom {g {2\pi  = 220}}} \right.
		\kern-\nulldelimiterspace} {2\pi  = 220}}$ MHz, which   has been   experimentally realized\cite{48}.  By assuming that  ${\Delta _r} = 2g$, a direct calculation depicts that the time to perform the three ${U_I}\left( t \right)$ gates is $0.013\mu s$. So, the time of scheme algorithm implementation in Fig.\ref{F44} is much shorter than the coherence time of the transmon system(${T_1}$ and ${T_2}$). Furthermore, we assume that  the resonator is driven by a classical 
microwave field with the frequency of Rabi ${{{\Omega _R}} \mathord{\left/
		{\vphantom {{{\Omega _R}} {2\pi  = 742.5}}} \right.
		\kern-\nulldelimiterspace} {2\pi  = 742.5}}$ MHz and should be  adjusted to meet the following condition ${{{\Omega _R}} \mathord{\left/
		{\vphantom {{{\Omega _R}} {\lambda  = {{\left( {2n + 1} \right)} \mathord{\left/
							{\vphantom {{\left( {2n + 1} \right)} 2}} \right.
							\kern-\nulldelimiterspace} 2}}}} \right.
		\kern-\nulldelimiterspace} {\lambda  = {{\left( {2n + 1} \right)} \mathord{\left/
				{\vphantom {{\left( {2n + 1} \right)} 2}} \right.
				\kern-\nulldelimiterspace} 2}}}$ mentioned above.
	\subsection{Grover's algorithm for three-qubit $X$-rotation gate}
	We here propose a simple scheme for implementing the three-qubit Grover's algorithm as shown in figure (\ref{F45}), based on the three-qubit $X$-rotation gate and phase gate. In the computational three-qubit basis, the phase oracle was  implemented by using   the Toffoli gate and the single-qubit X-rotation gate\cite{31}
	
	\begin{eqnarray}\label{E410}
	c{P_{111}}
	&=& {V_3}{U_3}V_3^{ - 1},
	\end{eqnarray}
	where ${U_3}$ plays a similar role to the quantum Toffoli gate, which was  achieved using a direct  implementation  for a superconducting transmon device in Ref. \cite{87},  and is given by 
	\begin{eqnarray}\label{E411}
	{U_3} = \left( {\begin{array}{*{20}{c}}
		1&0&0&0&0&0&0&0\\
		0&1&0&0&0&0&0&0\\
		0&0&1&0&0&0&0&0\\
		0&0&0&1&0&0&0&0\\
		0&0&0&0&1&0&0&0\\
		0&0&0&0&0&1&0&0\\
		0&0&0&0&0&0&0&{ - i}\\
		0&0&0&0&0&0&i&0
		\end{array}} \right).
	\end{eqnarray}
	and            ${V_3} = \left( {\begin{array}{*{20}{c}}
		1&0\\
		0&1
		\end{array}} \right) \otimes \left( {\begin{array}{*{20}{c}}
		1&0\\
		0&1
		\end{array}} \right) \otimes {W_1}$.
	\begin{figure}[H]
		\centering \includegraphics[scale=0.5]{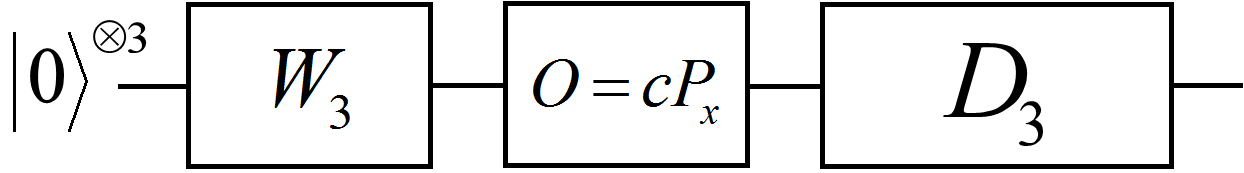}
		\caption{Schematic circuit   of the three-qubit  Grover  algorithm\cite{sakhouf}. }
		\label{F46}
	\end{figure} 
	According to reference \cite{39}, we can construct the seven
	other  quantum oracles  (see Eq. (\ref{E129}) in the  chapter \ref{Ch. 1}) as
	follows 
	\begin{eqnarray}\label{c20}
	c{P_{000}}& =& {\sigma _{x,3}}{\sigma _{x,2}}{\sigma _{x,1}}c{P_{111}}{\sigma _{x,1}}{\sigma _{x,2}}{\sigma _{x,3}},\nonumber\\
	c{P_{001}} &=& {\sigma_{x,2}}{\sigma _{x,1}}
	c{P_{111}}{\sigma _{x,1}}{\sigma _{x,2}},\nonumber\\
	c{P_{010}} &=& {\sigma _{x,3}}{\sigma _{x,1}}c{P_{111}}{\sigma _{x,1}}{\sigma _{x,3}},\nonumber\\
	c{P_{011}} &=&{\sigma _{x,1}}c{P_{111}}{\sigma _{x,1}},\nonumber\\
	c{P_{100}} &=& {\sigma _{x,3}}{\sigma _{x,2}}c{P_{111}}{\sigma _{x,2}}{\sigma _{x,3}},\nonumber\\
	c{P_{101}} &= &{\sigma _{x,2}}c{P_{111}}{\sigma _{x,2}},\nonumber\\
	c{P_{110}}& =& {\sigma _{x,3}}c{P_{111}}{\sigma _{x,3}}.
	\end{eqnarray}
	
	The quantum circuit of  Grover's algorithm shown in  Fig. (\ref{F46}) is read from left to right, and the three transmon systems are initially prepared in the state $\left| {000} \right\rangle $. We then apply the quantum gate ${W_3} = i{U_I}\left( {{t_R}} \right)$ [Eq.  \ref{E334}] to prepare these qubits in an equal superposition state.  This operation is followed by an oracle $c{P_x}$, which
	ﬂips the sign of the correct state’s amplitude. The diffusion
	transform ${D_3}$ is given  by
	\begin{eqnarray}\label{E413}
	{D_3} = -{U_I}\left( {{t_R}} \right)c{P_{111}}{U_I}\left( {{t_R}} \right)\sigma _x^{ \otimes 3}.
	\end{eqnarray}
	
	It is worth noting that to implement a three-qubit
	Grover’s search algorithm by using a decomposition into sequences of quantum operations requires nearly more than fifteen one and three-qubit gates\cite{1,31}. This makes the quantum algorithm complex, and consequently, the gate operation time is quite long and the coherence time of the system would get destroyed by the environmental noise. However, our proposed scheme for implementing Grover’s algorithm requires only the single-shot  X-rotation gate as well as a three-qubit phase gate. Consequently, it is faster compared to one containing the decomposition of single-qubit gates.
	\subsection*{Fidelity  and discussion}
\begin{figure}[H]
		\begin{center}
			\subfloat[\label{a} $O=c{P_{000}}$.]{%
				\includegraphics[width=0.5\columnwidth]{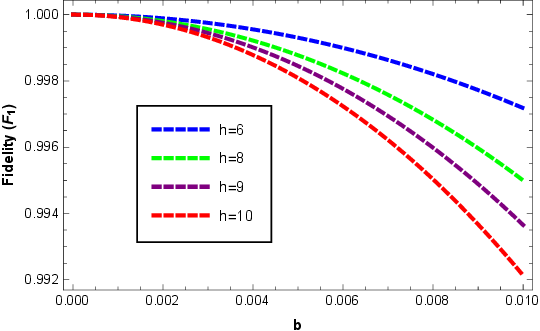}%
			}\hfill
			\subfloat[\label{b} $O=c{P_{001}}$,$c{P_{010}}$,$c{P_{101}}$.]{%
				\includegraphics[width=0.5\columnwidth]{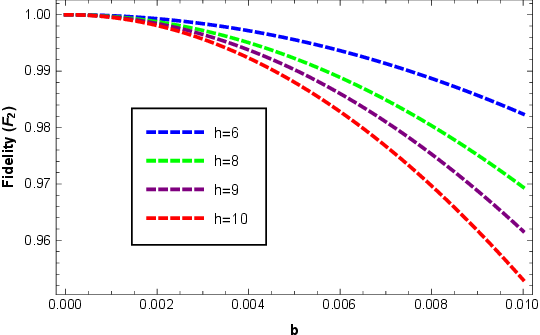}%
			}\hfill
			\subfloat[\label{c} $O=c{P_{011}}$,$c{P_{100}}$,$c{P_{110}}$.]{%
				\includegraphics[width=0.5\columnwidth]{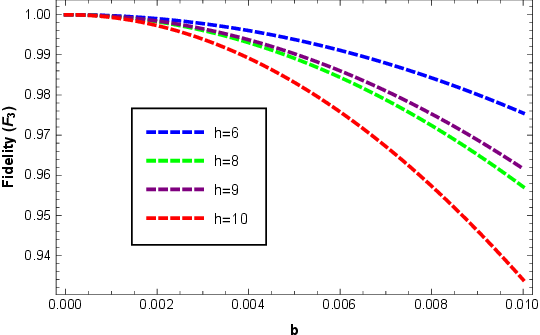}%
			}\hfill
			\subfloat[\label{d} $O=c{P_{111}}$.]{%
				\includegraphics[width=0.5\columnwidth]{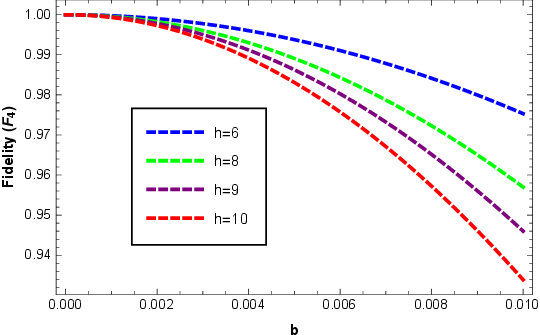}%
			}
			\caption{Numerical results of the algorithm  fidelity for the desired  state  $\left| {{1}{1}{1}} \right\rangle $   versus $b = 2\lambda t$ for different values of  $h = {{{\Omega _R}} \mathord{\left/
						{\vphantom {{{\Omega _R}} \lambda }} \right.
						\kern-\nulldelimiterspace} \lambda }$. Other parameters of the system  are referred to in the text. (a) Fidelity ${F_1}$  for $O=c{P_{000}}$. (b) Fidelity ${F_2}$ for  $O=c{P_{001}}$,$c{P_{010}}$,$c{P_{101}}$.(c) Fidelity ${F_3}$   for $c{P_{011}}$,$c{P_{100}}$,$c{P_{110}}$. (d) Fidelity ${F_4}$   for $O=c{P_{111}}$. }
			\label{F47}
		\end{center}
		
	\end{figure}

	In order to check  the algorithm  scheme as shown  in Fig. (\ref{F46}), we
	estimate  the algorithm  fidelity  by using equation (\ref{E322}), at which $\left| \psi  \right\rangle $ and $\rho $  are the target state and the fnal density matrix, respectively,  without  taking into  account system dissipation. We assume that  $\left| {111} \right\rangle $ the target state, the fidelity of the whole scheme  shown in Fig \ref{F46} for  all  possible oracles [Eqs. (\ref{E49}) and (\ref{E411})] are plotted in figure (\ref{F47}).
	From figure (\ref{F47}), it is clear that our algorithm achieved a high fidelity which decreases slightly while increasing the values of b and increasing values of h. The algorithm fidelity
	for the ratio ($h \le 10$) is greater than 93$\%$.
	Remarkably, from figures \ref{F47}(b) and (c) one can see  that the
	fidelities are nearly unchanged. For the ratio $h = 9$, we get  ${{{\Omega _R}} \mathord{\left/
			{\vphantom {{{\Omega _R}} {2\pi  \sim 200}}} \right.
			\kern-\nulldelimiterspace} {2\pi  \sim 200}}$ MHz, ${g \mathord{\left/
			{\vphantom {g {2\pi  \sim 60}}} \right.
			\kern-\nulldelimiterspace} {2\pi  \sim 60}}$ MHz and ${{{\Delta _r}} \mathord{\left/
			{\vphantom {{{\Delta _r}} {2\pi  \sim 80}}} \right.
			\kern-\nulldelimiterspace} {2\pi  \sim 80}}$ MHz , which were experimentally achieved \cite{117}. These results indicate  that the quantum  circuit 
	consisting of  the direct implementation  of the three-qubit  gate is efficient  than those using  a combination of many one and two-qubit gates. This is  caused  by  the gate  time which is quite short compared to  the lifetimes of the transmon  system.
	
		We brieﬂy discuss some additional though regarding the
		experimental feasibility of   our proposed scheme based on
		our chosen values for the cohrences  time  of transmon systems, the resonator  decay time and the Rabi frequency. The  interaction time of the three-qubit  $X$-rotation gate is ${t_R} = 0.62$ ns. The proposed  scheme in figure (\ref{F43}) includes  three gates  of type ${U_I}\left( {{t_R}} \right)$ and two phase gates. Since we can cancel the time of the one-qubit X-rotation gate compared to the Toffoli gate, the time required for the phase gate is the same as that of the Toffoli gate. This result was reported in Ref. [87] and measured to be $t_g = 26$ ns. In the following,   the implementation
		time of Grover’s search algorithm  for two iterations corresponds to $\sim 100$ ns, which  is shorter than the
		coherences time  of transmon systems  (${T_1}$ and ${T_2}$) and the resonator decay time used in our numerical simulations. Therefore, the experimental approach of our  proposed scheme would be an important step toward  quantum computing which is very complicated  with the circuits QED.
			\section{Characterization of the single-shot entangling gate by quantum process tomography }
		As an essential additional  characterization  to the results of the two- and three-qubit  entangling gate in the last  chapter, we here report a complete
		characterization of the entangling gate for two and three qubits using  quantum process tomography (QPT). 
		\subsection{ Quantum process tomography (QPT) }
		QPT is the method used to characterize a quantum  logic  gate,  or quantum  algorithm fully \cite{1}, it  takes an arbitrarily given input state  and transforms it into the output state ${\rho _{out}} = \varepsilon \left( {{\rho _{in}}} \right)$.   Reconstructing a completely positive linear map  $\varepsilon $ representing the process that acts on an arbitrary input state $\rho $,  is the idea behind QPT.  Any quantum process for an $n$-qubit system   can be expressed in terms of the Kraus operators as\cite{141,142}
		\begin{eqnarray}\label{E414}
		{\rho _{out}} = \varepsilon \left( {{\rho _{in}}} \right) = \sum\limits_j^{{4^n}} {{{\tilde {\rm A}}_j}{\rho _{in}}\tilde {\rm A}_j^ + } ,
		\end{eqnarray}
		where  ${{{\rm A}_j}}$ are the   "Kraus operators" of the process $\varepsilon $, which  is satisfied by the condition $\sum\limits_j^{{4^n}} {{{\tilde {\rm A}}_j}\tilde {\rm A}_j^ + }  = 1$. If we determine  these operators  on  an operator basis ${{\rm A}_a}$ such  that  ${{\rm A}_i} = \sum\limits_a^{{4^n}} {{\chi _{i,a}}{{\rm A}_a}} $,  then the equation  (\ref{E414})  can  be  rewritten  as   
		\begin{eqnarray}\label{	E415}
		{\rho _{out}} = \varepsilon \left( {{\rho _{in}}} \right) &= &\sum\limits_i^{{4^n}} {\sum\limits_a^{{4^n}} {{\chi _{i,a}}{{\rm A}_a}{\rho _{in}}\sum\limits_b^{{4^n}} {{\chi ^*}_{i,b}{{\rm A}^ + }_b} } }  \nonumber \\
		&=& \sum\limits_{a,b}^{{4^n}} {{{\rm A}_a}{\rho _{in}}{{\rm A}^ + }_b\sum\limits_i^{{4^n}} {{\chi _{i,a}}{\chi ^*}_{i,b}} } \nonumber \\
		& = &\sum\limits_{a,b}^{{4^n}} {{\chi _{a,b}}{{\rm A}_a}{\rho _{in}}{{\rm A}^ + }_b} ,
		\end{eqnarray} 
		where ${\chi _{a,b}} = \sum\limits_i^{{4^n}} {{\chi _{i,b}}{\chi ^*}_{i,b}} $. Whereas equation (\ref{	E415}) is the $\chi $-matrix representation of
		the  process in the basis {${{\rm A}_a}$},  containing all the information about the quantum  process.
		\subsection{ Two-qubit entangling gate characterization }
		\subsection*{ $\chi $-matrix process} 
		Starting  from the matrix  in equation  (\ref{E335}) describing the single-shot
		entangling  gate for two  qubits,   we  can show
		that at the interaction  time,  the matrix  in equation  (\ref{E335}) becomes
		\begin{eqnarray}\label{E416}
		{U_I}\left( {\frac{\pi }{{4\lambda }}} \right) = \frac{{{e^{ - i\frac{\pi }{4}}}}}{{\sqrt 2 }}\left( {\begin{array}{*{20}{c}}
			1&0&0&{ - i}\\
			0&1&{ - i}&0\\
			0&{ - i}&1&0\\
			{ - i}&0&0&1
			\end{array}} \right),
		\end{eqnarray} 
		where the overall phase ${{e^{ - i\frac{\pi }{4}}}}$ can be omitted. On the Pauli basis, the evolution operator ${U_I}\left( {{\pi  \mathord{\left/
					{\vphantom {\pi  {4\lambda }}} \right.
					\kern-\nulldelimiterspace} {4\lambda }}} \right)$  can  be decomposed as  ${U_I}\left( {{\pi  \mathord{\left/
					{\vphantom {\pi  {4\lambda }}} \right.
					\kern-\nulldelimiterspace} {4\lambda }}} \right) = \left( {{1 \mathord{\left/
					{\vphantom {1 {\sqrt 2 }}} \right.
					\kern-\nulldelimiterspace} {\sqrt 2 }}} \right)\left( {I \otimes I - iX \otimes X} \right)$.  leading to  readily obtaining  the nonzero elements that  correspond to the ${\chi _{ideal}}$ process matrix for the
		perfect ${U_I}\left( {{\pi  \mathord{\left/
					{\vphantom {\pi  {4\lambda }}} \right.
					\kern-\nulldelimiterspace} {4\lambda }}} \right)$  entangling  gate, which are
		\begin{eqnarray}\label{c21}
		\left\{ \begin{array}{l}
		\chi _{ideal}^{II,II} = \chi _{ideal}^{XX,XX} = {1 \mathord{\left/
				{\vphantom {1 2}} \right.
				\kern-\nulldelimiterspace} 2}\\
		\chi _{ideal}^{II,XX} =  - \chi _{ideal}^{XX,II} = {i \mathord{\left/
				{\vphantom {i 2}} \right.
				\kern-\nulldelimiterspace} 2}
		\end{array} \right..
		\end{eqnarray}
		\begin{figure*}
			\subfloat[\label{b1}]{%
				\includegraphics[width=0.5\columnwidth]{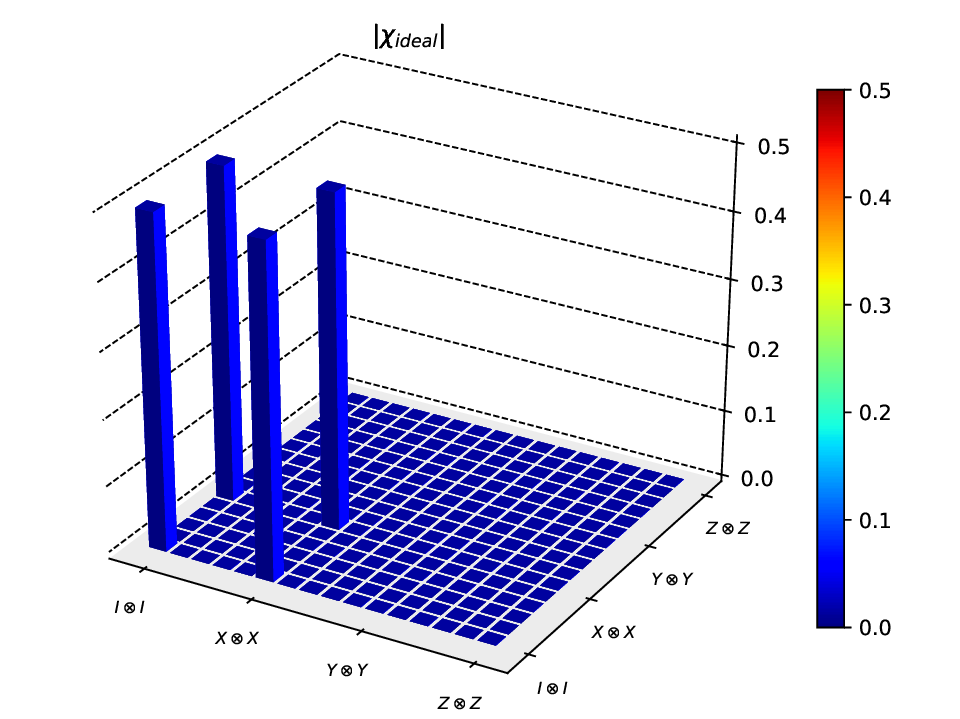}%
			}\hfill
			\subfloat[\label{b2} ]{%
				\includegraphics[width=0.5\columnwidth]{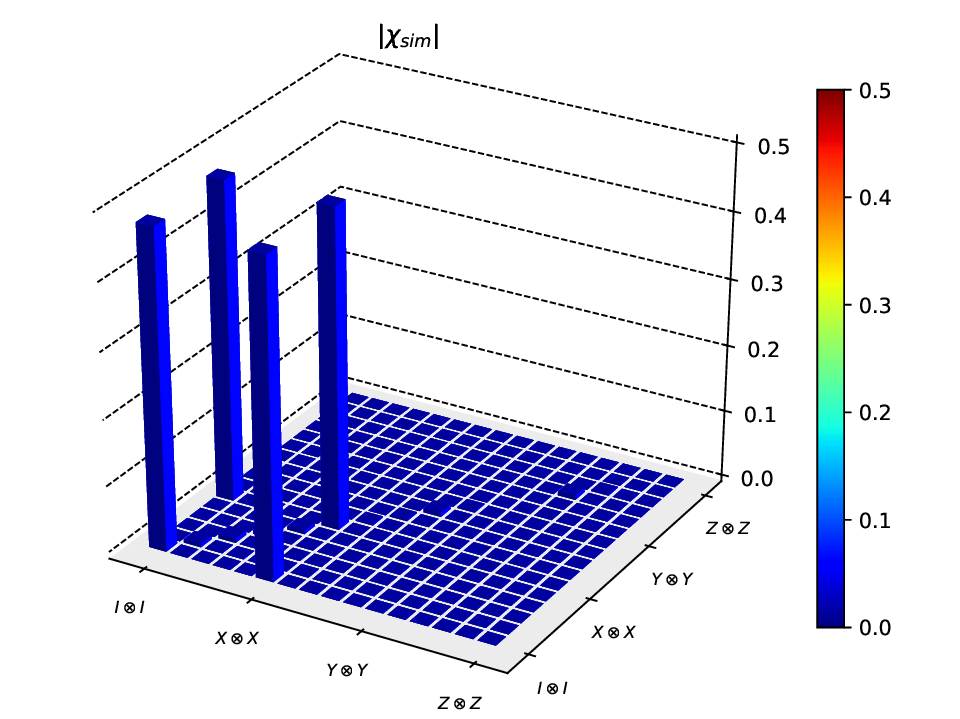}%
			}\hfill
			\caption{ $QPT$ of the single-shot two-qubit entangling gate. Absolute values of the matrix  elements of the  ${\chi _{ideal}}$ (a) and the  ${\chi _{sim}}$ (b). The matrix elements are shown  in the operator basis $\left\{ {I \otimes I,I \otimes X,I \otimes Y,...,Z \otimes X,Z \otimes Y,Z \otimes Z} \right\}$, where $I$ is identity matrix and $\left\{ {X,Y,Z} \right\}$ are the Pauli operators $\left\{ {{\sigma _x}, - i{\sigma _y},{\sigma _z}} \right\}$.}
			\label{F48}

		\end{figure*} 
		
		We note that this process matrix has only four nonzero elements on the Pauli basis, which is simpler compared to the process matrix of the  $CNOT$ or  $iSWAP$ gates, which have sixteen non-null elements on the same basis\cite{62,75}. As a comparison, we show the absolute values
		of the ideal and simulated $\chi$-process matrices in figure (\ref{F48}). Although, we can see some disagreement between the ${\chi _{ideal}}$ [Fig. (\ref{F48}a)] and ${\chi _{sim}}$[Fig. (\ref{F48}b)]   owing to the coupling of transmon systems to their environment.  Still,  the ${\chi _{sim}}$ process matrix for the entangling gate [Eq. (\ref{E416})] is in good agreement with the ${\chi _{ideal}}$. Similar results have been demonstrated with trapped ions\cite{80}.
		\subsection*{Process fidelity} 
		
		To quantitatively validate the performance of the two-qubit entangling gate [Eq. \ref{	E415}] it is necessary to calculate the process fidelity or mean fidelity using the equation (\ref{E322}),  with $\left| \psi  \right\rangle $ as the output state obtained from the perfect unitary map and $\rho  = \varepsilon \left[ {{\rho _{in}}} \right]$ of our matrix process which is obtained using numerical simulation. The process  fidelity corresponds to the results shown in figure (\ref{F48}) for $T=20$$\mu s$ is high 99.9$\%$ fidelity. 
		
		\subsection{Three-qubit entangling gate characterization }
		\subsection*{ $\chi $-matrix process} 
		Starting  from the evolution operator  in equation  (\ref{E336}) describing the single-shot entangling  gate for three  qubits,   one can show that at the operation   time $T$,  the matrix  in equation  (\ref{E336}) becomes
		\begin{eqnarray}\label{E418}
		M\left( {T} \right) = \frac{{{e^{ - i\theta }}}}{{\sqrt 2 }}\left( {\begin{array}{*{20}{c}}
			1&0&0&0&0&0&0&{ - i}\\
			0&1&0&0&0&0&{ - i}&0\\
			0&0&1&0&0&{ - i}&0&0\\
			0&0&0&1&{ - i}&0&0&0\\
			0&0&0&{ - i}&1&0&0&0\\
			0&0&{ - i}&0&0&1&0&0\\
			0&{ - i}&0&0&0&0&1&0\\
			{ - i}&0&0&0&0&0&0&1
			\end{array}} \right)
		\end{eqnarray}
		where the factor  $\frac{{{e^{ - i\theta }}}}{{\sqrt 2 }}$ denotes the overall phase which can be omitted. This operator in the Pauli  basis yields to  ${U_I}\left( {{T}} \right) = \left( {{1 \mathord{\left/
					{\vphantom {1 {\sqrt 2 }}} \right.
					\kern-\nulldelimiterspace} {\sqrt 2 }}} \right)\left( {I \otimes I \otimes I - iX \otimes X \otimes X} \right)$. So,  therefore,  by plugging this expression in Eq. (\ref{E414}) for three qubits systems, we readily obtain  the nonzero  elements which correspond to the ${\chi _{ideal}}$  matrix for the ideal entangling  gate, which are
		\begin{eqnarray}\label{E419}
		\left\{ \begin{array}{l}
		\chi _{ideal}^{III,III} =  \chi _{ideal}^{XXX,XXX} = {1 \mathord{\left/
				{\vphantom {1 2}} \right.
				\kern-\nulldelimiterspace} 2}\\
		\\
		\chi _{ideal}^{III,XXX} =  - \chi _{ideal}^{XXX,III} = {{ i} \mathord{\left/
				{\vphantom {{ - i} 2}} \right.
				\kern-\nulldelimiterspace} 2}
		\end{array} \right..
		\end{eqnarray}
		
\begin{figure*} 
			\subfloat[\label{b1}]{%
				\includegraphics[width=0.5\columnwidth]{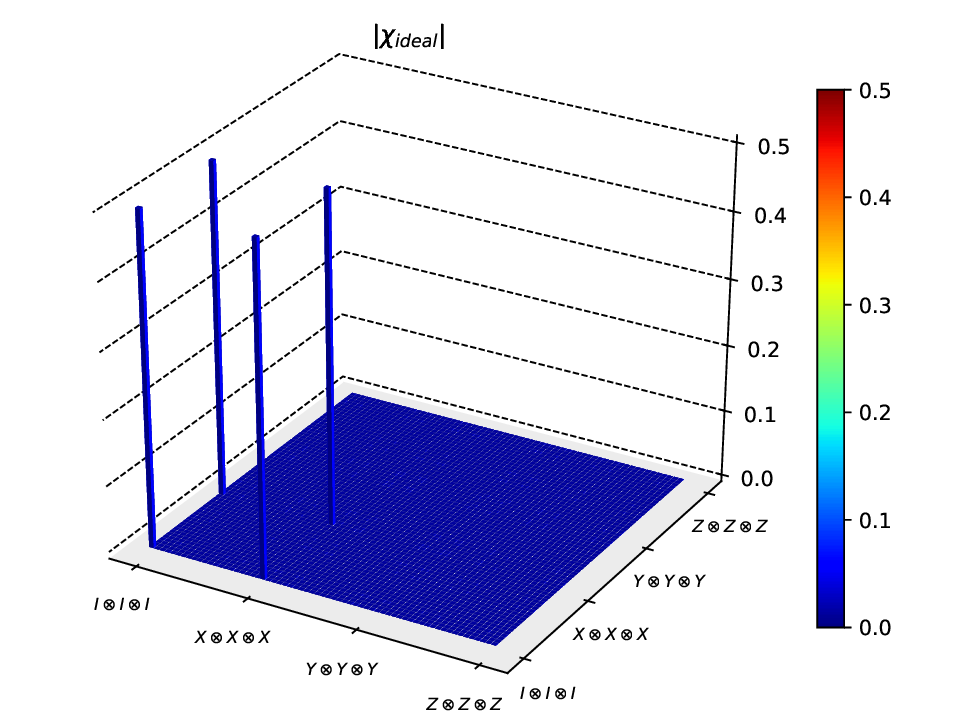}%
			}\hfill
			\subfloat[\label{b2} ]{%
				\includegraphics[width=0.5\columnwidth]{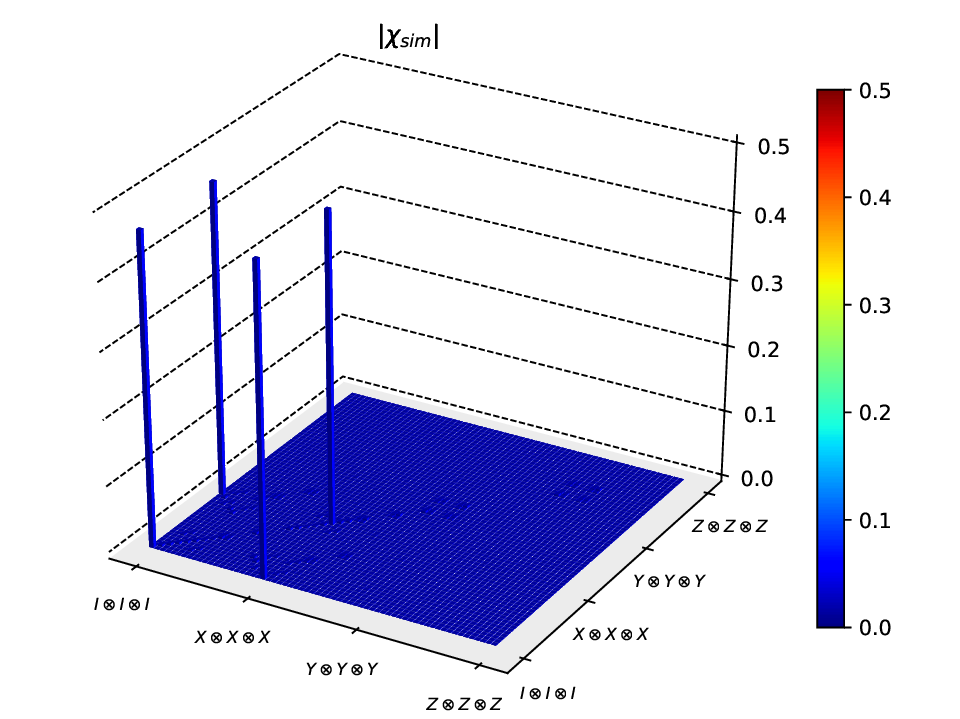}%
			}\hfill
			\caption{ $QPT$ of the single-shot  entangling gate for three qubits. Absolute values of the matrix  elements of the  ${\chi _{ideal}}$ (a) and the  ${\chi _{sim}}$ (b). The matrix elements are shown  in the operator basis $\left\{ {I \otimes I \otimes I,I \otimes I \otimes X,...,Z \otimes Z \otimes Y,Z \otimes Z \otimes Z} \right\}$, where $I$ is identity matrix and $\left\{ {X,Y,Z} \right\}$ are the Pauli operators $\left\{ {{\sigma _x}, - i{\sigma _y},{\sigma _z}} \right\}$.}
			\label{F49}

		\end{figure*}

		This operation matrix contains only four non-zero matrix elements, which  is similar  to a two-qubit process case in the last subsection and considerably efficient and faster in terms of simplicity  compared to  the process analyzed of  GHZ-protocol generated  from the decomposition  of  one- and two-qubit operations with  nearly  sixty-four non-null elements.  By comparing the simulated ${\chi _{sim}}$ with the ideal ${\chi _{ideal}}$ matrix process, we plot the absolute values corresponding to these metrics which are depicted in figure (\ref{F49}). It is clear that the ${\chi _{sim}}$ process matrix of the entangling gate for three qubits  [Fig. (\ref{F49}a)] is in good agreement with the ${\chi _{ideal}}$ matrix process [Fig. (\ref{F49}b)] with appearing  some disagreement which is caused by  the coupling of transmon  qubits to their environment.
		\subsection*{ Process fidelity} 
\begin{figure}[H]
			\begin{center}
				\includegraphics[width=0.5 \columnwidth]{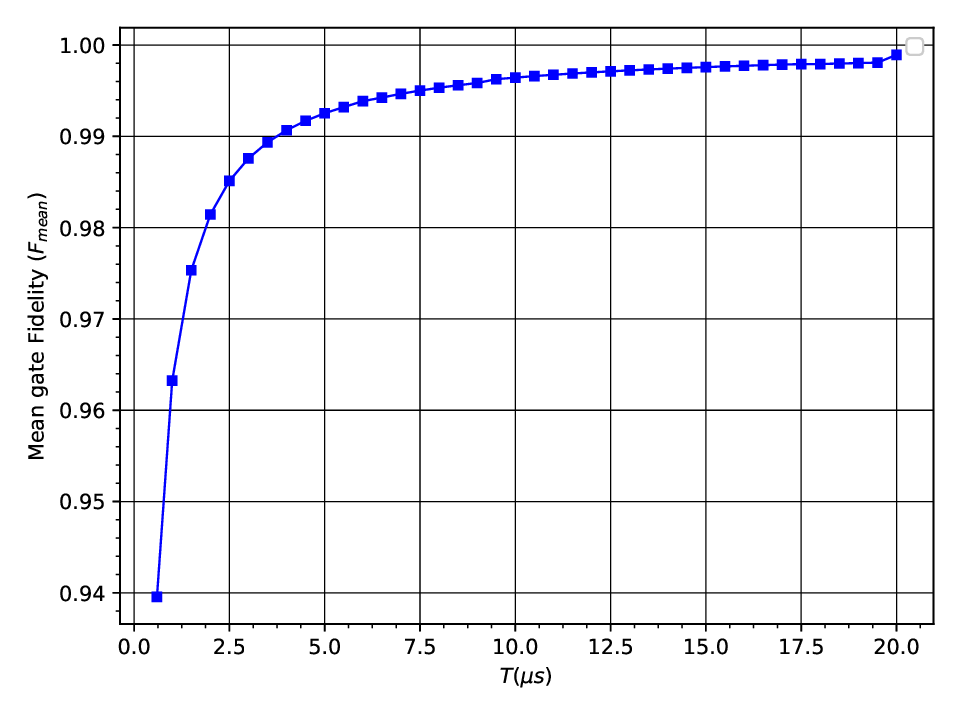}
				
				\caption{Performance of the single-shot entangling gate for three systems.  Numerical simulation of the gate fidelity $F_{mean}$ versus the coherence times $T \ge 0.6\mu s$. Assuming that  $T=T_1=T_2$, while  the other system parameters are the same as in figure \ref{F36}.}
				\label{F410}
			\end{center}
			
		\end{figure}

		In order to estimate the performance of the three-qubit entangling gate $M\left( {{T_m}} \right)$ in equation (\ref{E418}), we calculate the mean gate fidelity using the same as in the last subsection. The numerical simulations of the process fidelity versus the coherence time $T$ are plotted in Fig. (\ref{F410}),  we obtain the fidelity ${F_{mean}} = 93\%$ which corresponds to the results shown in Fig. (\ref{F49}) for  $T=0.6 \mu s$ and can be improved as the coherence time increases.
		\section{Chapter summary } 
		
		In summary, we have generally discussed Grover's search algorithm using the $X$-rotation gate instead Hadamard gate, this algorithm still meets the requirements of the full Grover algorithm in Ref. \cite{39} as we have demonstrated by the numerical simulations. Using  $X$-rotation gate realized with a transmon-resonator system and requiring only one-step operations as well as the phase gate,  we have proposed a practical implementation scheme of a two- and three-qubit Grover’s algorithm with an operation time significantly shorter than the coherence time of the used superconducting qubits and resonator decay time.  The advantage of this scheme is that it  does not need an ancillary qubit and  the series of single-qubit gates is also not required.  In addition, we have numerically  demonstrated the complete characterization of the entangling gate for two  and three qubits using a QPT and showed that with higher fidelity.

	\chapter{General Conclusion and Future Directions}
Quantum information processing and quantum computation  have been a promising and very active 
area of research, especially in the few years, in quantum science and technology, since the physicist Richard Feynman
proposed the idea of simulating quantum systems to build a quantum computer\cite{3}. Recent progress in circuit QED, in which superconducting qubits play the role of atoms in cavity QED,
makes it one of the most promising candidates for the realization of quantum information processing (QIP).   While to this day,  quantum computing with superconducting circuits has been experimentally achieved based on basic single and two-qubit gates, for instance,  two-qubit quantum algorithms\cite{16,113},  generation and detection of three and multi-qubit entanglement\cite{48,49,50}, the implementation of a Toffoli gate for three qubits\cite{63}, and realization of three- and multi-qubit quantum error correction\cite{26,27,28}. Experimentally, these kinds of realization and implementation approaches led to lower fidelities and required longer execution time due to they are mainly limited by the coherence time of the quantum system, although these fidelities are good to demonstrate simple quantum operations often with up to three qubits, they are too low to allow experiments with more than three qubits to be carried out.

To reduce the excitation time and experimental complications, an efficient approach that will be an alternative is a direct implementation which is an interesting and important topic of a multiqubit gate and algorithms.  In this context,  the implementation of a Three-qubit  Toffoli gate using a one-step approach is achieved with a high fidelity with up 99$\%$\cite{87,88,90}.  In this thesis, we have focused on different aspects of quantum computation and, more particularly, realizing quantum gates and algorithms with transmon-type superconducting qubits in open quantum systems using an efficient and alternative approach which  is   one-step operation.  

Before going into the detailed discussion of realizing quantum gates and algorithms with our transmon-resonator system,  we have dedicated chapter \ref{Ch. 1} of this dissertation to give the basic concepts of quantum information theory.  In  Chapter \ref{Ch. 2},    we reviewed some superconducting qubits, especially giving a detailed description of the transmon qubit used in this work. We have discussed the basis of the coupling between the transmission line or  LC oscillator with transmon-type superconducting qubits in circuit  QED.  Such circuit QED was  associated with a number of key concepts from cavity QED, including some regimes such as the dispersive coupling regime, which is useful for QIP.  Furthermore,  we also  have highlighted the realization of  single-qubit gates  as 
well  as  universal quantum gates for two qubits in the QED   circuit needed to build algorithms and create  maximally entangled states such as Bell and GHZ states.

In chapter \ref{Ch. 3} of this thesis, we have   derived  the time evolution operator describing   our  quantum  device which involves multi-transmon-type superconducting qubits coupled to one  resonator driven by a
microwave feld. Then, using it, we have suggested a simple and more efficient scheme to realize  $X$-rotation and  entangling gates for two and three qubits  using a one-step approach.  We have demonstrated  the robustness of these schemes  by simulating    the system  dynamics in a realistic situation   using  the master  equation for the transmon-resonator system,  taking  into account  the effect of the decoherence of the transmon  qubits as well  as  the resonator decay and   a high fidelity is achieved  with up to  $99\%$. Finally,  these schemes have the following advantages:  (i) do not  require the combination  of the basic  gates, for instance,  single- and two-qubit gates, (ii)  the resonator   does not require to be initially prepared   in the vacuum state, (iii) our scheme is immune to cavity decay, which  makes practical experiments easier.  Therefore, we have exploited these schemes  to  implement  Grover's algorithm and we  also  have  fully  characterized the single-shot  entangling  gates using  QPT in Chapter \ref{Ch. 4}. 

As an essential goal in the future,  we will try to implement a high-fidelity universal and multi-qubit gate and also create multiqubit GHZ  state based on quantum devices, not only for  ones consisting of superconducting qubits  coupled to one resonator\cite{16,26,87,88,90,113},  but for many multi-level systems described as distributed in many cavity systems\cite{143,144},  using the presented approach in this thesis.   In addition, this perspective also goes to quantum information processing and its implementation with other physical systems such as superconducting circuits, ion traps, and quantum dots,  in cases where the information is encoded in the quantum states of atoms in cavity QED, etc. 

Another essential building block of quantum computers is the realization of quantum error correction. In this context,  many experiments have been made by realizing schemes of the most basic quantum error correction\cite{1,22,23,24,25}. One of the most promising fault-tolerant quantum error correction schemes was done using superconducting qubits\cite{145,146}, which are based on two-dimensional surface codes.  Recently,  quantum error detection code has been experimentally demonstrated based on a physical device consisting of a grid of superconducting coplanar waveguide resonators and four transmon qubits coupled to each of them\cite{26,27,28}.   The working principle of surface codes would be an interesting step forward for trying to demonstrate simple realizations of such transmon-resonator grid structures.
	
	Finally, we hope that the analysis of the proposed approaches followed throughout this dissertation and their validation using strictly numerical simulations have proven helpful to stimulate experimental activities in the near future in quantum science and technology.

\newgeometry{left=1.5cm,right=1.5cm,lines=60,top=1in}
\setstretch{0.9}
\AddToShipoutPicture*{
	\unitlength=0.5cm
	\put(2,5){
		\parbox[b][\paperheight]{\paperwidth}{%
			\vfill
			\centering
			{\transparent{0.5}\includegraphics[width=1.4\textwidth]{logo.jpg}}
		}
	}
}
\vspace{2cm}
\begin{tikzpicture}[overlay,remember picture]
	\draw [line width=2pt,rounded corners=7pt]
	($ (current page.north west) + (.2cm,-.2cm) $)
	rectangle
	($ (current page.south east) + (-.2cm,.2cm) $);
	\draw [line width=1pt,rounded corners=7pt]
	($ (current page.north west) + (.3cm,-.3cm) $)
	rectangle
	($ (current page.south east) + (-.3cm,.3cm) $);
\end{tikzpicture}
	\vspace{-.5cm}
\begin{figure}[H]
	\begin{center}
		\includegraphics[width=0.5\linewidth]{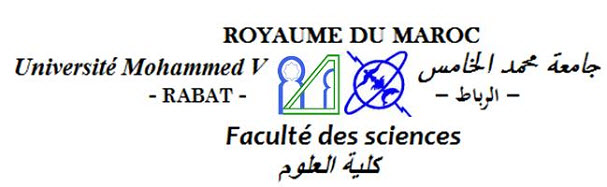}
	\end{center}
\end{figure}
\vspace{-1.5cm}
\begin{center}
	\begin{minipage}{17cm}
		\begin{center}
			{\textcolor{blue}{\fontfamily{pnc}{\selectfont
						{ \textit{CENTRE D’ETUDES DOCTORALES - SCIENCES ET TECHNOLOGIES}}
						\vskip .2cm
						\noindent\hrule height 2pt\vskip 0.2ex\nobreak
						{\textcolor{green}{
								\noindent\hrule height 2pt \vskip 0.2ex}}
			}}}
		\end{center}
	\end{minipage}
\end{center}
\vspace{-.8cm}
\begin{center}
	{\Large \textbf{Abstract}}
\end{center} 
\vspace{-.3cm}
\lettrine[lines=2]{}   This thesis focuses on quantum information processing using the superconducting device, especially, on realizing quantum gates and algorithms in open quantum systems. Such a  device is constructed by transmon-type superconducting qubits coupled to a superconducting resonator. For the realization of quantum gates and algorithms, a one-step approach is used. We suggest faster and more efficient schemes for realizing     $X$-rotation and entangling gates for two and three qubits. During these operations, the resonator photon number is canceled owing to the strong microwave field added. They do not require the resonator to be initially prepared in the vacuum state and the scheme is insensitive to resonator decay.  Furthermore, the robustness of these operations is demonstrated by including the effect of decoherence of transmon systems and the resonator decay in a master equation, high fidelity will be achieved on quantum simulation.  In addition,  Using the implemented x-rotation gates as well as phase gates, we present an alternative way for implementing Grover’s algorithm for two and three qubits, which does not require a series of single gates. As well, we demonstrate by numerically simulating the use of quantum process tomography to fully characterize the performance of a single-shot entangling gate for two and three qubits and obtaining the process fidelities greater than 93$\%$.  These gates are used to create   Bell and Greenberger-Horne-Zeilinger (GHZ) entangled states.

\underline{\bf Keywords:} Quantum  information  processing  and computation, Quantum  gates, Grover's search  algorithm , Entangling  gates, Bell and GHZ states , High  fidelity, superconducting  circuits.\\
\vspace{-1cm}
\begin{center}
	{\textcolor{black}{\fontfamily{pnc}{\selectfont
				\noindent\hrule height 1.1pt\vskip 0.2ex\nobreak
	}}}
\end{center}
\begin{center}
	{\Large \textbf{Résumé}}
\end{center}
Cette thèse se concentre sur le traitement de l'information quantique à l'aide d'un dispositif supraconducteur, en particulier sur la réalisation de portes quantiques et d'algorithmes dans des systèmes quantiques ouverts. Un tel dispositif est construit par des qubits supraconducteurs de type transmon couplés à un résonateur supraconducteur. Pour la réalisation des portes quantiques et des algorithmes, une approche en une seule  étape est utilisée. Nous proposons des schémas plus rapides et plus efficaces pour réaliser des portes de $X$-rotation  et des portes d'enchevêtrement pour deux et trois qubits. Au cours de ces opérations, le nombre de photons du résonateur est annulé en raison du fort champ de micro-ondes ajouté. Elles ne nécessitent pas que le résonateur soit initialement préparé dans l'état de vide et les schémas est insensible à la désintégration du résonateur.  En outre, la robustesse de ces opérations est démontrée en incluant l'effet de la décohérence des systèmes transmon et la désintégration du résonateur dans une équation maîtresse, ce qui permet d'obtenir une grande fidélité dans la simulation quantique.  En outre, en utilisant les portes de $X$-rotation  mises en œuvre ainsi que les portes de phase, nous présentons une autre façon de mettre en œuvre l'algorithme de Grover pour deux et trois qubits qui ne nécessite pas une série de portes simples. En outre, nous démontrons en simulant numériquement l'utilisation de la tomographie des processus quantiques pour caractériser pleinement la performance d'une porte d'enchevêtrement à un seul coup pour deux et trois qubits et nous obtenons des fidélités de processus supérieures à 93$\%$.  Ces portes sont utilisées pour créer des états intriqués de Bell et de Greenberger-Horne-Zeilinger (GHZ).

\underline{\bf Keywords:} Quantum  information  processing  and computation, Quantum  gates, Grover's search  algorithm , Entangling  gates, Bell and GHZ states , High  fidelity, superconducting  circuits.\\
	\vspace{.7cm}
\begin{center}
	\begin{minipage}{17cm}
		\begin{center}
			{\textcolor{black}{\fontfamily{pnc}{\selectfont
				{ Année Universitaire : {2022/2023}}
				\vspace{0.2cm}
				\noindent\hrule height 1.79pt\vskip 0.2ex\nobreak
				}}}
		\end{center}
	\end{minipage}

	{ \XBox} Faculté des Sciences, avenue Ibn Battouta, BP. 1014 RP, Rabat –Maroc\\
	\phone \hspace*{0.2cm}00212(0) 37 77 18 76, \hspace*{0.1cm} {\large \bell}Fax:\hspace*{0.1cm} 00212(0) 37 77 42 61 ; http://www.fsr.um5.ac.ma
\end{center}

\end{document}